%% file: MLG-24-002_temp.tex
\pdfoutput=1
\documentclass[11pt,twoside,a4paper,cmspaper,final,collab]{cms-tdr}
\def\svnVersion{812b30e}\def\svnDate{2026/05/15}\def\cmsCernNoTag{CERN-EP-2025-209}\def\cmsCernDate{\today}\def\cmsMessage{Published in Machine Learning: Science and Technology as \href{http://dx.doi.org/10.1088/2632-2153/ae6168}{\doi{10.1088/2632-2153/ae6168}.}}
\begin{document}\cmsNoteHeader{MLG-24-002}

\newcommand{\wnae}{WNAE\xspace}
\newcommand{\ptheta}{\ensuremath{p_{\theta}}\xspace}
\newcommand{\pdata}{\ensuremath{p_{\text{data}}}\xspace}
\newcommand{\ptd}{\ensuremath{\pt^{\mathrm{D}}}\xspace}
\newcommand{\ctwo}{\ensuremath{C_2^{(0.5)}}\xspace}
\newcommand{\tautwo}{\ensuremath{\tau_{2}}\xspace}
\newcommand{\tauthr}{\ensuremath{\tau_{3}}\xspace}
\newcommand{\Nc}{\ensuremath{N_{\text{c}}^\text{dark}}\xspace}
\newcommand{\Nf}{\ensuremath{N_{\text{f}}^\text{dark}}\xspace}
\newcommand{\lambdaDark}{\ensuremath{\Lambda_{\text{dark}}}\xspace}
\newcommand{\rinv}{\ensuremath{r_{\text{inv}}}\xspace}
\newcommand{\PMed}{\ensuremath{\Phi}\xspace}
\newcommand{\mMed}{\ensuremath{m_{\PMed}}\xspace}
\newcommand{\Pqdark}{\ensuremath{\chi}\xspace}
\newcommand{\mqdark}{\ensuremath{m_{\Pqdark}}\xspace}
\newcommand{\pidark}{\ensuremath{\pi_{\text{dark}}}\xspace}
\newcommand{\pidarkdm}{\ensuremath{\pi_{\text{dark}}^{\text{DM}}}\xspace}
\newcommand{\rhodark}{\ensuremath{\rho_{\text{dark}}}\xspace}
\newcommand{\rhodarkdm}{\ensuremath{\rho_{\text{dark}}^{\text{DM}}}\xspace}
\newcommand{\xxi}{\ensuremath{x_{i}}\xspace}
\newcommand{\xxrecoi}{\ensuremath{\hat{x}_{\theta,i}}\xspace}
\newcommand{\xxj}{\ensuremath{x^{(j)}}\xspace}
\newcommand{\xprime}{\ensuremath{x^{\prime}}\xspace}
\newcommand{\lreco}{\ensuremath{l_{\theta}}\xspace}
\newcommand{\lossAE}{\ensuremath{\mathcal{L}_{\text{AE}}}\xspace}
\newcommand{\lossNAE}{\ensuremath{\mathcal{L}_{\text{NAE}}}\xspace}
\newcommand{\PYTORCH}{\textsc{PyTorch}\xspace}
\newcommand{\spaceInput}{\ensuremath{\mathcal{X}}\xspace}
\newcommand{\spaceLatent}{\ensuremath{\mathcal{Z}}\xspace}
\newcommand{\Rm}{\ensuremath{\mathbb{R}^{m}}\xspace}
\newcommand{\Rd}{\ensuremath{\mathbb{R}^{d}}\xspace}
\newcommand{\fEnc}{\ensuremath{f_{e}}\xspace}
\newcommand{\fDec}{\ensuremath{f_{d}}\xspace}
\newcommand{\supSig}{\ensuremath{\mathcal{S}}\xspace}
\newcommand{\supBkg}{\ensuremath{\mathcal{B}}\xspace}
\newcommand{\supEx}{\ensuremath{\mathcal{E}}\xspace}
\newcommand{\pOne}{\ensuremath{p_{1}}\xspace}
\newcommand{\pTwo}{\ensuremath{p_{2}}\xspace}
\newcommand{\NMCMC}{\ensuremath{N_{\text{MCMC}}}\xspace}
\newcommand{\Eplus}{\ensuremath{E_+}\xspace}
\newcommand{\Eminus}{\ensuremath{E_-}\xspace}
\newcommand{\Etheta}{\ensuremath{E_\theta}\xspace}
\newcommand{\Otheta}{\ensuremath{\Omega_\theta}\xspace}
\newcommand{\deltheta}{\ensuremath{\nabla_\theta}\xspace}
\newcommand{\delx}{\ensuremath{\nabla_x}\xspace}
\newcommand{\Expdata}{\ensuremath{\mathbb{E}_{x\sim \pdata}}\xspace}
\newcommand{\Exptheta}{\ensuremath{\mathbb{E}_{x\sim \ptheta}}\xspace}
\newcommand{\Expptheta}{\ensuremath{\mathbb{E}_{\xprime\sim \ptheta}}\xspace}

\cmsNoteHeader{MLG-24-002}
\title{Wasserstein normalized autoencoder for anomaly detection}

\date{\today}

\abstract{
A novel anomaly detection algorithm is presented. The Wasserstein normalized autoencoder (WNAE) is a normalized probabilistic model that minimizes the Wasserstein distance between the learned probability distribution---a Boltzmann distribution where the energy is the reconstruction error of the autoencoder---and the distribution of the training data. This algorithm has been developed and applied to the identification of semivisible jets---conical sprays of visible standard model particles and invisible dark matter states---with the CMS experiment at the CERN LHC. Trained on jets of particles from simulated standard model processes, the WNAE is shown to learn the probability distribution of the input data in a fully unsupervised fashion, such that it effectively identifies new physics jets as anomalies. The model exhibits stable, convergent training and recovers strong classification performance for a wide range of signals against the selected background process, for which a standard autoencoder fails because of outlier reconstruction. In addition, the model improves upon standard normalized autoencoders while remaining fully agnostic to the signal. The WNAE directly tackles the problem of outlier reconstruction, a common failure mode of autoencoders in anomaly detection tasks.
}

\hypersetup{%
pdfauthor={CMS Collaboration},%
pdftitle={Wasserstein normalized autoencoder for anomaly detection},%
pdfsubject={CMS},%
pdfkeywords={CMS, anomaly detection, autoencoder, dark matter, semivisible jets}} % limit six total

\maketitle 

\section{Introduction}
\label{sec:introduction}

Unsupervised machine-learning algorithms have proven to be powerful tools in the search for new physics at the LHC~\cite{Heimel:2018mkt, Farina:2018fyg}. They can effectively separate events from known standard model (SM) processes (background) from potential signal events arising from interactions beyond the SM (BSM), without relying on the simulation of specific signal hypotheses. Autoencoders (AEs) are frequently used for unsupervised learning tasks. One of the main challenges when using AE-based unsupervised machine-learning algorithms for anomaly detection is preventing the reconstruction of outliers---examples that are not part of the training data. A well-known example of outlier reconstruction called complexity bias~\cite{Finke:2021sdf} occurs when outliers are less complex than the training data. In this paper, we introduce the Wasserstein normalized autoencoder (\wnae), an evolution of the normalized autoencoder (NAE) algorithm~\cite{pmlr-v139-yoon21c}, to solve these challenges. The identification of semivisible jets (SVJs)~\cite{Cohen:2015toa} in the CMS experiment~\cite{CMS:2008xjf} is used as a case study to demonstrate the effectiveness of the \wnae in separating signal from background events.

The CMS apparatus~\cite{CMS:2008xjf,CMS:2023gfb} is a multipurpose, nearly hermetic detector, designed to trigger on~\cite{CMS:2020cmk,CMS:2016ngn,CMS:2024aqx} and identify electrons, muons, photons, and (charged and neutral) hadrons~\cite{CMS:2020uim,CMS:2018rym,CMS:2014pgm}. A global ``particle-flow'' algorithm~\cite{CMS:2017yfk} aims to reconstruct all individual particles in an event, combining information provided by the all-silicon inner tracker and by the crystal electromagnetic and brass-scintillator hadron calorimeters, operating inside a 3.8\unit{T} superconducting solenoid, with data from the gas-ionization muon detectors embedded in the flux-return yoke outside the solenoid. The reconstructed particles are used to build jets~\cite{CMS:2016lmd}---conical sprays of particles---and missing transverse momentum~\cite{CMS:2019ctu}---the imbalance of visible momentum in the transverse plane. Events of interest are selected using a two-tiered trigger system. The first level, composed of custom hardware processors, uses information from the calorimeters and muon detectors to select events at a rate of around 100\unit{kHz} within a fixed latency of 4\mus~\cite{CMS:2020cmk}. The second level, known as the high-level trigger, consists of a farm of processors running a version of the full event reconstruction software optimized for fast processing, and reduces the event rate to a few kHz before data storage~\cite{CMS:2016ngn,CMS:2024aqx}.

The paper is organized as follows. Section~\ref{sec:ae_ad} discusses the use and limitations of AEs as anomaly detection tools, focusing on the outlier reconstruction problem, with a brief review of the NAE algorithm. In Section~\ref{sec:motivation_wnae}, the AE and NAE algorithms are applied to the search for SVJs, and we introduce a metric based on the Wasserstein distance~\cite{Kantorovich:1939,Vaserstein:1969,villani_optimal_2009} to quantify the degree of outlier reconstruction without relying on the signal in any form during training. Section~\ref{sec:wae} presents the major advance in this paper: the use of this metric as the direct objective for the autoencoder training to build a more robust anomaly detection algorithm, the \wnae. The advantages of the \wnae over the NAE are discussed. Finally, Section~\ref{sec:summary} discusses the implications of the findings and possible future directions.

\section{Autoencoder-based anomaly detection}
\label{sec:ae_ad}

\subsection{Autoencoders}
\label{sec:ae}

Autoencoders~\cite{KRAMER1992313} are a class of neural networks that are intended to learn a compressed representation of the input data. This is usually achieved by mapping the input feature space $\spaceInput \subset \Rd$ to a lower-dimensional latent space $\spaceLatent \subset \Rm$ via an encoder network $\fEnc: \Rd \mapsto \Rm$ ($m < d$) and mapping this latent space to an output space with the same dimension as the input space via a decoder network $\fDec: \Rm \mapsto \Rd$.
Autoencoders are typically trained by minimizing a loss function that penalizes the difference between the input and output, referred to as the reconstruction error of the AE. The latter is denoted \lreco in the following, where $\theta$ represents the parameters of the AE network.

Crucially, when AEs are used for anomaly detection, the network is only exposed to non-anomalous (background) examples during training. If the network is able to learn a compressed representation of these examples, it can reconstruct them with low error. However, since the anomalous (signal) examples are not part of the training set, the AE is expected not to reconstruct them well, resulting in a larger average reconstruction error. 

While this reasoning holds for many practical applications of AEs for anomaly detection, it does not follow in general that achieving low reconstruction error on the background examples is sufficient for an AE to effectively identify anomalies, as discussed below. This shortcoming is addressed by the \wnae algorithm introduced in this paper.

\subsection{The problem of outlier reconstruction}
\label{sec:ood}

Given a data distribution $\pdata$, an example datum $x \in \spaceInput$ is an outlier if it lies in a region with probability density lower than a given threshold $\tau$: $\pdata(x) \leq \tau$. In the extreme case where $\tau = 0$, the outlier lies outside of the support of the data distribution.

\begin{figure}[htbp]
    \centering
    \includegraphics[width=0.9\textwidth]{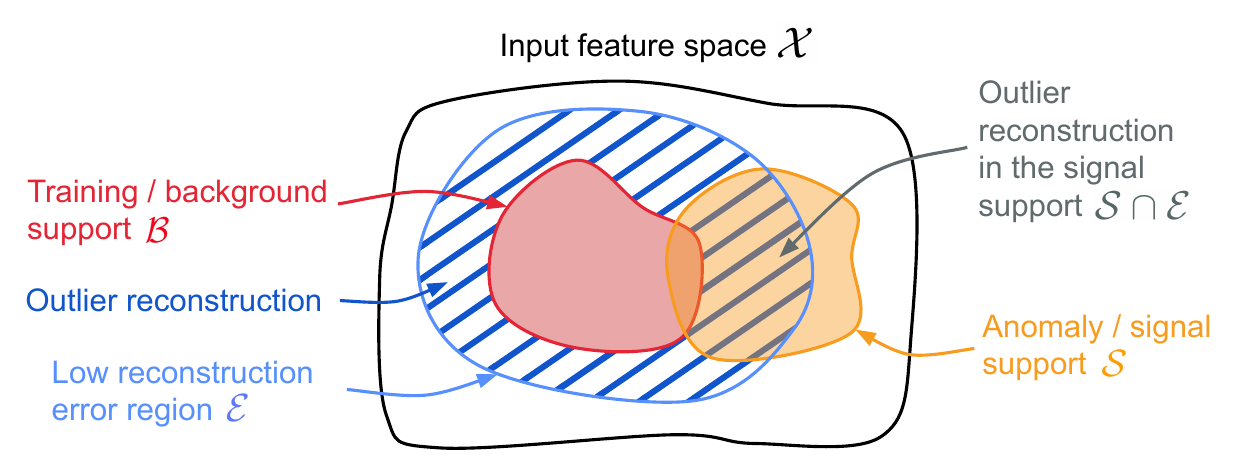}
    \caption{Schematic visualization of the outlier reconstruction failure mode. Signal samples drawn from the hatched area are reconstructed well by the AE, despite not being part of the training set, and thus are not separated from the background. The AE training is assumed to have converged such that the background is reconstructed well.}
    \label{fig:outlier_reconstruction}
\end{figure}

Outlier reconstruction occurs when an AE, while learning to reconstruct the training examples, also learns to reconstruct outlier examples. 
More formally: let \supBkg and \supSig be the supports of the background and signal distributions in the input feature space \spaceInput, respectively. The AE is trained to achieve minimum reconstruction error on background samples $x_b\in\supBkg$. Let \supEx be the region where the AE achieves reconstruction error below some threshold. Since there is no constraint on the reconstruction error of signal samples $x_s\in\supSig$, there can be a sizable overlap $\supSig\cap \supEx$. This overlap can in general be at least partly disjoint from the training data: $(\supSig\cap \supEx) - \supBkg\neq \emptyset$. Examples drawn from this region of phase space are assigned low reconstruction error by the AE, even though they are outside the training data. Thus, these examples and the background both have low reconstruction error, decreasing the performance of the reconstruction error as a discriminator. This is illustrated in Fig.~\ref{fig:outlier_reconstruction}.

Ideally, the reconstruction error \lreco should be a function of the probability density of the training data: the lower the probability density of the training data, the higher the reconstruction error. Thus, examples $x$ with $\pdata(x) \leq \tau$ must have reconstruction error above a threshold $\tau_r$: $\lreco(x) \geq \tau_r$. The reconstruction error itself can accordingly be used as a summary statistic to find anomalies. The objective of the autoencoder is therefore to learn the distribution of the training data (the ``inlier'' distribution), rather than simply to minimize the reconstruction error.

\subsection{The normalized autoencoder paradigm}
\label{sec:nae}

The NAE~\cite{pmlr-v139-yoon21c} is based on a proposed mechanism to learn the probability distribution of the training data. In analogy with energy-based models~\cite{ebm1,ebm2,ebm3}, often used for generative modeling, the NAE is assigned a normalized probability distribution \ptheta, parametrized by the AE network weights $\theta$. Following the maximum entropy principle~\cite{Jaynes:1957zza}, the probability distribution is chosen as the Boltzmann distribution from statistical mechanics:
\begin{equation}
    \ptheta(x) = \frac{1}{\Otheta}\exp{\left(-\Etheta(x)\right)},
\end{equation}
where \Etheta is the energy of the system and \Otheta is the partition function:
\begin{equation}
    \Otheta = \int \rd x \exp(-\Etheta(x)).
\end{equation}
The energy of the system is defined as the reconstruction error of the AE:
\begin{equation}
    \Etheta(x) = \lreco(x).
\end{equation}
The problem is then recast as a maximum likelihood estimation problem, where the goal is to find the weights $\theta$ that maximize the likelihood of the training data, with probability distribution denoted \pdata. As is common, the negative log likelihood is minimized instead:
\begin{equation}
    \Expdata[-\log \ptheta(x)] = \Expdata[\Etheta(x)] + \log \Otheta.
\end{equation}
Calculating the partition function $\Omega_\theta$ is in general intractable, but the problem can be circumvented when using gradient descent to find the optimum, since the gradient of the partition function can be calculated:
\begin{align}
   \deltheta \log \Otheta &= \frac{1}{\Otheta}\deltheta \Otheta = \frac{1}{\Otheta}\int_\mathcal{B} \rd x \deltheta \exp(-\Etheta(x)) \\
   &= \int \rd x \frac{1}{\Otheta}\exp(-\Etheta(x))\deltheta (-\Etheta(x)) \\
   &= \int \rd x \ptheta(x)\deltheta (-\Etheta(x)) \\
   &= -\Exptheta[\deltheta \Etheta(x)],
\end{align}
which is the expectation value of the gradient of the AE reconstruction error over the probability distribution \ptheta. The gradient of the loss can thus be calculated by sampling from the probability distribution \ptheta. This can be done via a Markov chain Monte Carlo (MCMC) algorithm. The application of the MCMC in the NAE training procedure is discussed in more detail in Section~\ref{sec:nae_training} and Appendix~\ref{sec:mcmc_tuning}. The loss function to be minimized is then:
\begin{equation}
    \lossNAE = \Expdata[\Etheta(x)] - \Expptheta[\Etheta(\xprime)].
\label{eq:nae_loss}
\end{equation}
Both terms in Eq.~\eqref{eq:nae_loss} are expectation values of the reconstruction error of the AE. The first term is the expectation value over the training data $x$ (the usual AE loss), referred to as the positive energy. The second term is the expectation value over samples \xprime drawn from the probability distribution \ptheta, referred to as the negative energy. The NAE loss is thus the difference between the positive and negative energy, denoted \Eplus and \Eminus, respectively:
\begin{equation}
    \lossNAE \equiv \Eplus - \Eminus.
    \label{eq:nae_loss_2}
\end{equation}
In analogy, the training examples are referred to as positive examples, and examples drawn from \ptheta are referred to as negative examples. In this way, the NAE is not trained just to reconstruct background examples well, but rather to learn the probability distribution of the training data. The NAE is therefore penalized for having low reconstruction error in regions with low probability density of the training data, hence suppressing the reconstruction of examples outside the background support $\mathcal{B}$. It is worth emphasizing that the whole procedure is still fully unsupervised. The NAE is trained only on background examples, and the signal is not used in any form during training.

\subsection{Markov chain Monte Carlo}
\label{sec:mcmc}

The key step in the normalized autoencoder paradigm is to estimate the autoencoder probability \ptheta. This is achieved by running an MCMC algorithm to obtain a set of examples following the \ptheta distribution. The Markov chain is only needed to train the network and is not used for inference. The Langevin Monte Carlo algorithm is employed in this work. First, a starting set of points $\left\{x\right\}_0$ is drawn from an initial distribution, following the procedure laid out in Section~\ref{sec:nae_training}. The negative samples are obtained by repeating a fixed number of steps, each step proceeding as follows:
\begin{equation}
    \label{eq:langevin_dynamics}
    x_{i+1} = x_{i} + \lambda \delx \log{\ptheta\left(x_{i}\right)} + \sigma \epsilon_{i},
\end{equation}
where $\lambda$ is the step size, $\sigma$ is the noise coefficient, $\epsilon_{i}$ is an independent draw from a normal distribution on $\Rd$ with mean 0 and covariance matrix equal to the $d \times d$ identity matrix, and $d$ is the number of dimensions of the input feature space.
The Markov chain has therefore the following dependence on the reconstruction error of the autoencoder:
\begin{equation}
    \label{eq:mcmc_eq_with_reco_error}
    x_{i+1} = x_{i} - \lambda \delx \Etheta\left(x_{i}\right) + \sigma \epsilon_{i}.
\end{equation}
With $\rho_{i}$ denoting the probability distribution of the sample $\left\{x\right\}_i$, choosing:
\begin{equation}
    \label{eq:mcmc_theoretical_relation}
    \sigma = \sqrt{2\lambda},
\end{equation}
and assuming an infinite number of steps, the probability $\rho_{i}$ converges towards the target probability \ptheta: $\lim_{i\to\infty} \rho_{i} = \ptheta$. In practice, \ptheta needs to be estimated at every gradient descent step, once per batch, while training an NAE. For that reason, the number of Markov chain steps cannot be too large, and the probability \ptheta can only be approximated. In order to enhance the gradient term in Eq.~\eqref{eq:langevin_dynamics}, the temperature $T$ is introduced:
\begin{equation}
    \label{eq:langevin_dynamics_with_T}
    x_{i+1} = x_{i} + \frac{\lambda}{T} \delx \log{\ptheta\left(x_{i}\right)} + \sigma \epsilon_{i}.
\end{equation}
Temperatures $T<1$ result in an effective increase of the gradient term.

\subsection{Related work}
\label{sec:related_work}

There has long been interest in applying autoencoders to model a data distribution, mostly for generative tasks.
Variational autoencoders (VAE)~\cite{Kingma2014} make use of the Kullback-Leibler divergence to impose a Gaussian prior in the latent space \spaceLatent and the decoder \fDec learns to model the data distribution from this Gaussian prior.
Adversarial autoencoders (AAE)~\cite{44904} minimize the reconstruction error and enforce a Gaussian prior in the latent space by training the network in an adversarial way, similarly to generative adversarial networks (GANs)~\cite{NIPS2014_f033ed80}.
The encoder \fEnc plays the role of the generator network and a discriminator network estimates the probability that a latent sample arises from the prior distribution rather than being the output of the generator $\fEnc(x)$, where $x$ is a training datum.
In Wasserstein autoencoders (WAE)~\cite{tolstikhin2018wasserstein}, the loss function is composed of two terms: a divergence between the latent space \spaceLatent and a (Gaussian) prior, and the Wasserstein distance between the training data distribution and that learned by the AE (the output of the decoder \fDec).
The Wasserstein distance between two probability distributions \pOne and \pTwo is defined as:
\begin{equation}
    \label{eq:emd_primal}
    W(\pOne, \pTwo) = \inf_{\gamma \in \Pi(\pOne, \pTwo)} \mathbb{E}_{(x, \xprime)\sim \gamma}\left[\|x - \xprime\|\right],
\end{equation}
where $\Pi(\pOne, \pTwo)$ denotes the set of all joint distributions whose marginal distributions are \pOne and \pTwo and $\inf$ indicates the infimum over this set. This distance measures the cost of morphing \pOne into \pTwo. 
Using the Kantorovich-Rubinstein duality~\cite{kantorovich1958space}, this expression can be written in the dual form as:
\begin{equation}
    \label{eq:emd_dual}
    W(\pOne, \pTwo) = \sup_{f \in \mathcal{F_L}} \mathbb{E}_{x\sim \pOne}\left[f(x)\right] - \mathbb{E}_{x\sim \pTwo}\left[f(x)\right],
\end{equation}
where the supremum is over all 1-Lipschitz functions $\mathcal{F_L}$.
The Wasserstein distance in Eqs.~\eqref{eq:emd_primal} and~\eqref{eq:emd_dual} is typically approximated by relaxing the constraint on the coupling distribution $\gamma$ or the function $f$ and replacing it with a simpler constraint or regularizing the loss function. In Wasserstein GANs~\cite{Arjovsky:2017fjr}, the 1-Lipschitz constraint is approximated by implementing $f$ as a neural network and clipping its weights. In WAEs, the constraint on the coupling distribution $\gamma$ is relaxed and the loss function is regularized by adding this term to the loss function: $\lambda [D_f(\gamma_1, \pOne) + D_f(\gamma_2, \pTwo)]$, where $\lambda$ is a positive constant, $\gamma_1$ and $\gamma_2$ are the marginal distributions of $\gamma$, and $D_f$ denotes an $f$-divergence~\cite{tolstikhin2018wasserstein}.
An exact solution to the optimal transport problem in Eq.~\eqref{eq:emd_primal} is implemented in Ref.~\cite{flamary2021pot}, but does not scale well to large data sets. Nevertheless, computing the Wasserstein distance on minibatches of the entire data set is efficient and was found to provide an unbiased estimator of the gradients~\cite{fatras2019learning}.

As explained in Section~\ref{sec:ood}, an ideal outlier detection algorithm must learn the training data distribution.
For that reason, and as further highlighted in Section~\ref{sec:ae_training}, the vanilla autoencoder defined in Section~\ref{sec:ae} is not a suitable algorithm for anomaly detection.
While the reconstruction error of VAEs can directly be used as an anomaly score, VAEs have been shown to be prone to outlier reconstruction~\cite{nalisnick2018do,xiao2020likelihood}.

Normalizing flows present a similar shortcoming. This class of networks learns to model complex probability distributions by mapping a simple distribution to a chosen complex distribution via a series of invertible and differentiable transformations. This procedure can be used to compute the likelihood of the data, which can serve as an anomaly detection score. However, normalizing flows can assign high likelihoods to outliers, hindering their ability to be used as anomaly detection algorithms~\cite{nalisnick2018do,kirichenko2020why}. Though recent work has tried mitigating this issue~\cite{kamkari2024geometric,Cook2024Feature}, a general solution is not known.

Algorithms based on GANs, often constructed similarly to AAEs, need to be modified to provide an anomaly score. Generative probabilistic novelty detection~\cite{pidhorskyi2018generative}, GANomaly~\cite{akcay2018ganomaly}, and AnoGAN~\cite{schlegl2019f} follow different approaches, but all are challenging to train because of the iterative adversarial procedure~\cite{NIPS2014_f033ed80}.

The NAE, described in Section~\ref{sec:nae}, is based on a different principle: it relies on an MCMC algorithm to sample data points from the probability distribution learned by the network.
Though it is a normalized probabilistic model, its loss function takes the form of a minimax rule, which leads to numerical instability and therefore requires ad-hoc regularization~\cite{NEURIPS2019378a063b,pmlr-v139-yoon21c}. In addition, as detailed in Section~\ref{sec:nae_training}, the NAE has a failure mode that can cause outlier reconstruction.

Since the first results using AEs to search for new physics at the LHC~\cite{Heimel:2018mkt, Farina:2018fyg}, there has been active research on this topic. The CMS experiment has implemented two realtime anomaly detection algorithms in the level-1 trigger to select new physics signatures: one based on a VAE using global trigger information, and another based on a convolutional AE using calorimeter trigger information~\cite{Gandrakota:2024yqs}. The CMS experiment has recently conducted a search for new physics in dijet resonances with anomalous jet substructure using five different anomaly detection algorithms, two of which are based on AEs~\cite{CMS:2024nsz}. The ATLAS experiment looked for new physics in the invariant mass distribution of pairs of SM objects consisting of one light-flavor or \PQb~jet, and a lepton, a photon, or a second light-flavor or \PQb~jet, using an AE to select anomalous data~\cite{ATLAS:2023ixc}. NAEs were first used in high-energy physics in the context of jet tagging for new physics searches~\cite{Dillon:2022mkq}.

\section{Case studies}
\label{sec:motivation_wnae}

\subsection{Data set generation}
\label{sec:svj}

Astronomical observations~\cite{Rubin:1980zd,Persic:1995ru,Clowe:2006eq,Chang:2017kmv,Planck:2018vyg} provide compelling evidence for dark matter (DM) in the universe. However, particles that could constitute the DM are not found in the SM of particle physics. If DM particles can be produced in high-energy collisions at the LHC by new physics processes, their existence can be inferred from the signature they leave in detectors. This work considers a specific class of models of new physics, described below.

Hidden valley models~\cite{Strassler:2006im} are new physics scenarios in which the SM is expanded with a dark sector that has its own particles and forces. The dark sector particles are not charged under any SM gauge group and hence are not directly detectable. Similarly, SM particles are not charged under the dark gauge group. Hidden valley theories may, under certain conditions, result in SVJs as an experimental signature~\cite{Cohen:2015toa}, where some of the particles in the final state are not detectable. If the dark sector has a new non-Abelian $SU(N)$ confining interaction, it would exhibit showering and hadronization qualitatively similar to the SM quantum chromodynamics (QCD). In analogy with the SM sector, the new confining force and the new particles charged under it are referred to as dark QCD and dark quarks, respectively. More specifically, we consider a dark QCD interaction $SU(\Nc)$, where \Nc denotes the number of dark colors, with a confinement scale \lambdaDark and \Nf flavors of dark quark \Pqdark with a mass \mqdark, which communicates with the SM via a mediator \PMed with mass \mMed. In such a model, QCD-like showers occur when $\mMed \gg \lambdaDark$ and $\mqdark \sim \lambdaDark$. The dark quarks therefore form bound states called dark hadrons. The unstable dark hadrons, \pidark and \rhodark, decay back to SM particles to produce a visible signature, depending on the symmetries of the theory. The stable dark hadrons, \pidarkdm and \rhodarkdm, are invisible and escape detection, giving rise to missing transverse momentum. The ensemble of stable dark hadrons and the decay products of unstable dark hadrons from the dark QCD shower constitutes an SVJ. The fraction of stable dark hadrons in the dark QCD shower can be parametrized by the invisible fraction \rinv:
\begin{equation}
    \rinv = \left\langle \frac{\text{Number of stable dark hadrons}}{\text{Total number of dark hadrons}} \right\rangle.
\end{equation}

\begin{figure}[t!]
    \centering
    \includegraphics[width=0.49\textwidth]{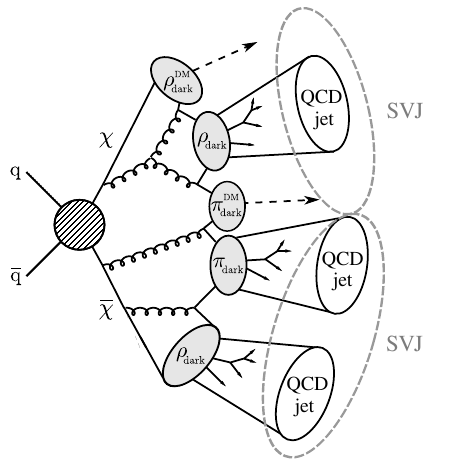}
    \caption{An illustration of collider SVJ production. The dashed black arrows indicate stable, undetectable DM candidate particles. Adapted from Ref.~\cite{Bernreuther:2019pfb}.}
    \label{fig:svj_illustration}
\end{figure}

Figure~\ref{fig:svj_illustration} illustrates the production of SVJs. The CMS and ATLAS Collaborations have published searches for resonant production of pairs of SVJs~\cite{CMS:2021dzg,ATLAS:2025kuz}. The production of SVJs via $t$-channel interactions through a bifundamental scalar mediator, charged under both SM QCD and dark QCD, is also possible~\cite{Cohen:2017pzm}, with one published search from ATLAS~\cite{ATLAS:2025kuz}. The latter production mode is considered in this paper.

The SVJs may have different radiation patterns than SM jets, which can be exploited to identify them. The details depend on the exact spectrum of the dark hadrons, which in turn depends on \Nc, \Nf, \lambdaDark, the masses \mqdark, and the nonperturbative physics in the dark sector, all of which are unknown. Therefore, anomaly detection, which does not require a detailed simulation of every target signal hypothesis in this large parameter space, is a powerful complement to traditional, supervised searches for SVJs, such as Ref.~\cite{CMS:2021dzg}.

Simulated events are used to train and evaluate the performance of the machine-learning models discussed in this paper.
The background considered in this study consists of top-antitop quark production (\ttbar) with additional jets, where each top quark decays either hadronically or leptonically.
Background and SVJ signal events are generated at the parton level with \MGvATNLO 2.6.5~\cite{Alwall:2014hca}, final-state partons are hadronized with \PYTHIA 8.240~\cite{Sjostrand:2014zea}, and the interactions of the resulting particles with the CMS detector~\cite{CMS:2008xjf} are simulated using \GEANTfour~\cite{Agostinelli:2002hh}. The NNPDF3.1 parton distribution functions are used for all processes~\cite{NNPDF}. The hadronization in the dark sector during the signal generation follows the procedure described in Ref.~\cite{CMS:2021dzg}, and all dark hadrons have a fixed mass, 20\GeV. Because of the double hadronization, first in the dark sector and then in the SM sector, the substructure of SVJs is expected to be different from that of SM jets. In particular, SVJs are expected to be wide, and therefore jets are reconstructed using the anti-\kt algorithm~\cite{Cacciari:2008gp, Cacciari:2011ma} with radius parameter $R=0.8$. Each jet is represented by the following features, describing its substructure:
\begin{itemize}
    \item the major and minor axes~\cite{CMS-PAS-JME-13-002}, which characterize the elliptical shape of the jet in the $\eta$-$\phi$ plane (where $\eta$ is the pseudorapidity and $\phi$ is the azimuthal angle);
    \item the first energy flow polynomial (EFP1)~\cite{Komiske:2017aww}, a multiparticle energy correlator that directly results from infrared and collinear safety;
    \item the \ctwo energy correlation function~\cite{Larkoski:2013eya}, a ratio of two ratios of energy correlation functions that can be used to determine if a jet has two hard subjets;
    \item the transverse momentum dispersion \ptd~\cite{CMS-PAS-JME-13-002}, which describes the spread of transverse momenta of the jet constituents;
    \item the softdrop mass~\cite{Larkoski:2014wba}, the mass of the jet after soft wide-angle radiation has been removed; and
    \item the 2- and 3-subjettiness \tautwo and \tauthr~\cite{Thaler:2010tr}, describing the compatibility of the jet with a two- or three-prong structure, respectively;
\end{itemize}
for a total of $d = 8$ inputs to the AE. These input features are chosen to exploit potential differences between SM jets and BSM jets in shape (major and minor axes), energy distribution (\ptd and EFP1), prong structure (\ctwo, \tautwo, \tauthr), and mass of the high-momentum core of the jet (softdrop mass).
The distributions of these features in the physical space, before any preprocessing, can be found in Appendix~\ref{sec:input_features}.
These features represent a space of physical variables that facilitates the classification task by providing some separation between signal and background, even though a single boundary in this space, unbiased across all possible signal models, may be difficult to define.
For each event, the two jets with highest transverse momenta are used for training and evaluation. In addition, among the two selected jets, only jets identified via generator-level information as originating from dark-sector particles are considered as anomalous signal jets when evaluating the network performance. The inputs are preprocessed to follow a normal distribution using quantile normalization, as implemented in the \textsc{scikit-learn} package~\cite{Pedregosa:2012toh}. 

\subsection{Training setup}

The architecture is a fully connected network with five hidden layers. All layers have ten nodes, except for the third one, which, as the bottleneck, has six.
The performance of the different models discussed in this paper was found to have little dependence on either the number of layers or the number of nodes per layer; the listed choices are found to produce a sufficiently expressive network for this anomaly detection task.
The reconstruction of the AE is parametrized by the network weights $\theta$. The reconstruction error \lreco for an example $x$ is defined as the mean squared error between the features \xxi of $x$ and the features \xxrecoi of the reconstructed output:
\begin{equation}
    \lreco(x) = \frac{1}{d}\sum_{i=1}^d \left(\xxi - \xxrecoi\right)^2.
\end{equation}
The loss function of the AE is the reconstruction error averaged over the batch:
\begin{equation}
    \lossAE = \frac{1}{n}\sum_{j=1}^n \lreco(\xxj),
\end{equation}
where \xxj denotes the example in position $j$ in the batch and $n$ is the number of training examples in the batch. The background data set is split into three independent parts: a training set on which the loss is minimized (70\%), a validation set used to monitor the training (15\%), and a test set used to evaluate the performance of the network (15\%). The background data set has a total of 47000 jets and the number of trainable parameters is 534. The AE is trained until the loss function evaluated on the validation set ceases to improve. The AE is then tested using multiple SVJ signal models, as well as the background test sample.

\subsection{Autoencoder training}
\label{sec:ae_training}

The issue of outlier reconstruction is found to be especially impactful in the case of the SVJ search. While previous studies have shown that AEs are an effective tool for the task of discriminating between the signal and the SM background from light-flavor jets~\cite{Canelli:2021aps}, the performance is found to degrade significantly when jets originating from \PQt quarks in the \ttbar process are included. This is shown by training an AE solely on the \ttbar background. The cause is understood to be outlier reconstruction, based on the nature of \ttbar events. Boosted jets from high-momentum top quarks, which contain multiple prongs and genuine missing transverse momentum from neutrinos, are the closest SM analog to SVJs. These events span multiple topologies, as they can also include boosted \PW boson jets, light-flavor quark jets, and gluon jets from initial- or final-state radiation.

\begin{figure}[t!]
    \centering
    \includegraphics[height=0.43\textwidth]{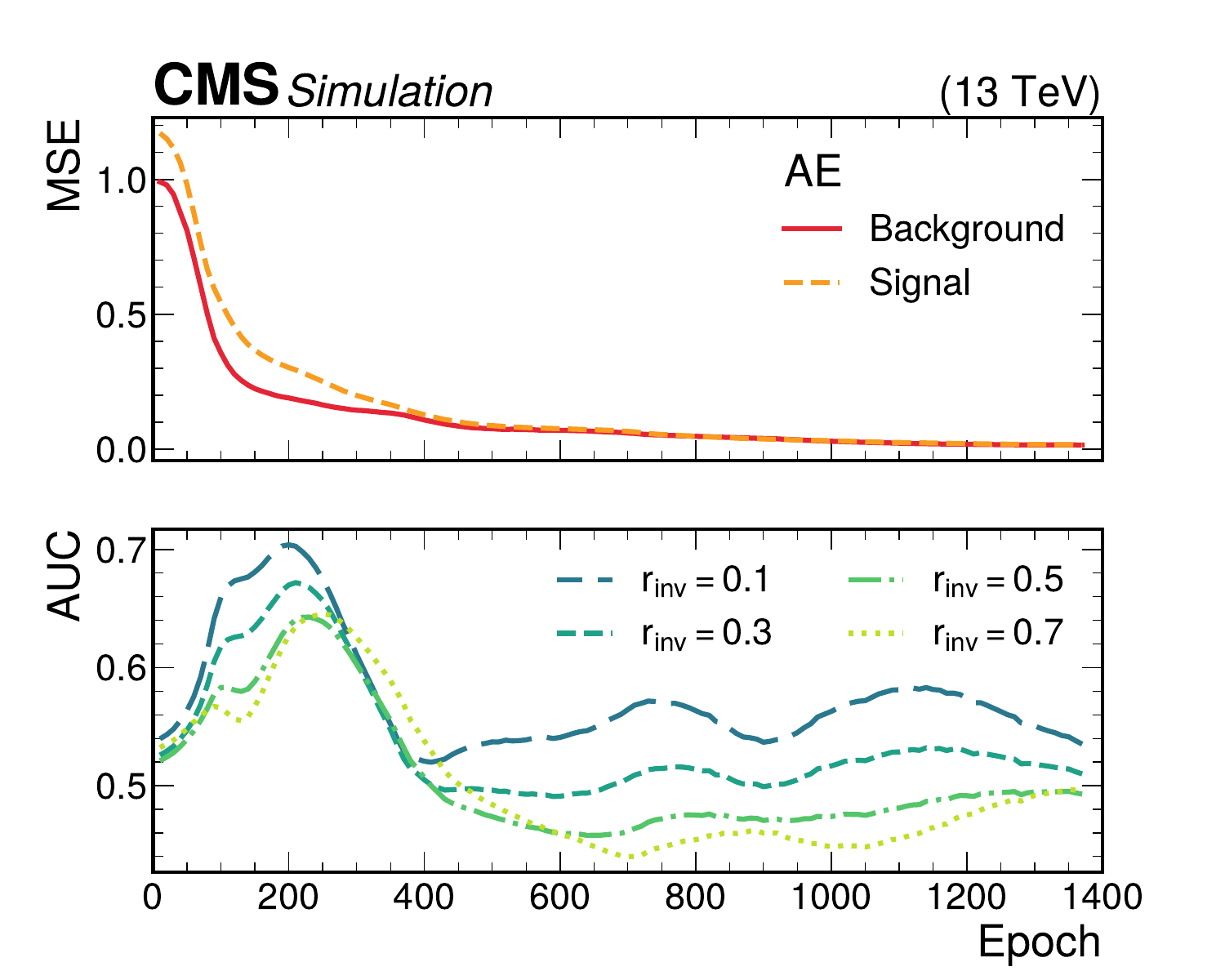}
    \includegraphics[height=0.43\textwidth]{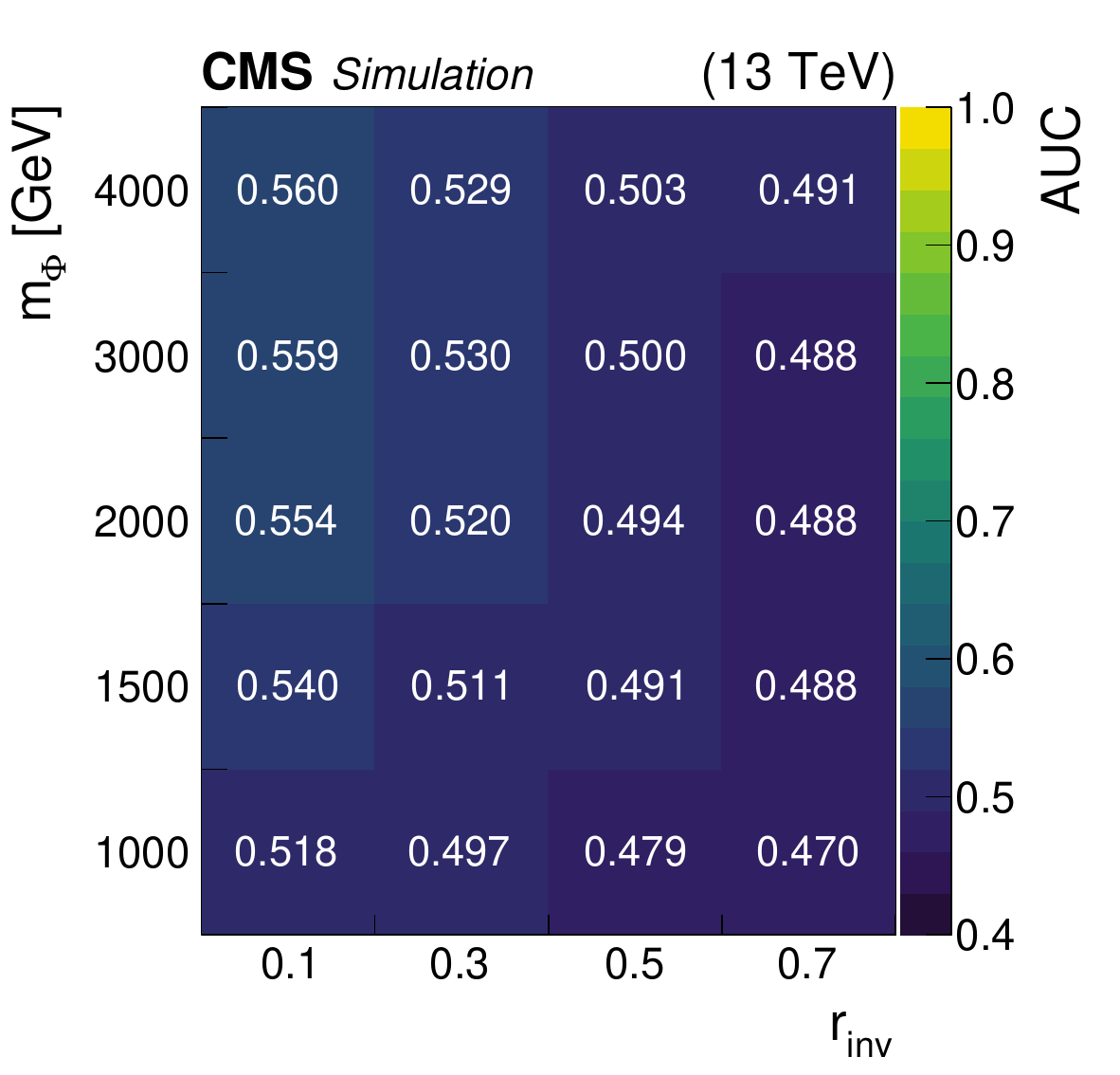}
    \caption{Left: the reconstruction error (upper panel) and the AUC scores (lower panel) for the AE trained on the \ttbar background, evaluated during each training epoch on \ttbar background jets and signal models with $\mMed = 2000\GeV$ and $\rinv = 0.3$ (upper) or $\rinv = 0.1, 0.3, 0.5, 0.7$ (lower). Right: The AUC scores for the same AE, evaluated for the epoch with the minimal background reconstruction error, for the classification of several SVJ signal hypotheses against the \ttbar background. The AUC scores are close to 0.5, indicating that the AE is unable to discriminate between the SVJ signal and the \ttbar background.}
    \label{fig:ae_auc_ttbar}
\end{figure}

The receiver operating characteristic curve compares the true- and false-positive rates for the AE reconstruction error used as a discriminator, and the area under the curve (AUC) is used to quantify the performance of this discriminator. As shown in Fig.~\ref{fig:ae_auc_ttbar}, when achieving minimum reconstruction error on the \ttbar background, the AE is unable to discriminate between the SVJ signal and the \ttbar background, and the AUC score is close to 0.5. In order to check that this behavior does not occur because the bottleneck was too large, the same experiment was repeated with a bottleneck of only five, four, three, or two nodes. The results were found to be consistent with a bottleneck size of six, indicating that the issue does not arise from insufficient compression of the data.

The AUC scores evolve in nontrivial ways throughout the training. As shown in Fig.~\ref{fig:ae_auc_ttbar}, the AUC score improves during the first stages of the training, indicating that the AE is performing as desired. However, after further training, the AUC score starts to degrade, and eventually converges to a value close to 0.5. This is a clear indication that the requirement to minimize the reconstruction error on the \ttbar background is too weak to achieve good discrimination; it does not prevent the AE from also reconstructing signal jets.

\subsection{Normalized autoencoder training}
\label{sec:nae_training}

The most delicate part of the NAE training is the tuning of the MCMC. A suboptimal MCMC would not accurately sample the distribution \ptheta, and in turn, the NAE would not be able to learn the training data probability distribution \pdata. In theory, the convergence of an MCMC algorithm does not depend on the random initial set of points from which the Markov chain starts. However, there are two motivations for using a nonrandom initial set of points. First, the Markov chain is not infinitely long, and so there is no guarantee of convergence starting from any initial distribution. Second, from one batch to another, the neural network weights and \ptheta both change only slightly, so a good initialization for batch $k$ usually also works for batch $k+1$.
In addition, tuning the hyperparameters of the Markov chain is a complex task since the \ptheta distribution is unknown.
The procedure to initialize the MCMC and tune its associated hyperparameters, which accounts for these considerations, is detailed in Appendix~\ref{sec:mcmc_tuning}.

\begin{figure}[htbp]
    \centering
    \includegraphics[width=0.49\textwidth]{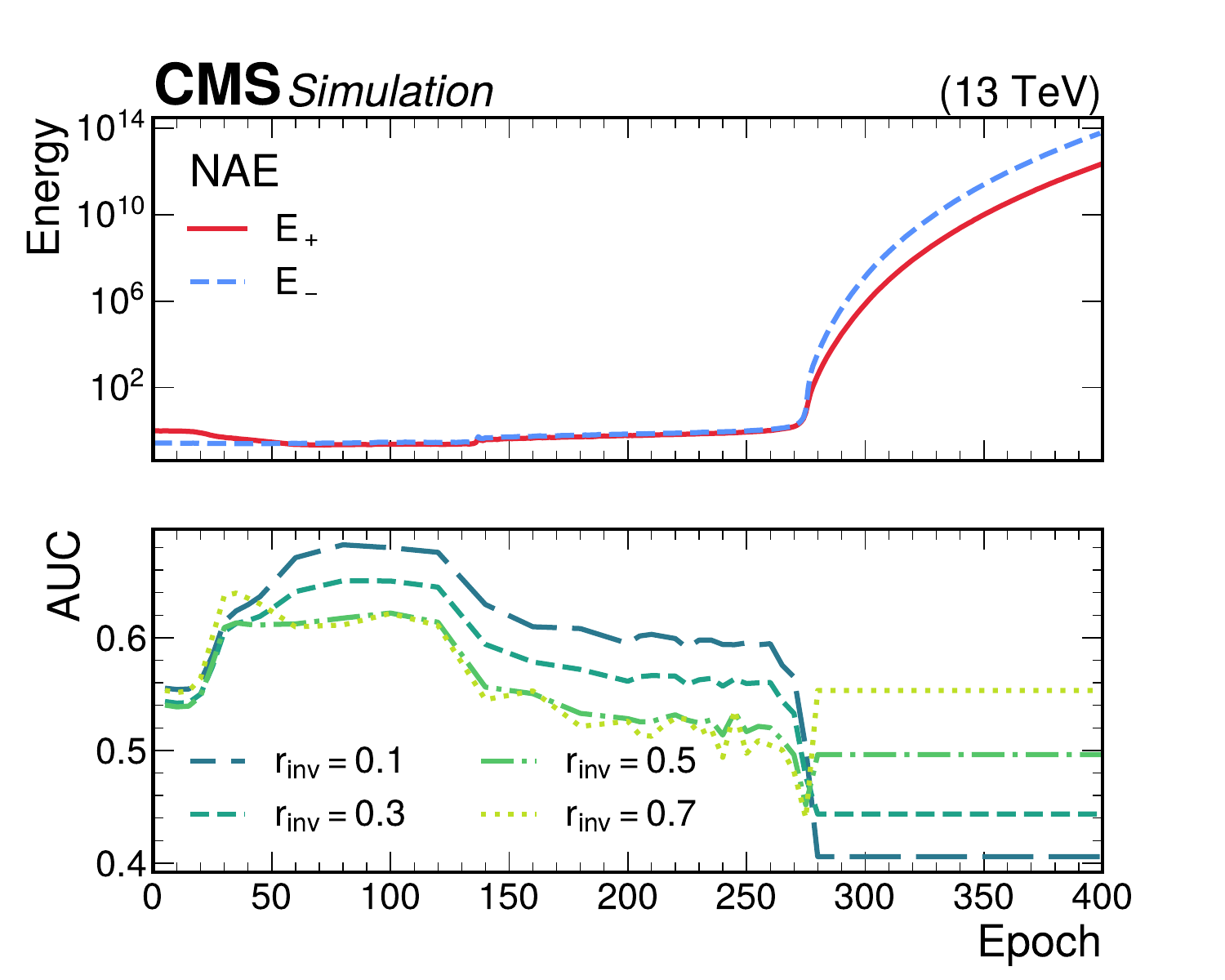}
    \includegraphics[width=0.49\textwidth]{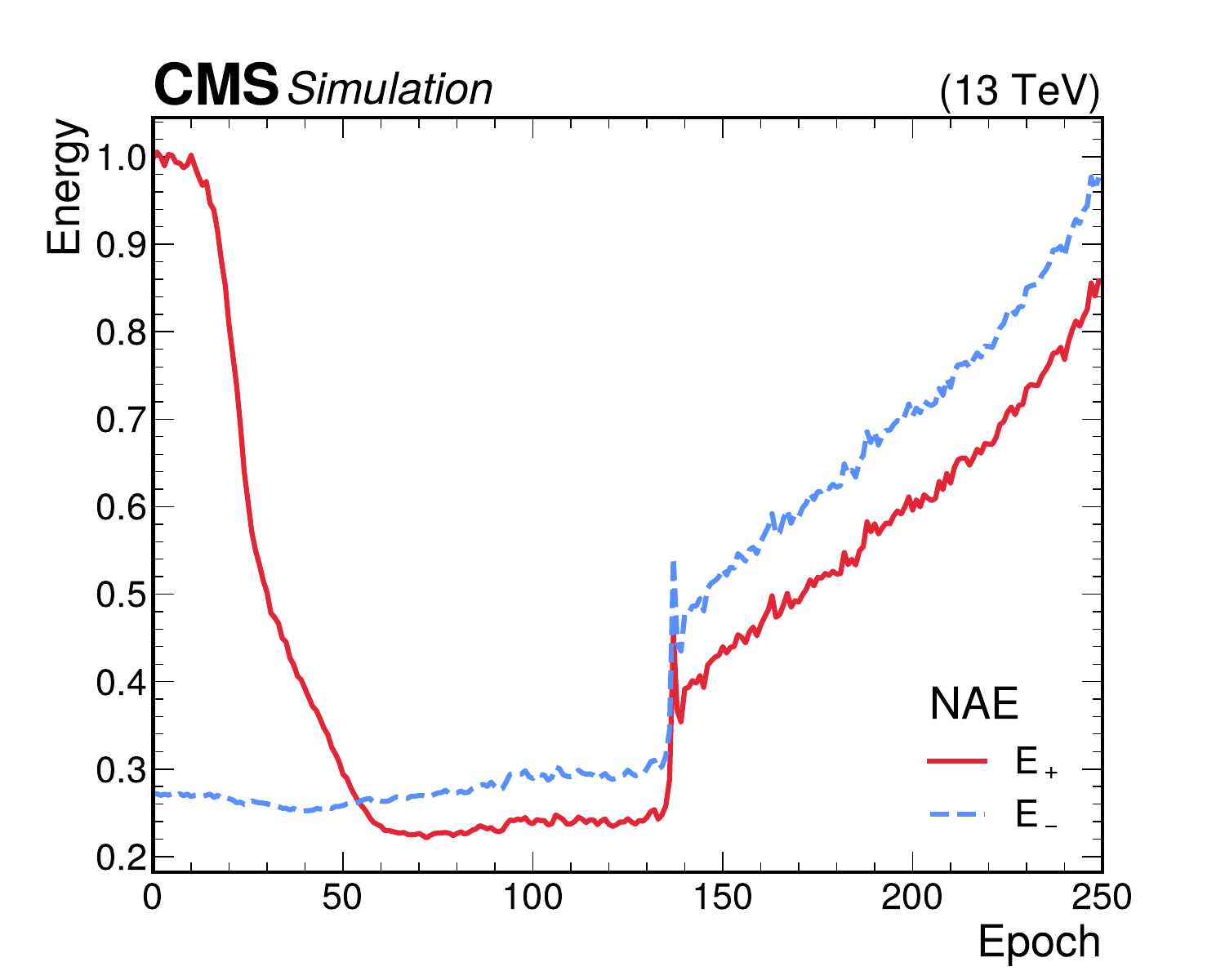}
    \caption{Left: NAE training showing the divergence of the loss function, in terms of positive and negative energy (upper panel), and the AUC for several signal hypotheses with fixed $\mMed = 2000\GeV$ and varying \rinv values (lower panel). Right: the positive and negative energies from the upper panel of the left plot, shown for $\text{epoch} < 250$---before the divergence---and $0.18 < \text{energy} < 1.04$ on a linear scale, to illustrate their differences.}
    \label{fig:nae_loss_divergence}
\end{figure}

After tuning the MCMC algorithm, the evolution of the positive and negative energies of an NAE with the loss function defined in Eq.~\eqref{eq:nae_loss_2} is presented in Fig.~\ref{fig:nae_loss_divergence}. It shows two failure modes: negative loss values, indicating the negative energy is higher than positive energy; and divergence of both energies.

Negative values of the energy difference occur when the network did not learn the background probability distribution, such that \pdata is higher than \ptheta in regions with low reconstruction error. This is a severe failure mode, illustrating that energy difference minimization is not sufficient to guarantee the intended behavior.

\begin{figure}[htb]
    \centering
    \includegraphics[width=.32\linewidth]{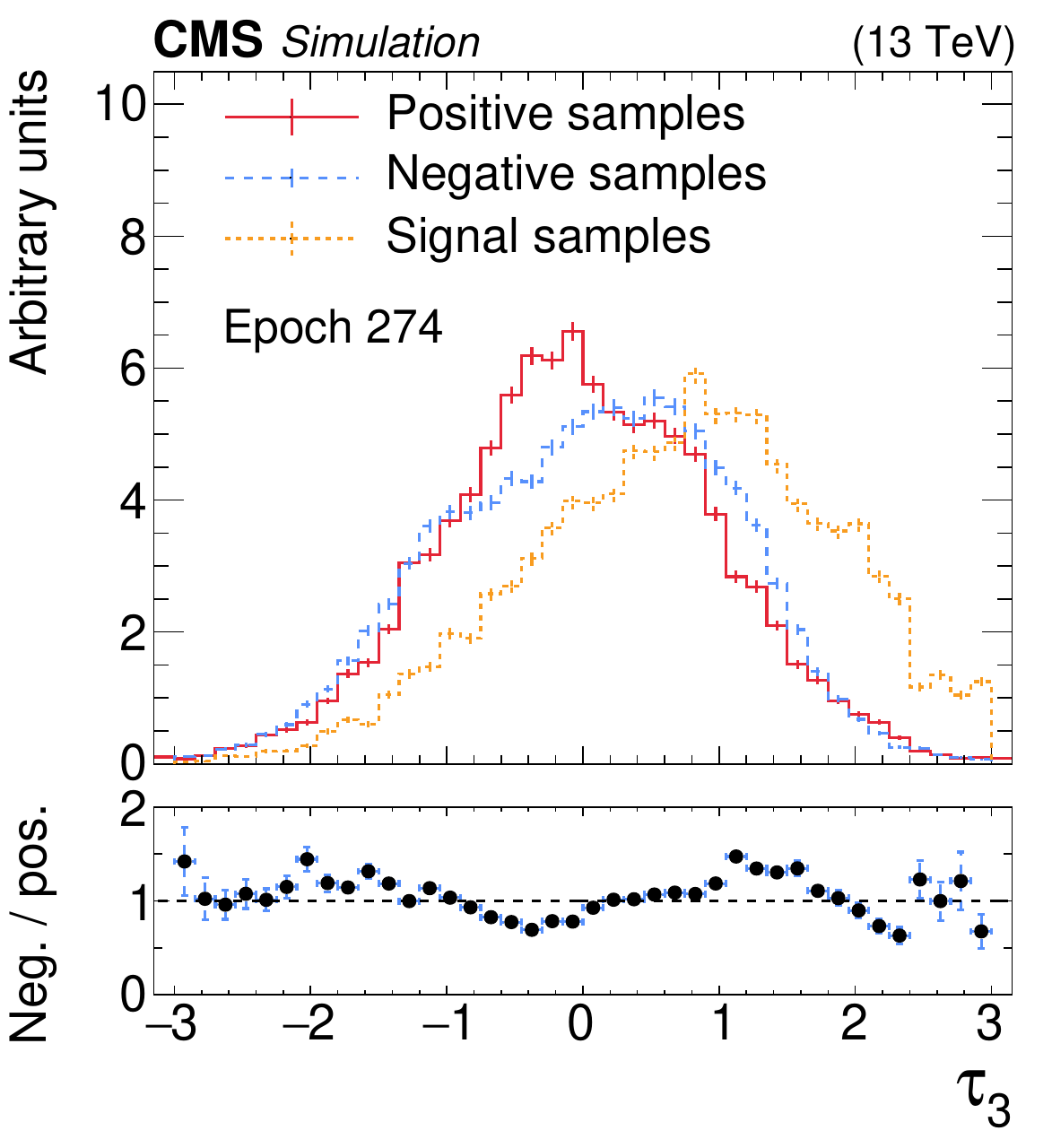}
    \includegraphics[width=.32\linewidth]{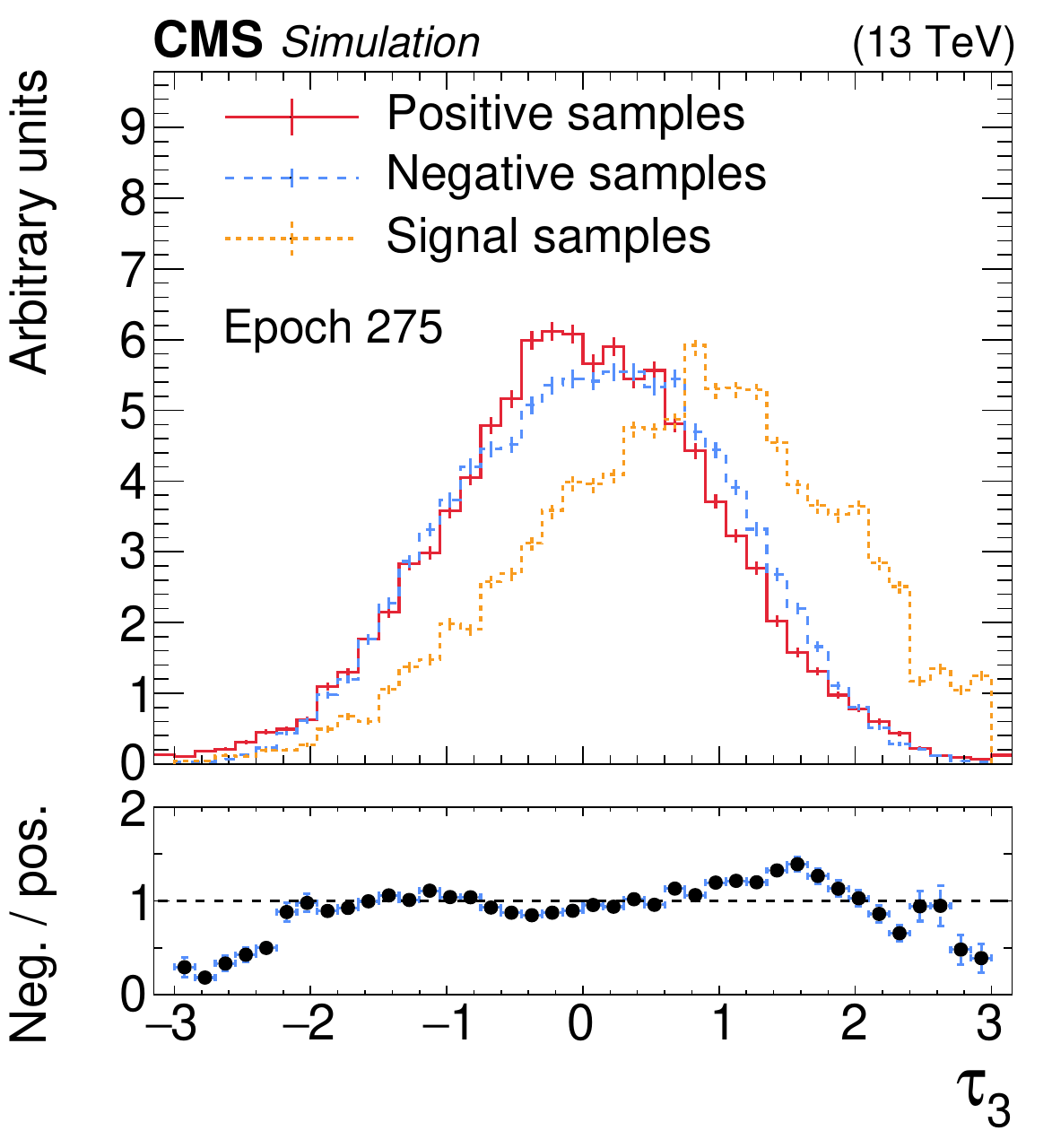}
    \includegraphics[width=.32\linewidth]{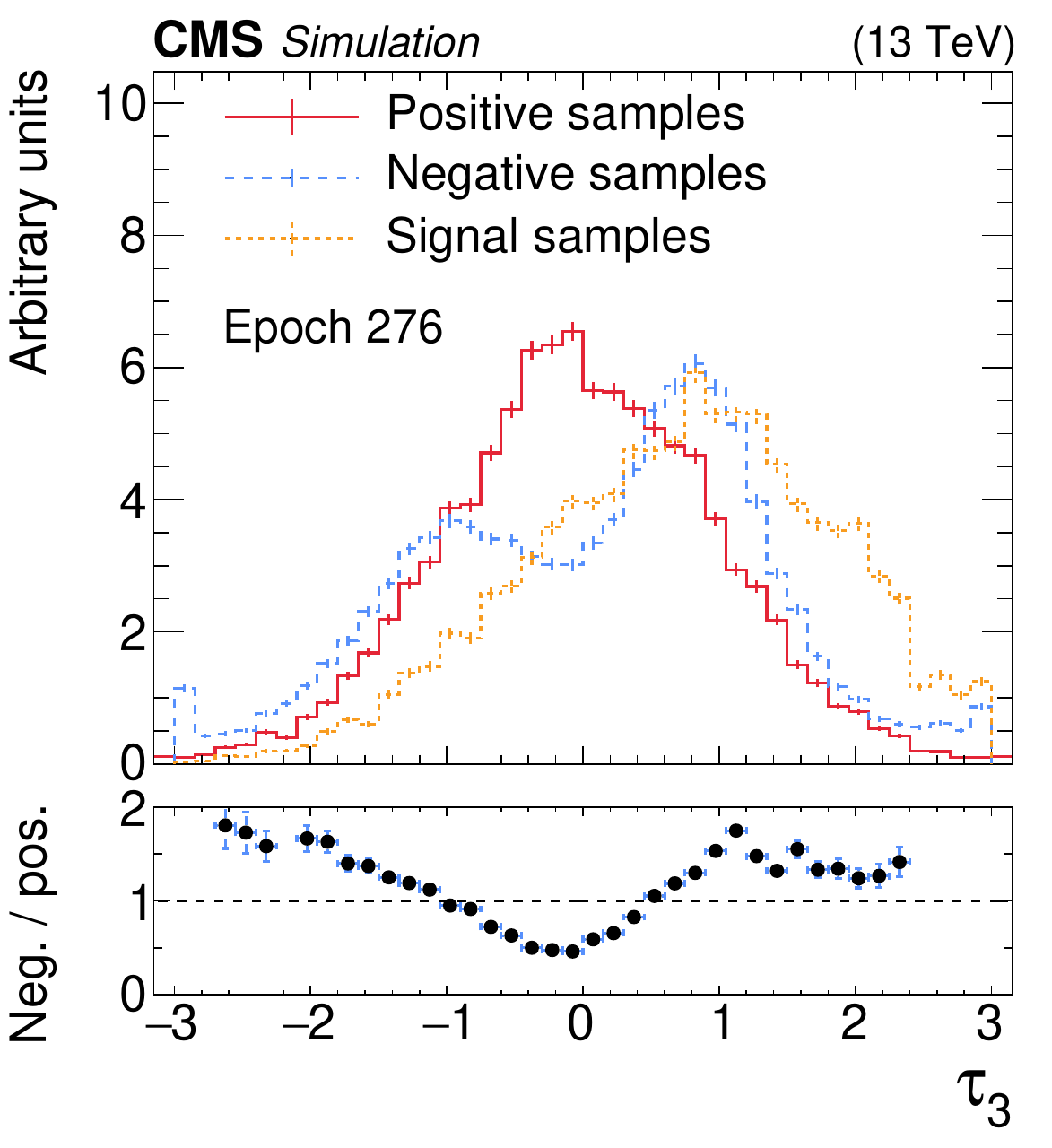}\\
    \includegraphics[width=.32\linewidth]{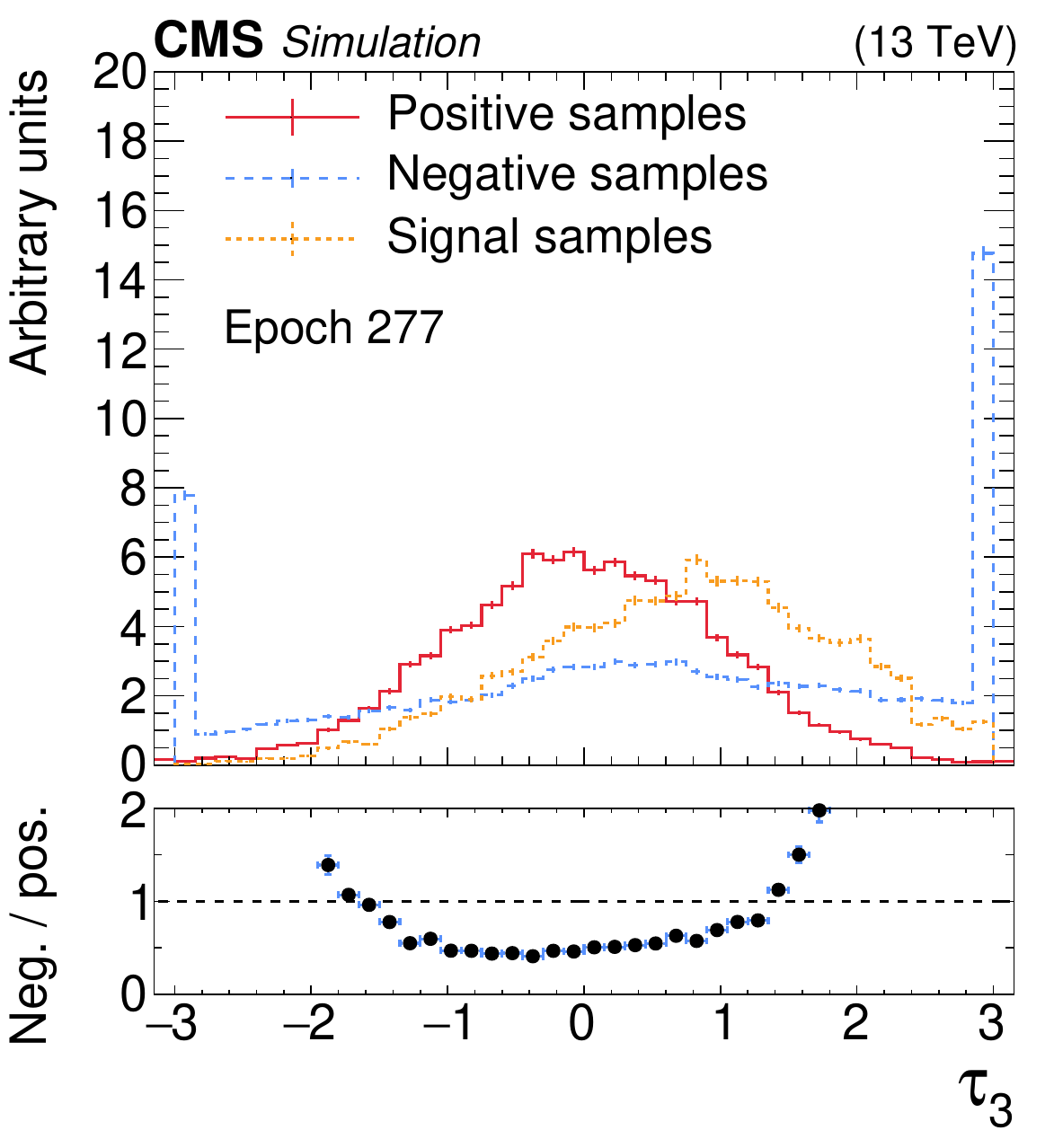}
    \includegraphics[width=.32\linewidth]{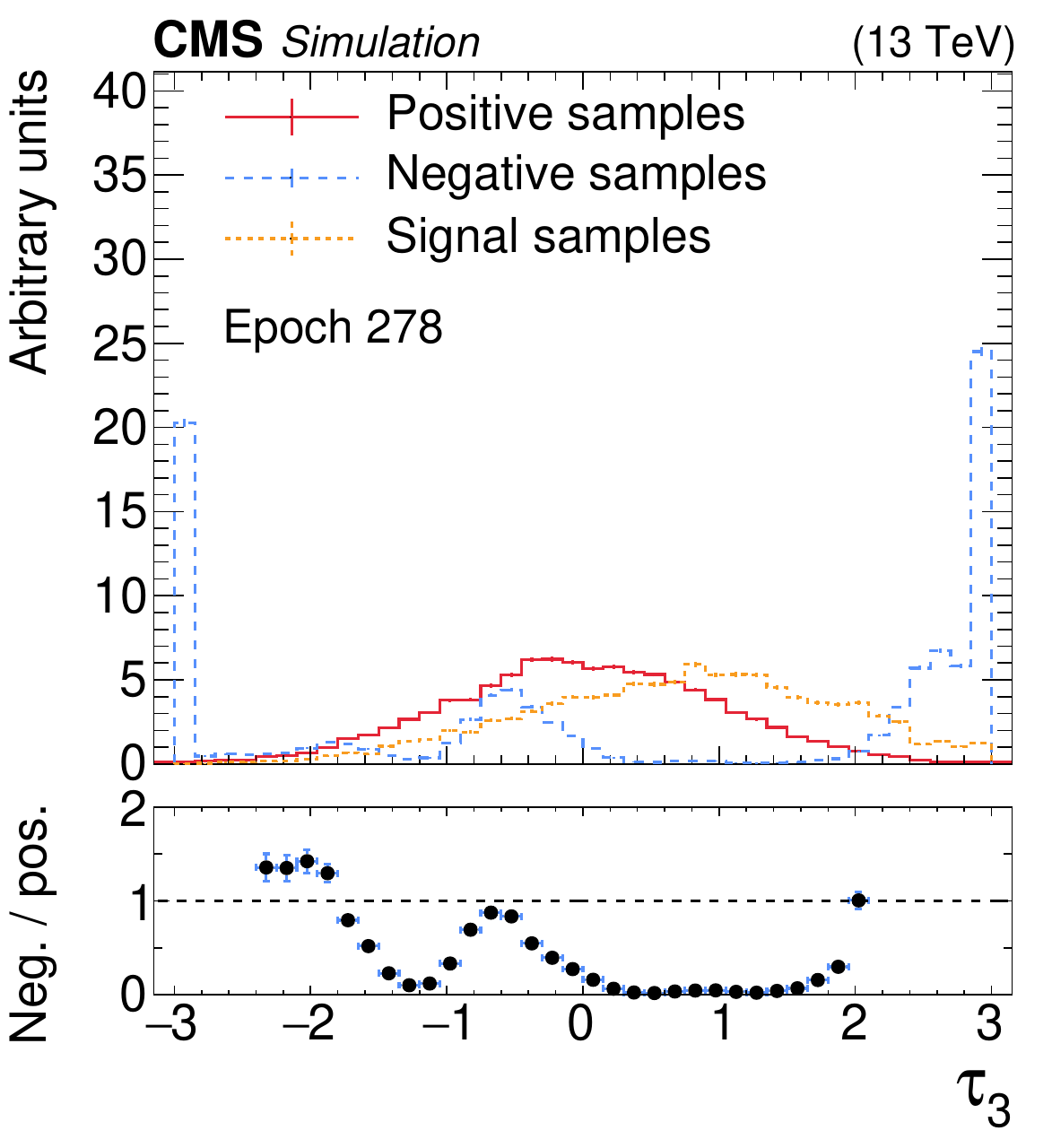}
    \includegraphics[width=.32\linewidth]{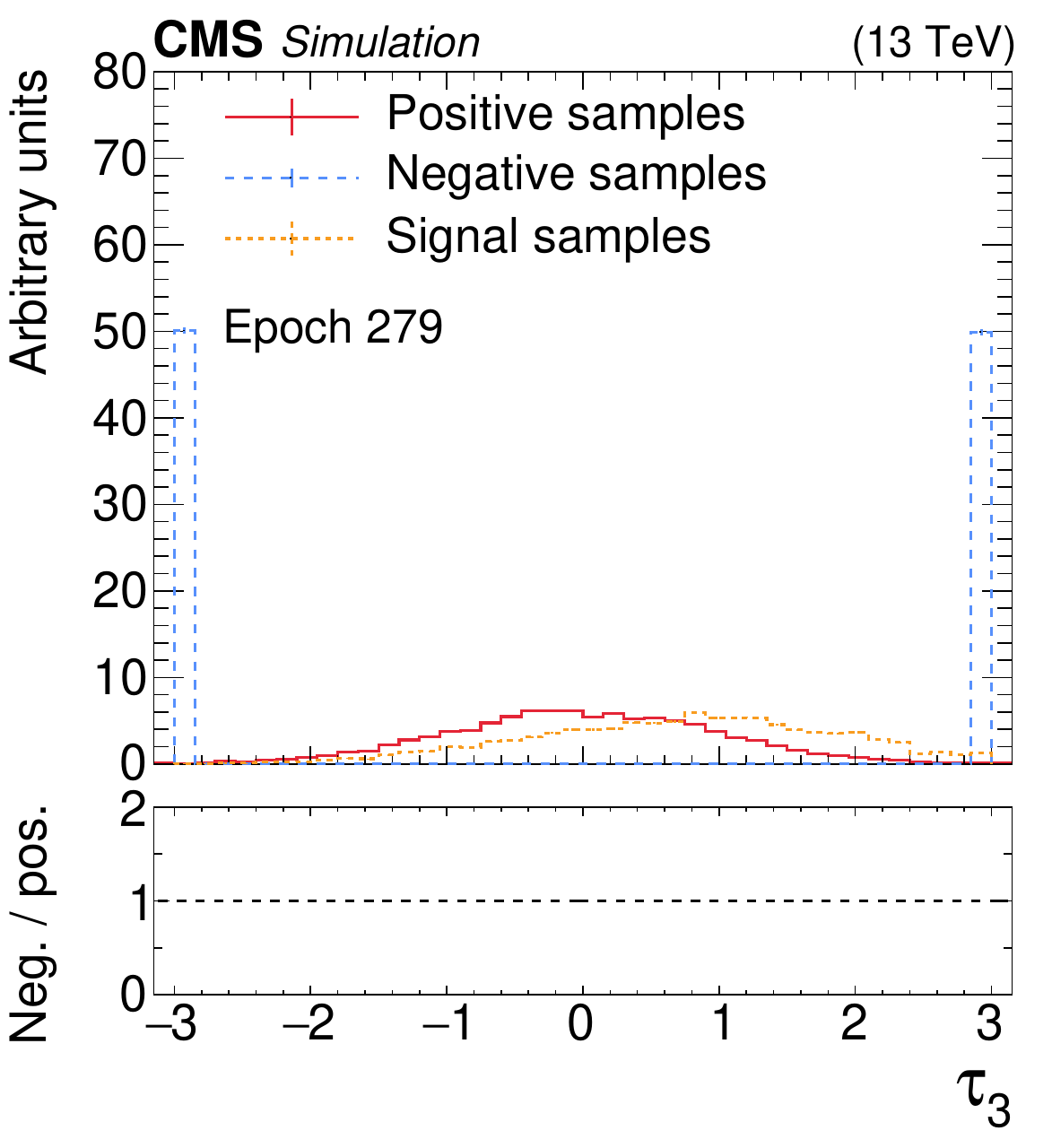}
    \caption{Distributions of the input feature \tauthr for positive, negative, and signal samples, before (epochs 274) and after (epochs 275--279) the start of the divergence of the NAE loss. The signal distributions are overlaid for illustration; signal samples are not used during the training. All distributions are normalized such that their integral is 100.}
    \label{fig:nae_loss_divergence_negative_samples}
\end{figure}

The divergence of the negative energy was also reported in previous work~\cite{pmlr-v139-yoon21c,Dillon:2022mkq}. As a consequence of the form of the loss function, after the positive energy has reached a minimal value, further minimization of the loss is achieved by increasing the negative energy, effectively leading to a divergence of both energy terms and the loss itself, as shown in Fig.~\ref{fig:nae_loss_divergence}.
The distributions of a selected input feature \tauthr for the positive, negative, and signal samples before and after the start of the divergence are provided in Fig.~\ref{fig:nae_loss_divergence_negative_samples}.
The large reconstruction error after the divergence leads to very large gradient values and thus large steps in the MCMC. This results in the negative samples reaching the boundaries of the sampling phase space and no longer being representative of \ptheta.

\begin{figure}[htbp]
    \centering
    \includegraphics[width=0.6\textwidth]{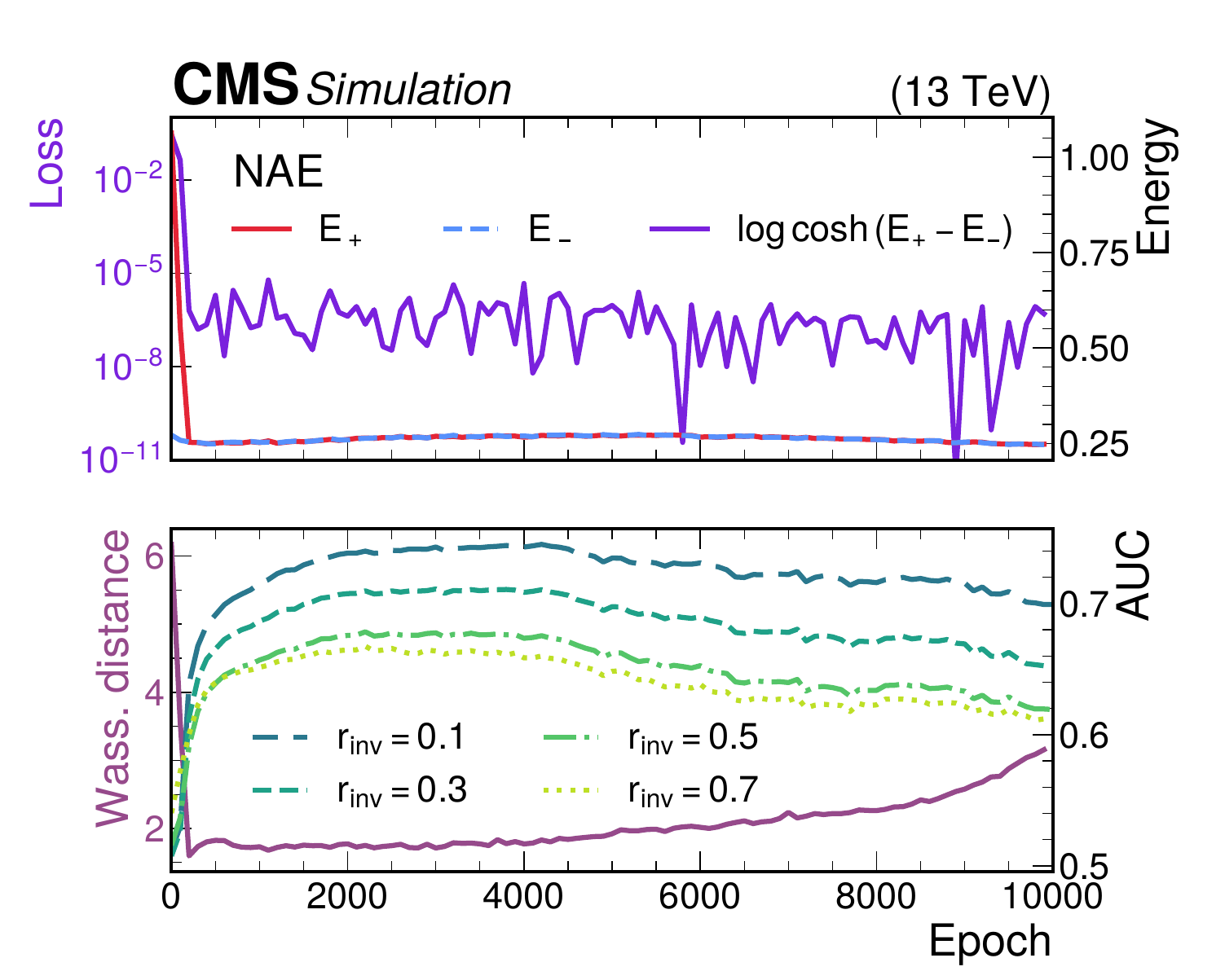}
    \caption{As a function of epoch during the training of an NAE with the loss function from Eq.~\eqref{eq:nae_loss_logcosh}: the positive and negative energies and the value of the loss function (upper panel); the Wasserstein distance between negative and positive samples and the AUC for several signal hypotheses with fixed $\mMed = 2000\GeV$ and varying \rinv (lower panel).}
    \label{fig:nae_logcosh}
\end{figure}

A simple modification of the NAE loss can mitigate these issues:
\begin{equation}
    \mathcal{L} = \log\cosh(\Eplus - \Eminus).
    \label{eq:nae_loss_logcosh}
\end{equation}
Because the $\log\cosh$ function is positive definite, this loss removes the incentive to favor negative values of the energy difference and leads to a more stable training procedure. The evolution of the energies, loss value, and AUC score for a training with this loss function are shown in Fig.~\ref{fig:nae_logcosh}.
Their behavior in this figure can be qualitatively understood as follows. There is a first stage in which the energy difference is minimized, giving the largest improvement in AUC. Once the difference between positive and negative energies is approximately zero, there is a second stage in which the network fine-tunes its knowledge of \ptheta, further increasing the AUC score until it reaches a plateau around epoch 4000. There is subsequently a third stage in which a new form of mode collapse is observed: the energy difference is still approximately zero and the positive energy is further minimized, but the AUC score begins to decrease. Crucially, it is not possible to avoid this degradation by relying solely on the positive and negative energies. On the other hand, making use of the AUC score as a stopping condition for the training would be in contradiction with the unsupervised nature of the NAE.
The evolution of the Wasserstein distance, also shown in this figure, is discussed in Section~\ref{sec:wae}.

\begin{figure}[htbp]
    \centering
    \includegraphics[width=0.49\textwidth]{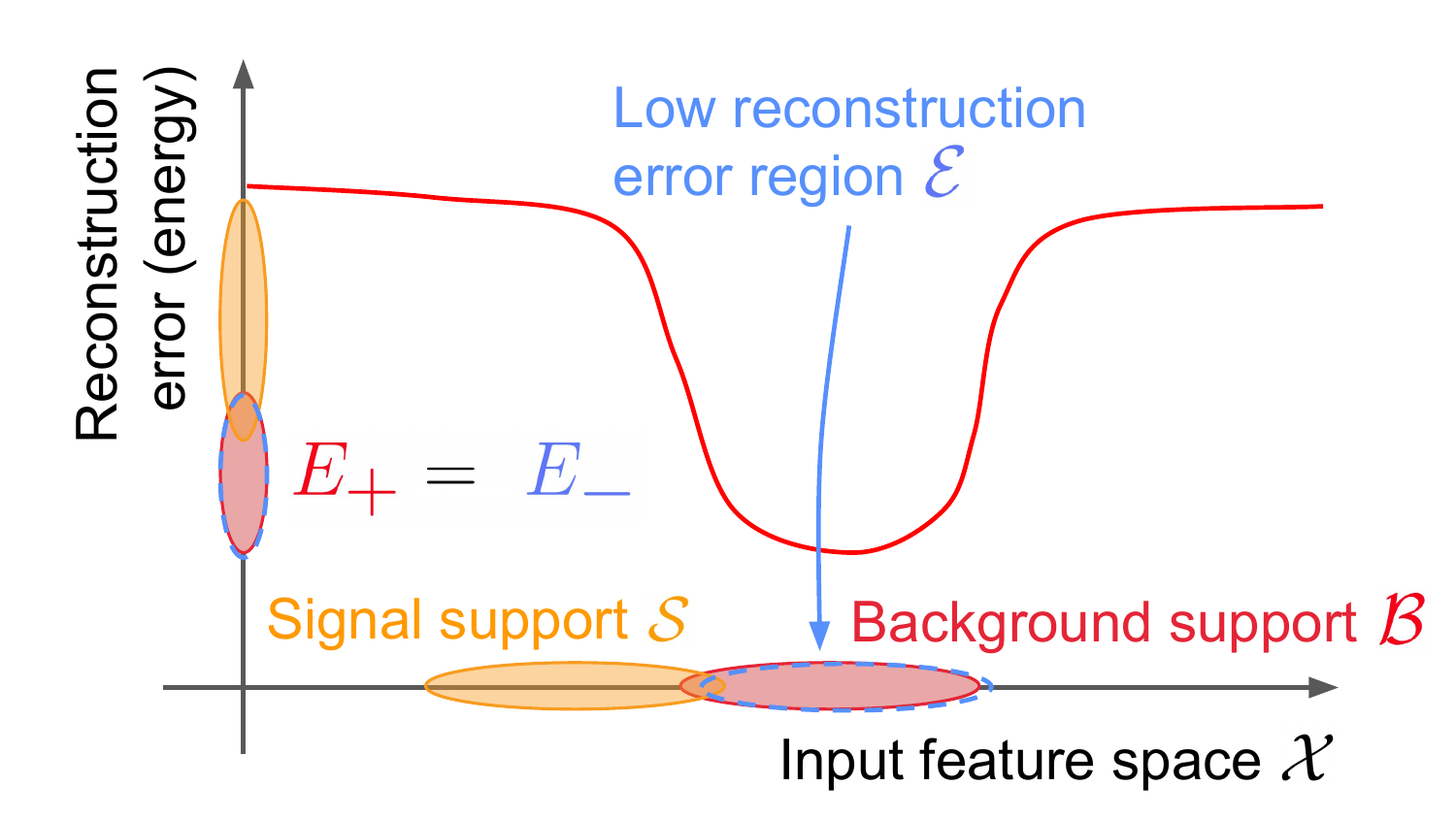}
    \includegraphics[width=0.49\textwidth]{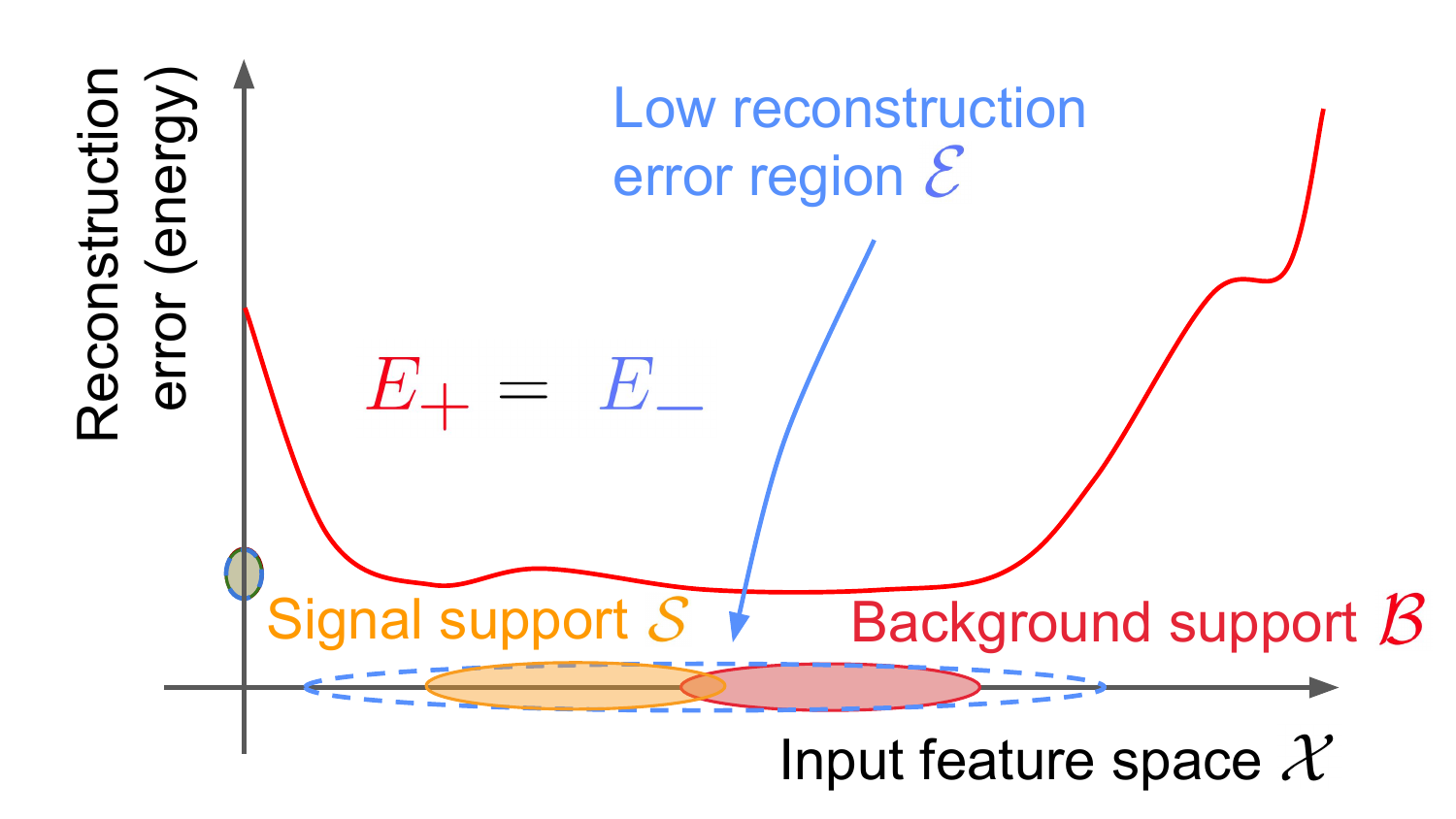}
    \caption{Schematic representation of the mode collapse when using the loss function described in Eq.~\eqref{eq:nae_loss_logcosh}. The energy landscape is shown before (left) and after (right) the mode collapse. On the right, the reconstruction errors for the signal and background supports are completely overlapping on the vertical axis. The symbols \Eplus and \Eminus denote the positive and negative energies, respectively.}
    \label{fig:nae_mode_collapse}
\end{figure}

This collapse arises as a form of overtraining of the network, leading to outlier reconstruction. While keeping the reconstruction error low, the network eventually achieves low reconstruction in a region \supEx that contains, but is larger than, the support of the training data \supBkg, while keeping $\mathbb{E}_{x\sim \supEx}[E(x)]$ comparable to $\mathbb{E}_{x\sim \supBkg}[E(x)]$. As the overlap of \supEx with the support of the signal $\mathcal{S}$ becomes larger, the AUC score decreases. This behavior is shown schematically in Fig.~\ref{fig:nae_mode_collapse}. Before the mode collapse, \supEx and \supBkg overlap, while \supEx and \supSig do not. After the mode collapse, \supEx expands and can partially include \supSig, reducing the difference in reconstruction error between background and signal and correspondingly lowering the AUC score. In both cases, the difference between the positive and the negative energies is zero.

\section{The Wasserstein normalized autoencoder}
\label{sec:wae}

\subsection{The Wasserstein distance as a measure of outlier reconstruction}
\label{sec:wassertein_distance_metric}

\begin{figure}[htbp]
    \centering
    \includegraphics[width=0.49\textwidth]{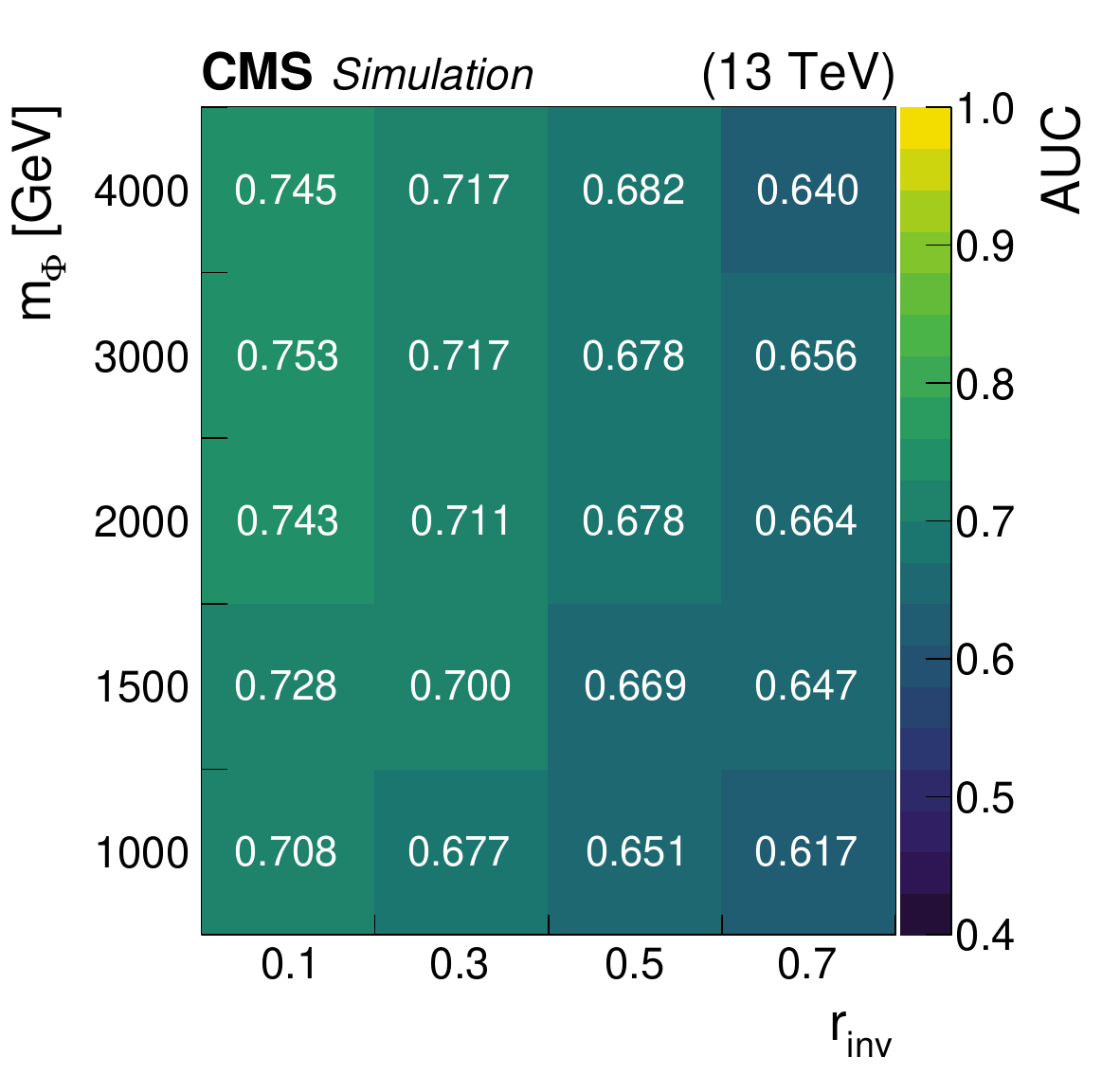}
    \includegraphics[width=0.49\textwidth]{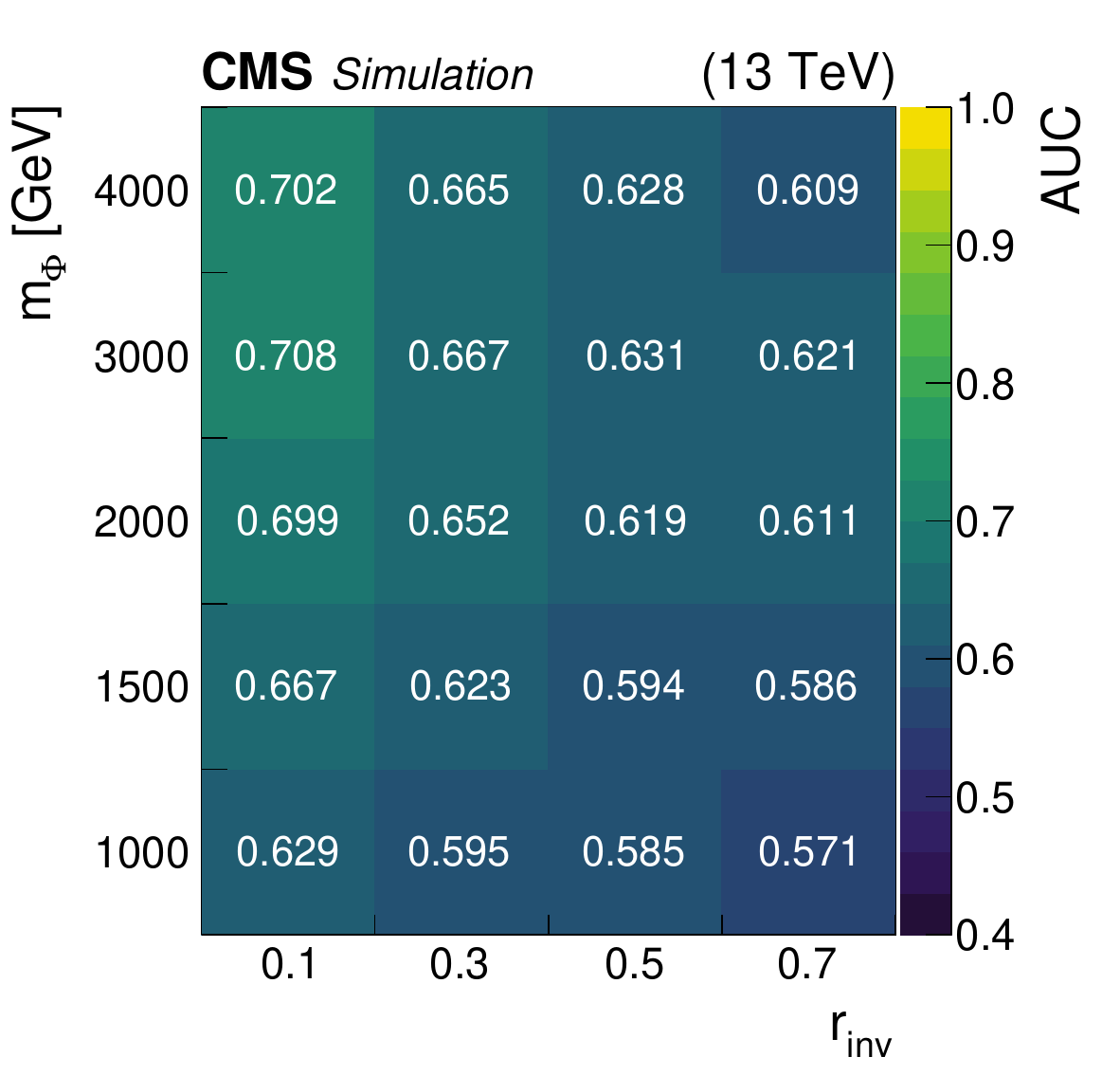}
    \caption{Left: the AUC scores for an NAE trained on the \ttbar background and tested against a grid of possible SVJ signals, before the increase of the Wasserstein distance (at epoch 3000). Right: the AUC scores for the same NAE after the increase in Wasserstein distance (at epoch 10000).}
    \label{fig:nae_errors_emd_auc_logcosh}
\end{figure}

The failure mode described at the end of Section~\ref{sec:nae_training} is a consequence of the fact that the energy difference is unable to distinguish the two cases shown in Fig.~\ref{fig:nae_mode_collapse}. In fact, no stopping condition based solely on combinations of positive and negative energies was found to be a good indicator of the start of this mode collapse. In other words, without explicitly monitoring the AUC during training, there is no clear stopping condition; choosing a stopping condition based on the AUC, however, violates the principle of unsupervised learning. In order to overcome this problem, a different metric is used: the 1-Wasserstein distance $W$ (also known as the earth mover's distance) from Eq.~\eqref{eq:emd_primal} between the probability distribution of the training data \pdata and the probability distribution learned by the network \ptheta, in analogy to what was applied successfully to generative adversarial networks~\cite{Arjovsky:2017fjr}:
\begin{equation}
    \label{eq:emd_loss}
    W(\pdata, \ptheta) = \inf_{\gamma \in \Pi(\pdata, \ptheta)} \mathbb{E}_{(x, \xprime)\sim \gamma}\left[\|x - \xprime\|\right].
\end{equation}
The 1-Wasserstein distance is chosen for consistency with other machine learning applications, which use this distance rather than any other $p$-Wasserstein distance because the Kantorovich-Rubinstein duality is only known to hold for $p = 1$.
Experiments using the 2-Wasserstein distance demonstrated similar performance.
This metric is expected to solve the problem of outlier reconstruction by measuring the distance between the \ptheta and \pdata distributions directly in the input feature space. By minimizing $W(\pdata, \ptheta)$, these two probability distributions are forced to match, which prevents feature space regions distinct from \pdata from having low reconstruction error, achieving the desired behavior illustrated in Fig.~\ref{fig:nae_mode_collapse} (left). 
The Wasserstein distance is, indeed, found to be a good indicator of the start of the mode collapse, as seen in Fig.~\ref{fig:nae_logcosh} around epoch 4000: the Wasserstein distance increases significantly when this collapse occurs and the AUC decreases.
The point where the Wasserstein distance increases, which can be quantified as the point where its numerical derivative exceeds an empirical threshold for the first time during the training, can then be identified as a stopping condition for the training of the NAE.
It is worth stressing that the Wasserstein distance, calculated from the positive and negative samples, is completely agnostic to the signal, and yet it is found to be closely correlated with the performance of the NAE as an outlier detector. With this prescription, the NAE is able to successfully learn the \ttbar distribution and effectively detect the signal, as shown in Fig.~\ref{fig:nae_errors_emd_auc_logcosh}. 

\subsection{The Wasserstein distance as a loss function}

Having shown that the Wasserstein distance is a good indicator of the performance of the NAE as an outlier detection algorithm, it is natural to use it directly as a loss function.
Compared to other common distances or divergences between probability distributions, such as the Kullback-Leibler or Jensen-Shannon divergences, the Wasserstein distance has better analytic properties.
In particular, if $g_{\theta}$ is a feedforward neural network parametrized by weights $\theta$ with probability distribution $p_{\theta}$, and $p$ is a fixed distribution, then the Wasserstein distance $W(\ptheta, p)$ is continuous everywhere and differentiable almost everywhere with respect to $\theta$.
This is not the case for the Kullback-Leibler and Jensen-Shannon divergences between $\ptheta$ and $p$.
A review and proof of these analytic properties can be found in Ref.~\cite{Arjovsky:2017fjr}.

\begin{figure}[htb]
    \centering
    \includegraphics[width=\textwidth]{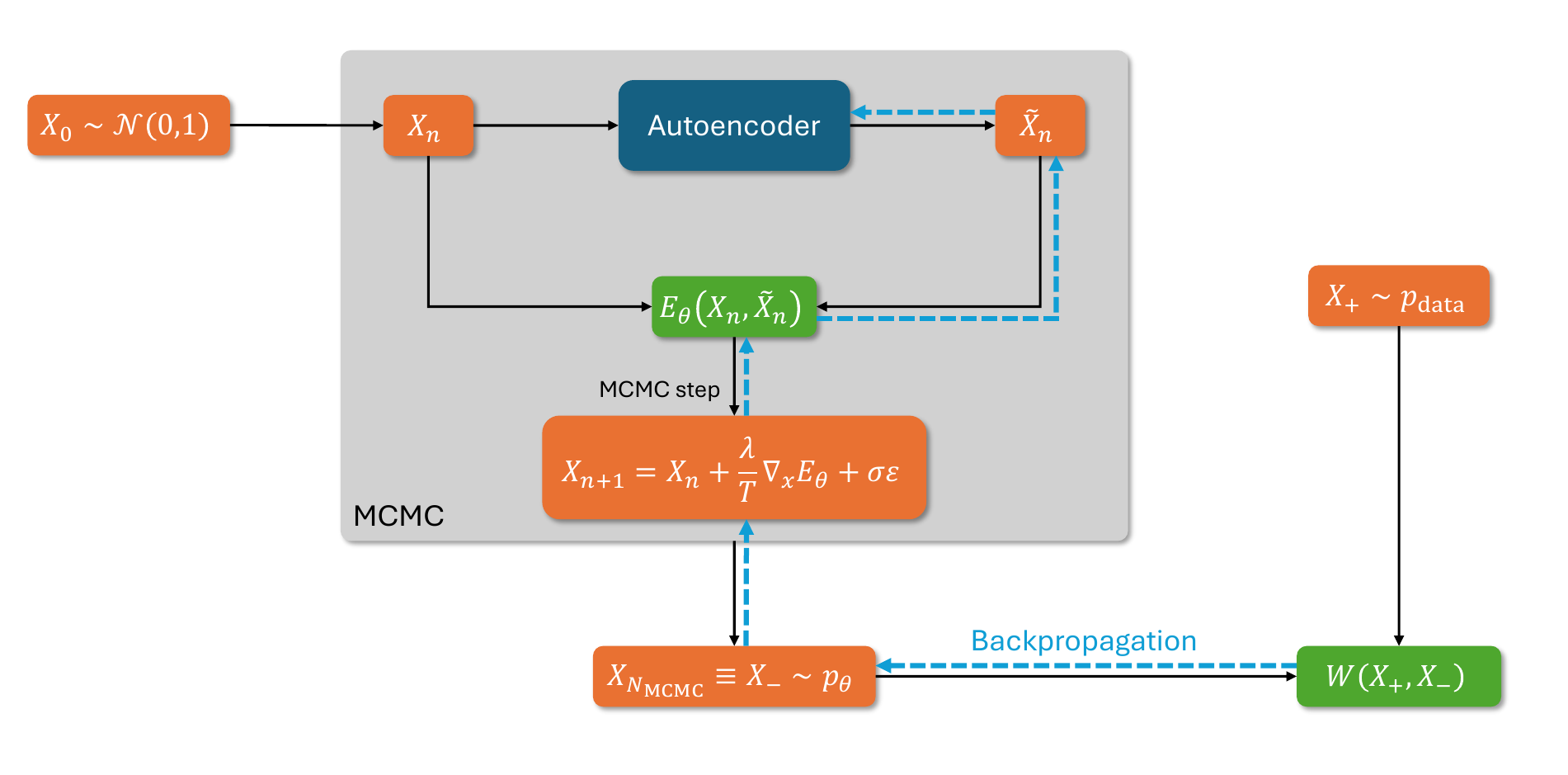}
    \caption{Flowchart of the Wasserstein normalized autoencoder training. The negative examples are generated via MCMC using the energy function of the model, which is the Boltzmann-distributed reconstruction error. The energy function is computed from random input feature values $X_n$ and the corresponding reconstructed feature values $\widetilde{X}_n$, obtained by passing the inputs through the autoencoder. The positive examples are compared to the negative examples through the Wasserstein distance. The gradients are backpropagated through the entire MCMC chain.}
    \label{fig:wae_flowchart}
\end{figure}

Correctly incorporating the Wasserstein distance in the neural network training requires two computational graphs: one for differentiation with respect to the input feature space during each step of the Markov chain, another for differentiation with respect to the network weights $\theta$ during gradient descent. The procedure to implement this is detailed in Appendix~\ref{sec:ae_weights_dependency_in_mcmc}.
The resulting model is called the Wasserstein normalized autoencoder (\wnae), with source code publicly available online~\cite{gitlab}. The flowchart for training the \wnae is shown in Fig.~\ref{fig:wae_flowchart}.

\subsection{Training the Wasserstein normalized autoencoder}
\label{sec:training_wnae}

The training is found to qualitatively proceed in two phases: first, a coarse adjustment in which the Wasserstein distance decreases sharply while the model adapts \ptheta to produce negative samples that resemble physical examples, and second, a fine-tuning phase in which the model learns the specifics of the training sample.
The use of a learning rate scheduler was found to decrease the training time while preserving the stability of the training. In the case of a constant learning rate, if the learning rate is too small, the time needed for the first phase increases significantly, while if it is too large, the training becomes unstable in the second phase.
Technical details about the learning rate scheduler can be found in Appendix~\ref{sec:scheduler}.

The computational bottleneck for training the \wnae is the calculation of the Wasserstein distance between the positive and negative samples, which takes approximately 90\% of the computation time for each training batch.
The complexity of the Wasserstein distance scales linearly with the dimensionality of the data $d$ and at worst cubically with the number of data points $n$~\cite{bonneel2011displacement}. In practice, for the data set used in this paper, the complexity is measured to scale quadratically with $n$, similar to Ref.~\cite{bonneel2011displacement}.
The second most impactful operation is the backpropagation through the MCMC chain, which takes approximately 7\% of the computation time. The time needed for the calculation of the Wasserstein distance heavily depends on the batch size, which was set to 4096 to allow for a robust estimation of the distance between positive and negative samples.
Lowering the batch size reduces the computation time at the expense of decreasing the stability of the loss function.
Depending on the batch size, the training of the \wnae can thus be significantly slower than that of the NAE, by up to an order of magnitude for large enough batch sizes. However, the Wasserstein distance is found to provide a much more reliable and stable stopping condition.
All of these computational considerations only impact training; at inference time, the \wnae behaves like a standard AE, with no additional computational overhead.

\subsection{Performance of the Wasserstein normalized autoencoder}
\label{sec:results}

\begin{figure}[htb]
    \centering
    \includegraphics[width=0.49\textwidth]{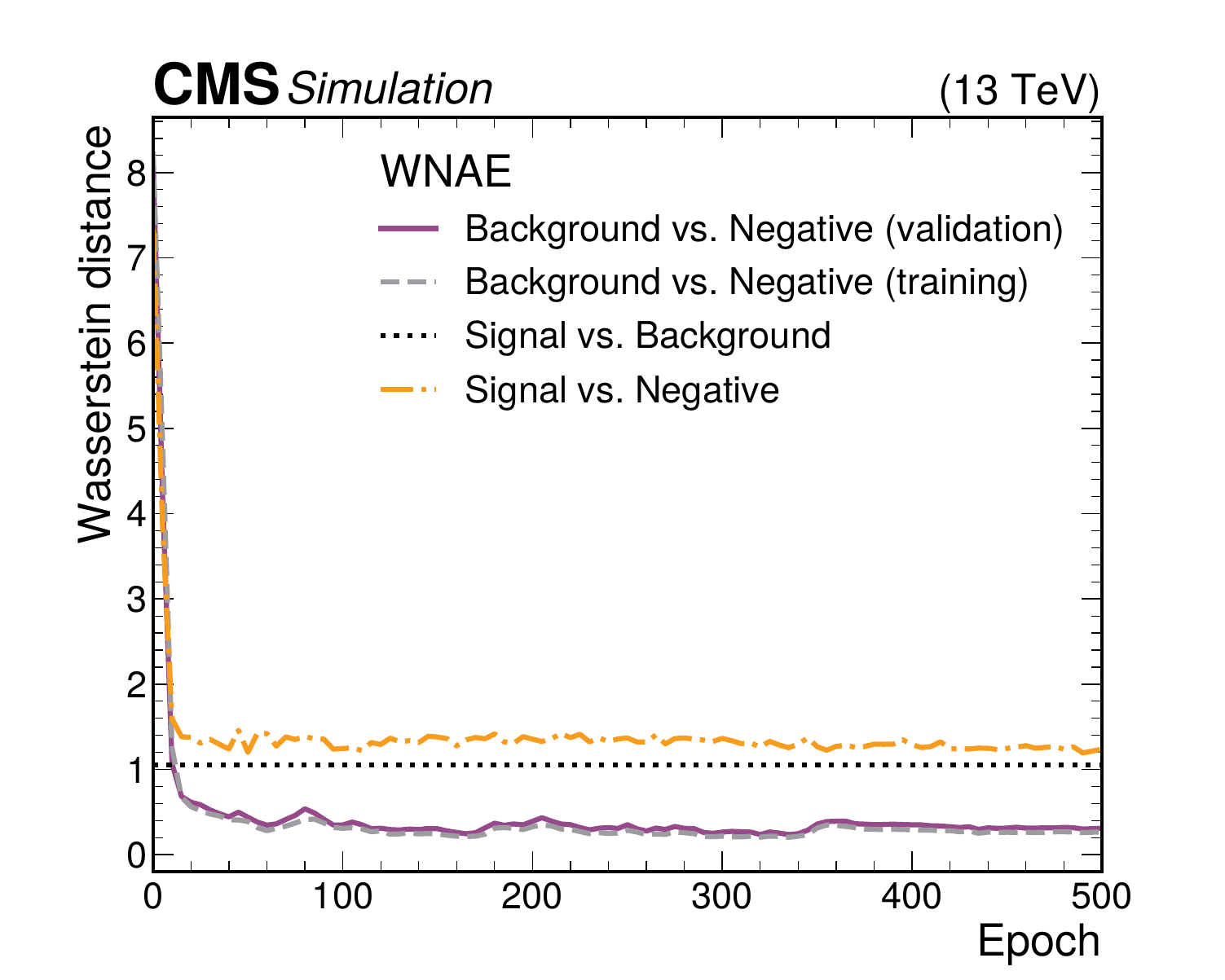}
    \includegraphics[width=0.49\textwidth]{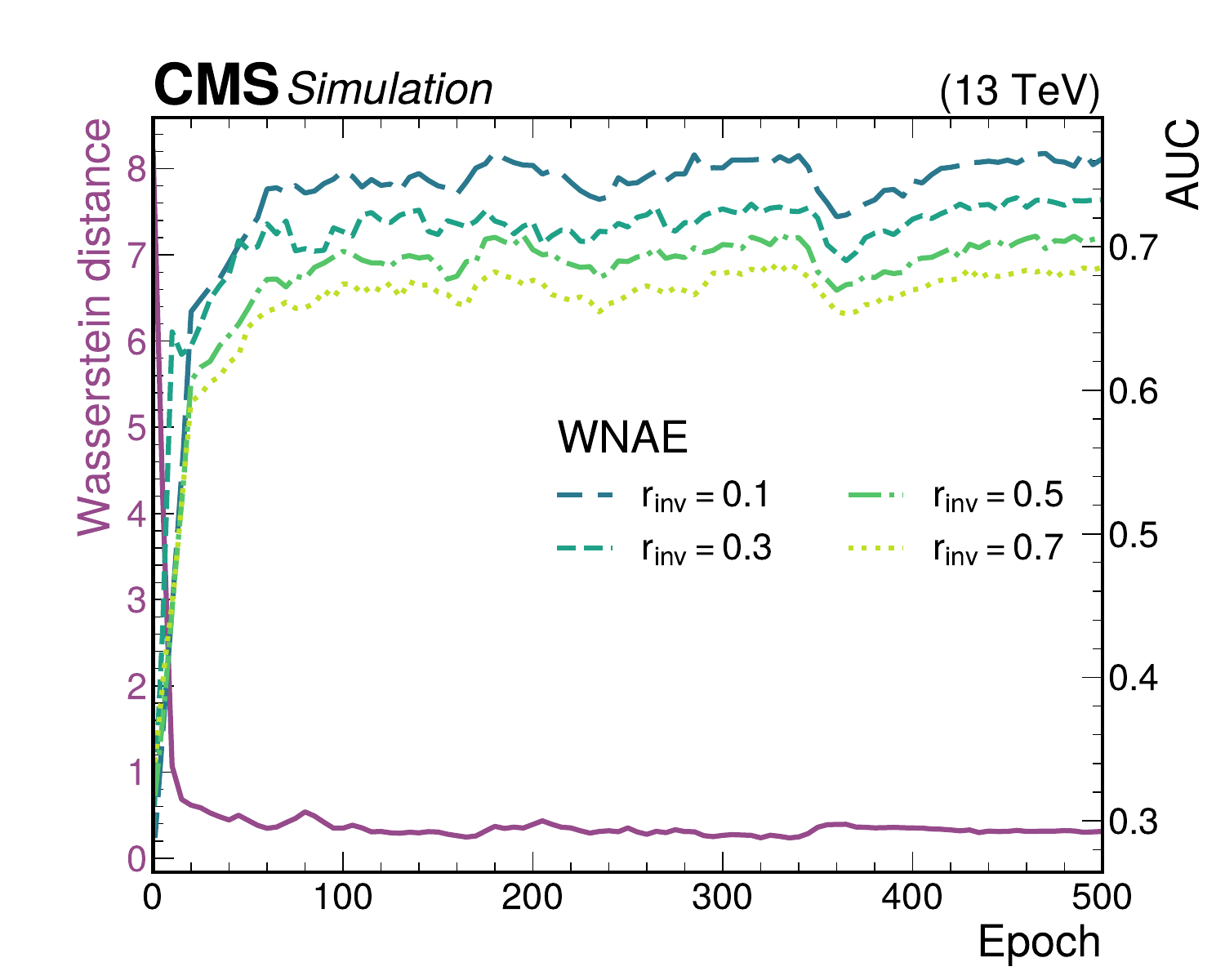}
    \caption{Left: the Wasserstein distance between pairs of the positive, negative, and signal samples during the \wnae training. Right: the Wasserstein distance between negative and positive samples and the AUC scores from the same \wnae for several signal hypotheses with fixed $\mMed = 2000\GeV$ and varying \rinv.}
    \label{fig:wae_roc_auc}
\end{figure}

\begin{figure}[htb]
    \centering
    \includegraphics[width=0.49\textwidth]{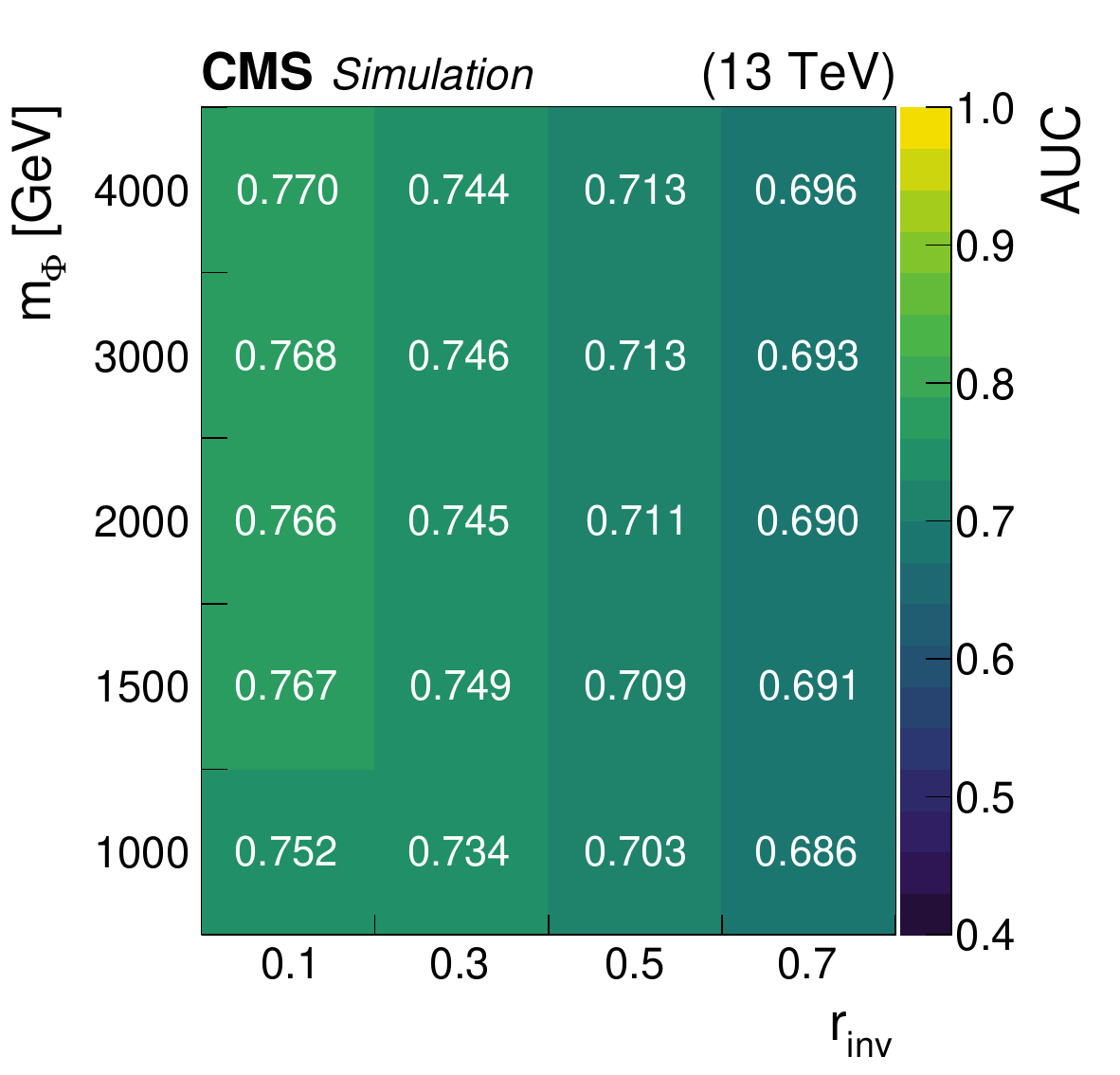}
    \caption{The AUC scores for a \wnae trained on the \ttbar background and tested on a grid of possible SVJ signal models. The statistical uncertainty in the AUC scores, evaluated by retraining the \wnae with different random initializations, is found to be less than 0.02.}
    \label{fig:wae_roc_auc_table}
\end{figure}

The \wnae is found to be significantly more stable in training, not showing any of the forms of divergence described previously in Section~\ref{sec:nae_training}. It is able to achieve strong anomaly detection performance while using the minimum of the validation loss, the Wasserstein distance evaluated on the validation set, as a clear stopping condition. Figure~\ref{fig:wae_roc_auc} shows the evolution of the Wasserstein distance and the AUC score for several SVJ signal hypotheses. The minimum of the Wasserstein distance corresponds to the maximum discrimination power. Figure~\ref{fig:wae_roc_auc_table} shows the AUC score evaluated for a grid of SVJ signals, showing good discrimination power across the parameter space, on par with or better than any of the NAE models trained above. The \wnae is thus found to be a more reliable, stable, and effective tool for anomaly detection in the context of the SVJ search. In addition, this algorithm can achieve the same performance with all hidden layers having higher dimension than the input feature space. The requirement to minimize the Wasserstein distance controls the behavior of the network without separately requiring an architecture with a lower-dimensional bottleneck.

\begin{figure}[tp!]
    \centering
    \includegraphics[width=0.24\textwidth]{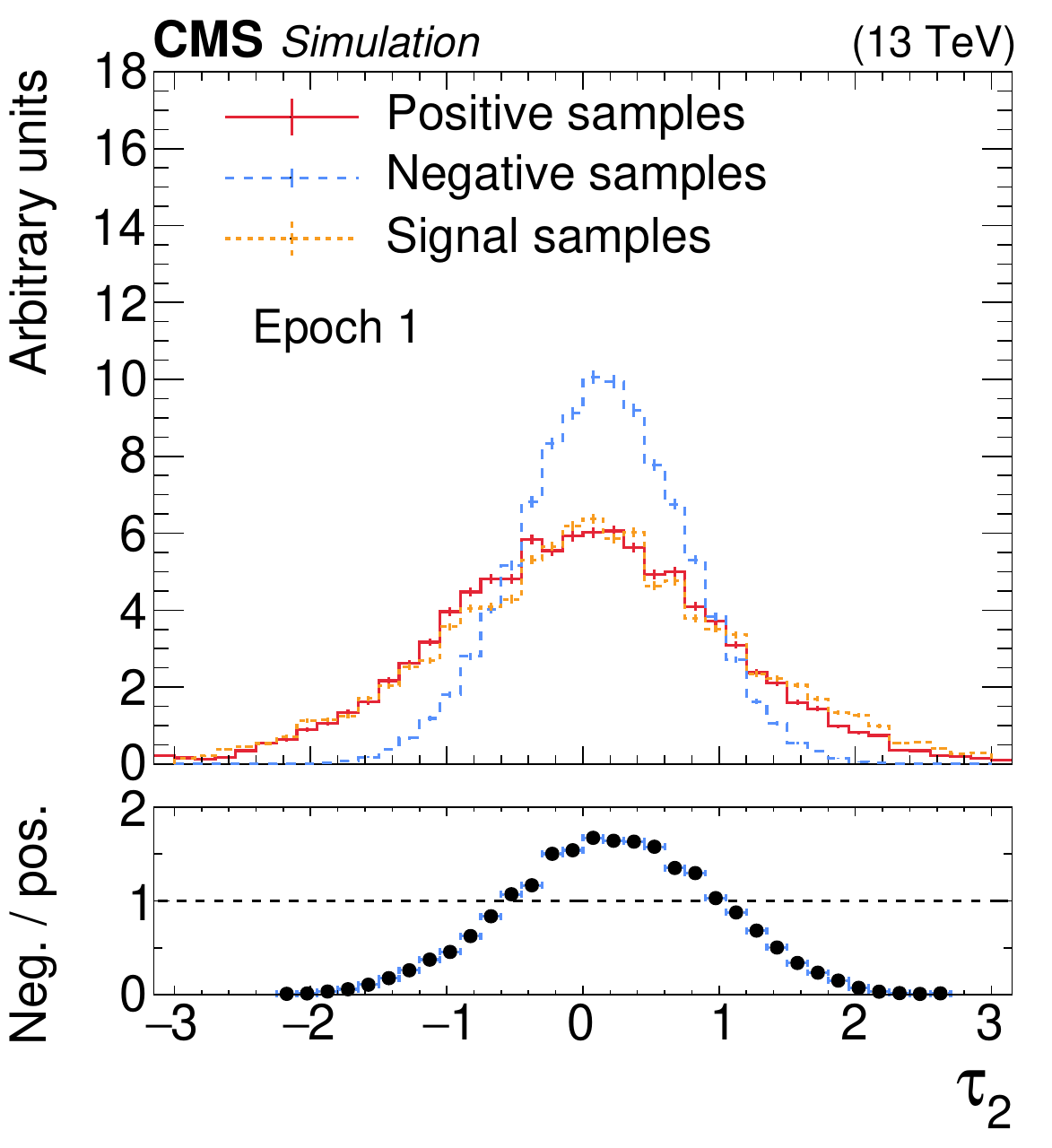}
    \includegraphics[width=0.24\textwidth]{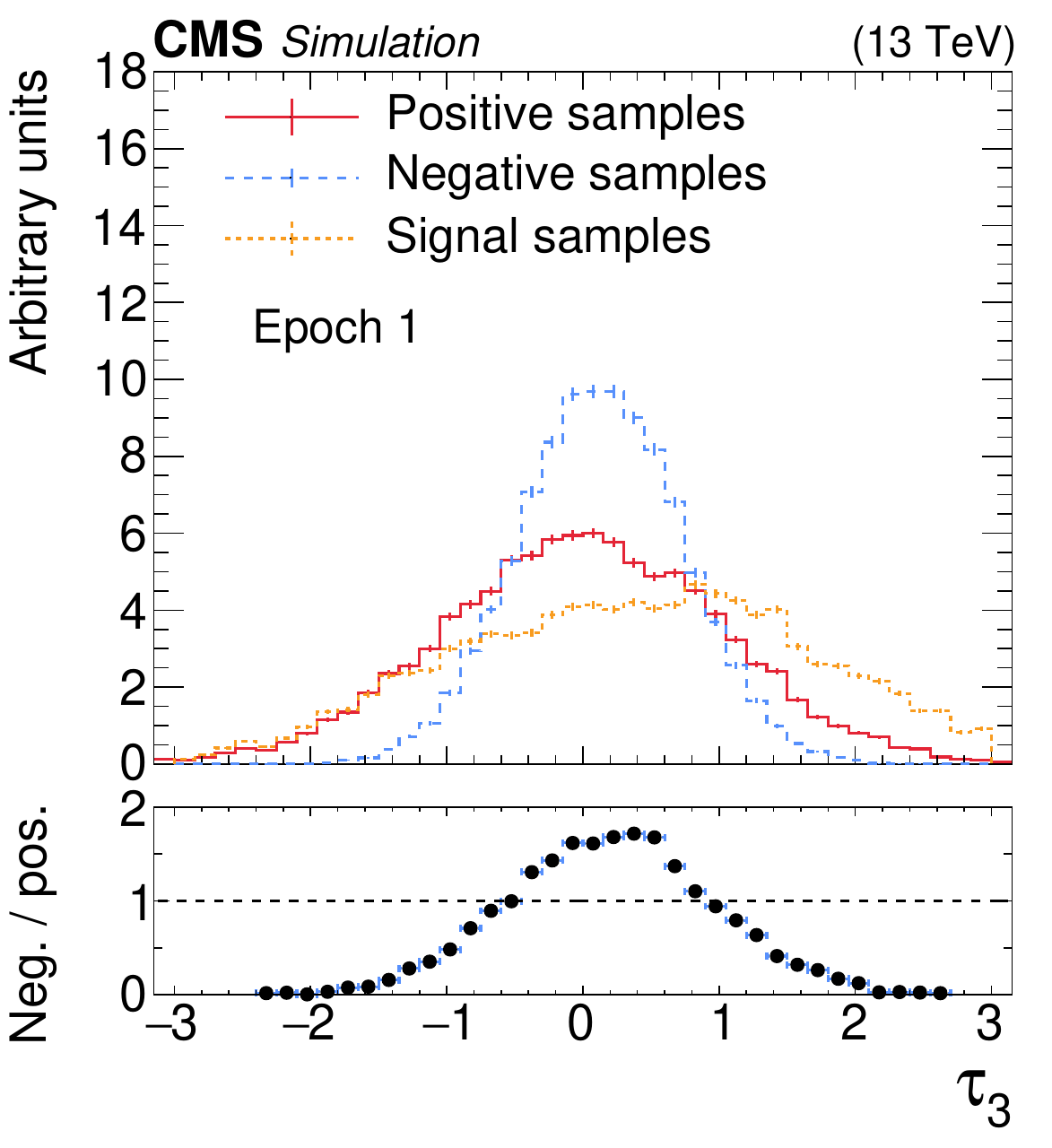}
    \includegraphics[width=0.24\textwidth]{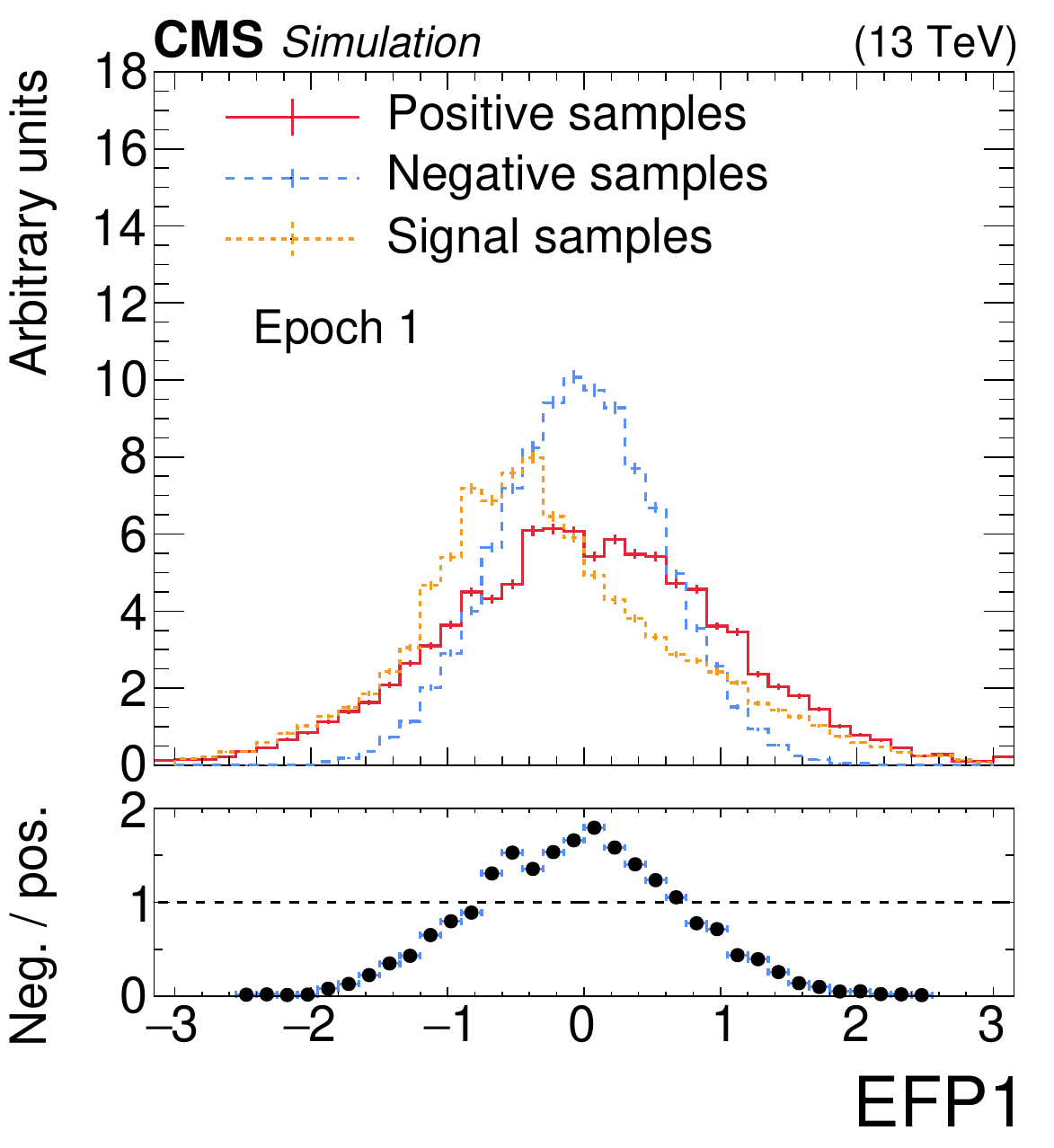}
    \includegraphics[width=0.24\textwidth]{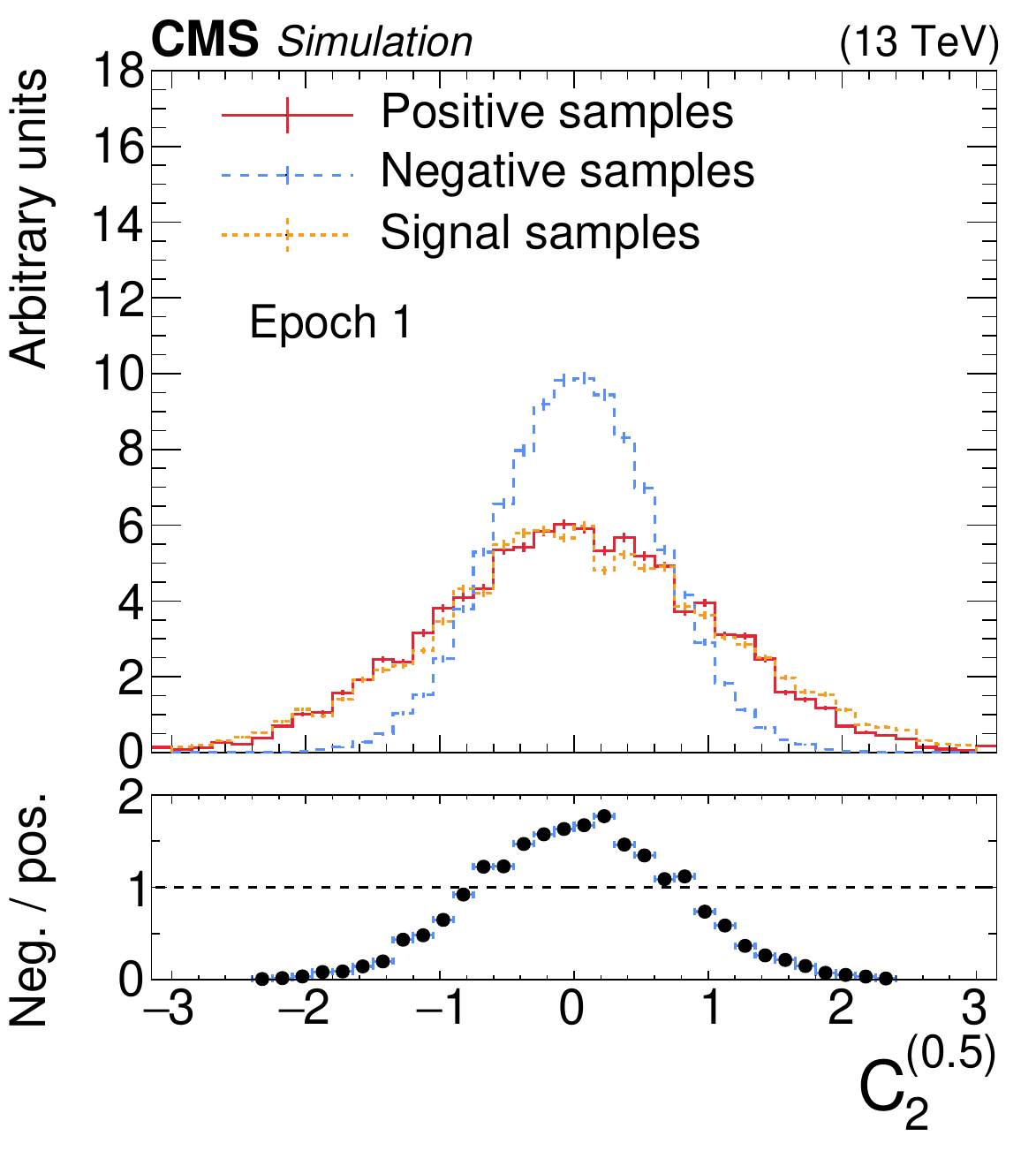}\\
    \includegraphics[width=0.24\textwidth]{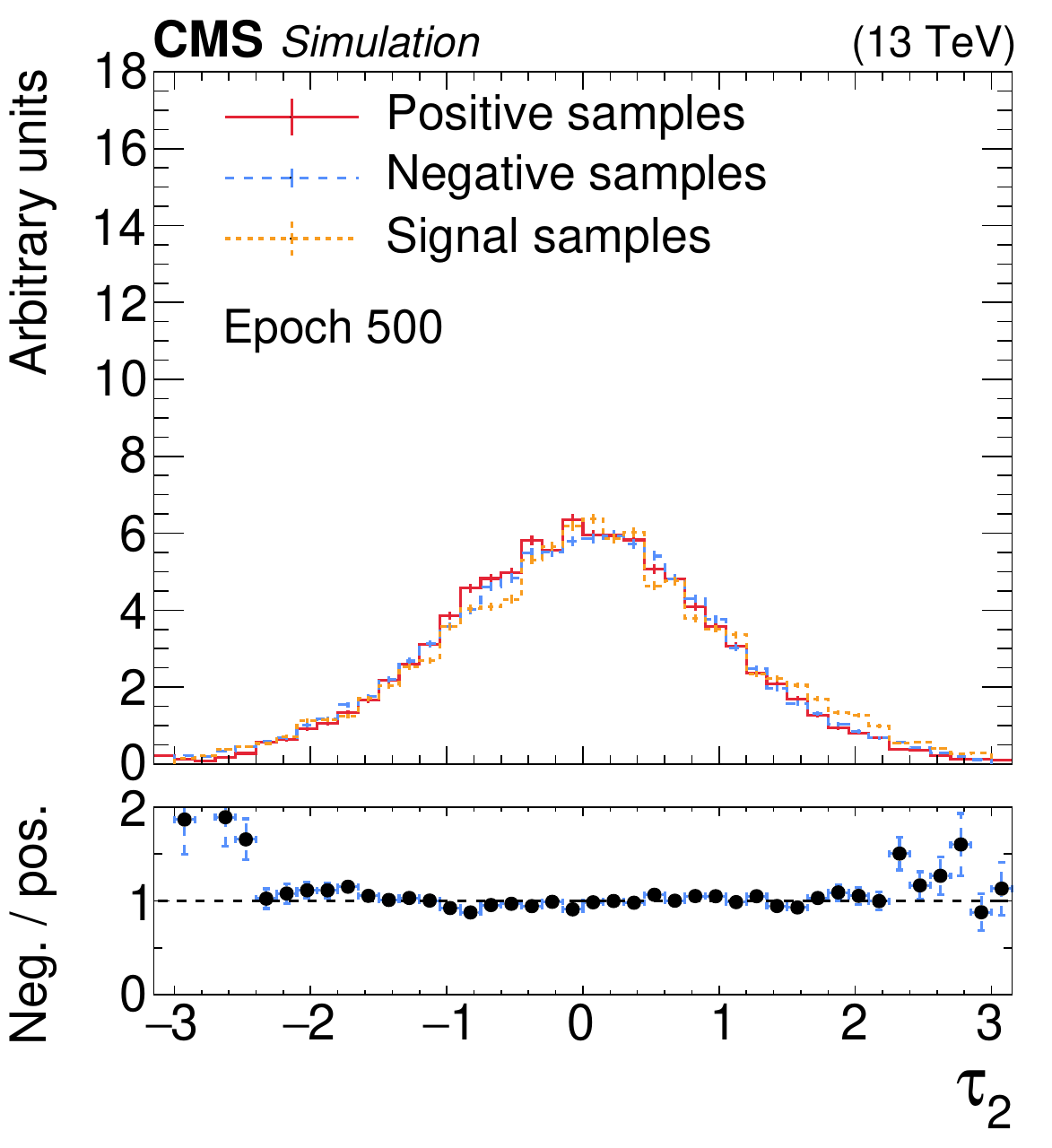}
    \includegraphics[width=0.24\textwidth]{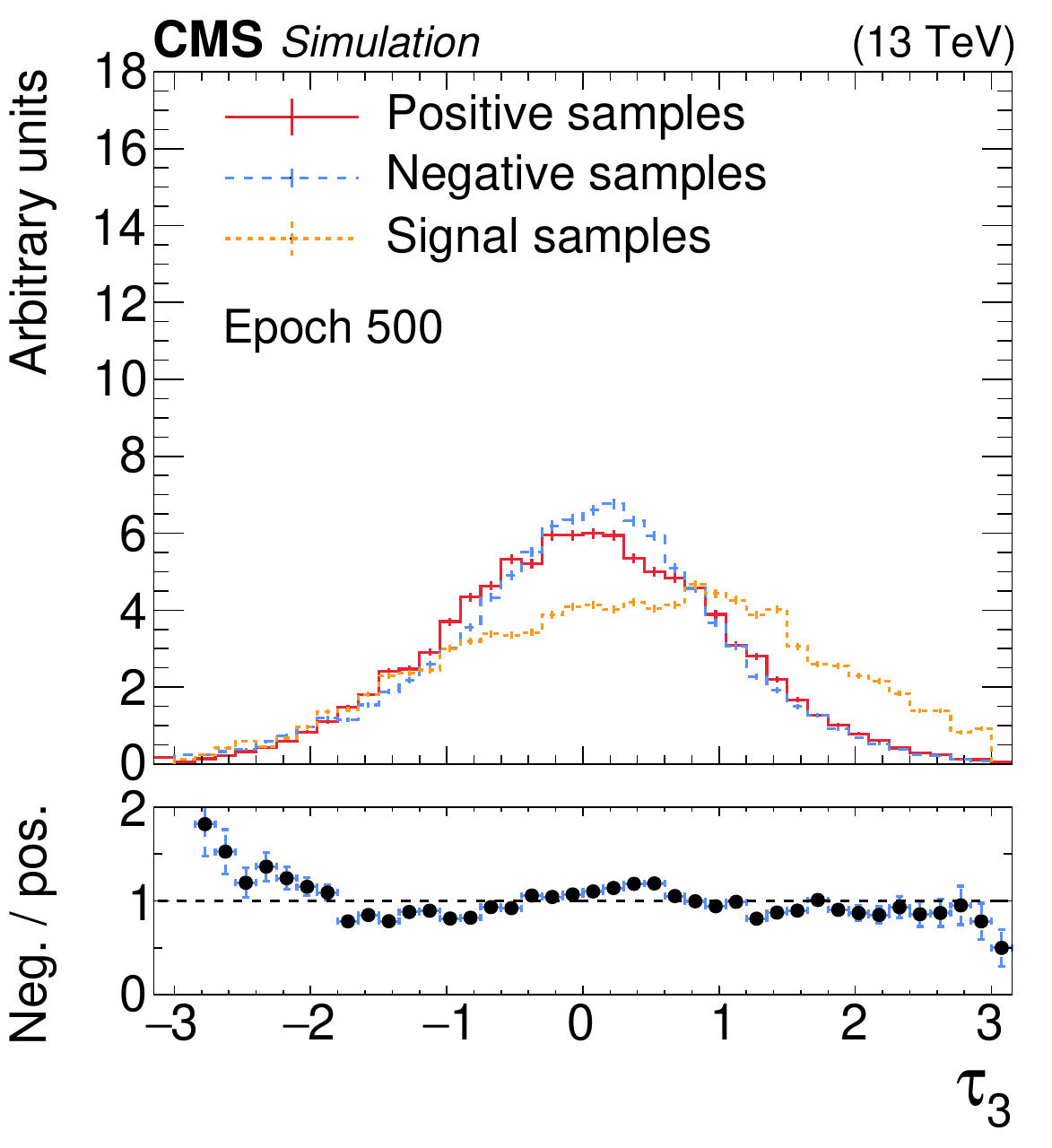}
    \includegraphics[width=0.24\textwidth]{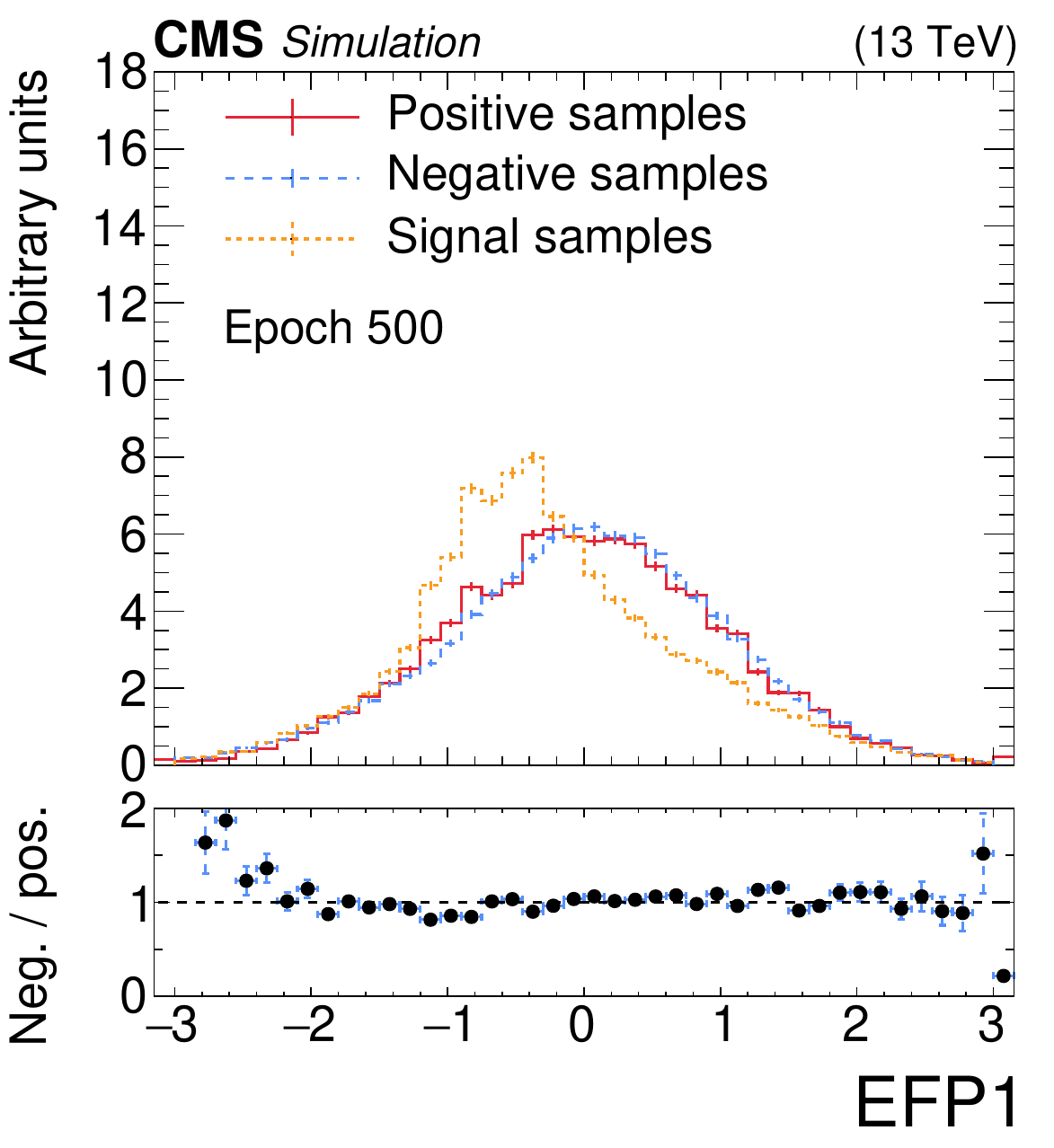}
    \includegraphics[width=0.24\textwidth]{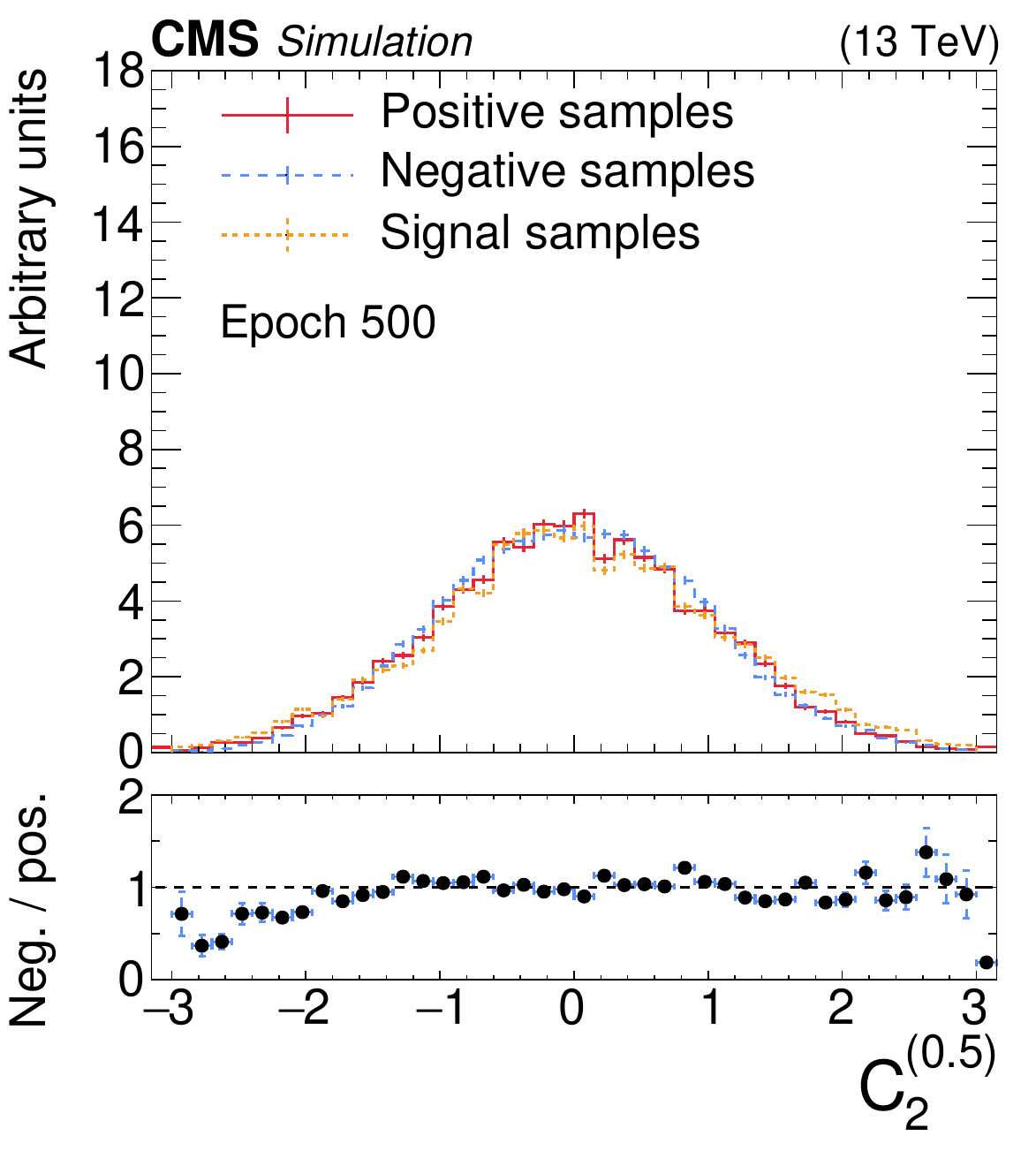}
    \caption{The distributions of half of the input variables, \tautwo, \tauthr, EFP1, and \ctwo, for the positive, negative, and signal samples, at the start (upper) and at the end (lower) of the \wnae training. The signal distributions are overlaid for illustration; signal samples are not used during the training. All distributions are normalized such that their integral is 100.}
    \label{fig:wae_input_features_epoch_1}
	\medskip
    \includegraphics[width=0.24\textwidth]{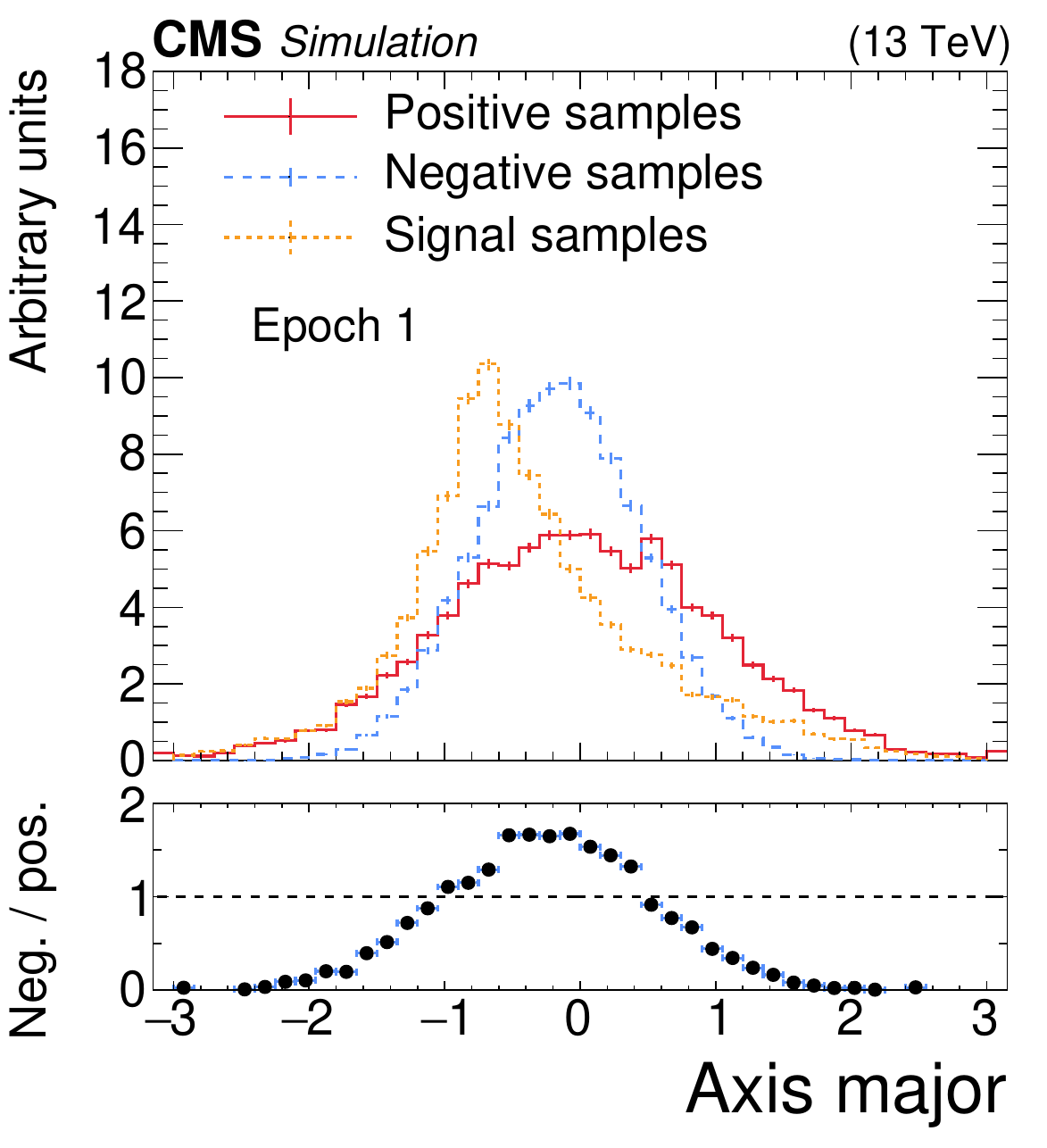}
    \includegraphics[width=0.24\textwidth]{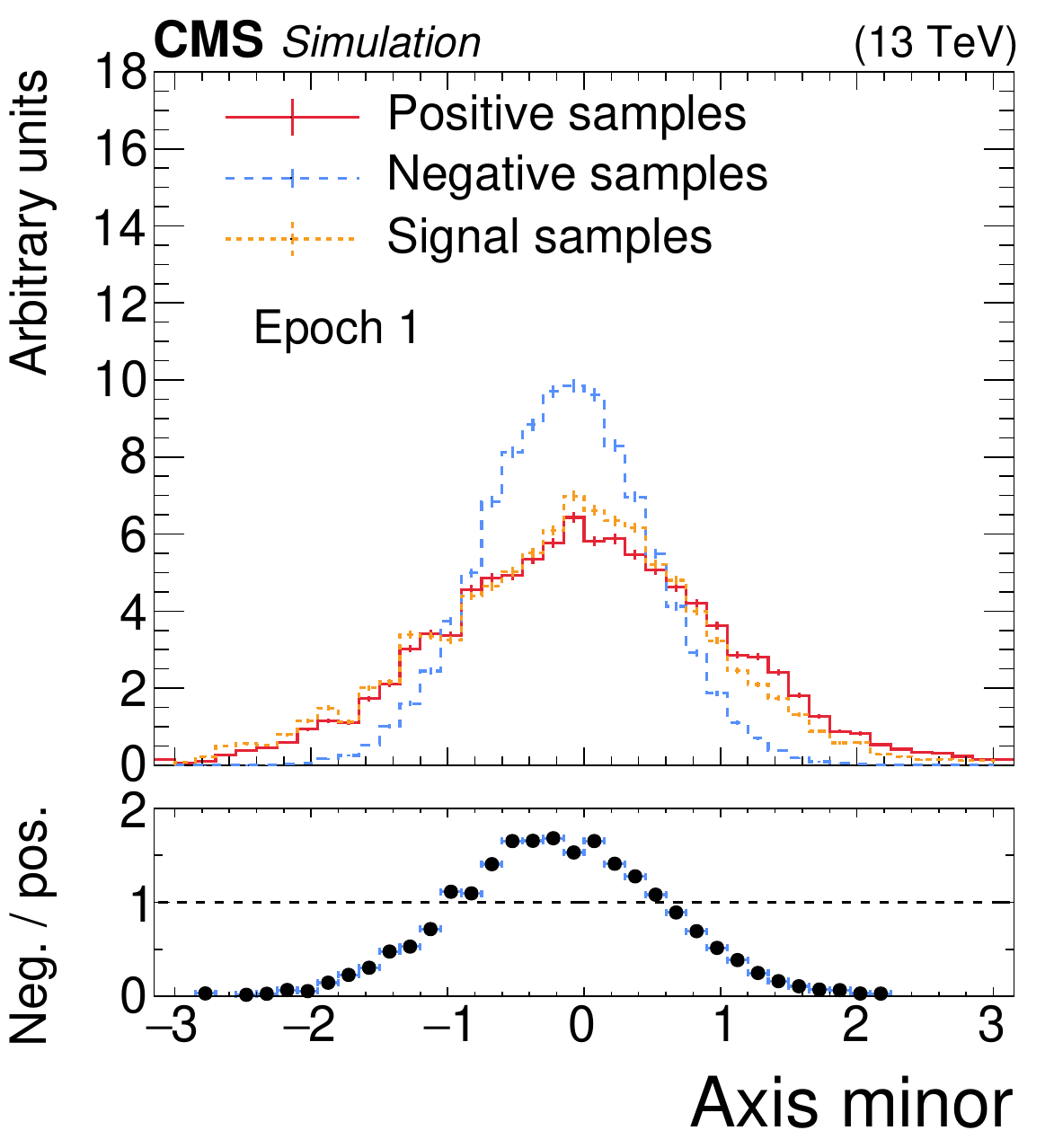}
    \includegraphics[width=0.24\textwidth]{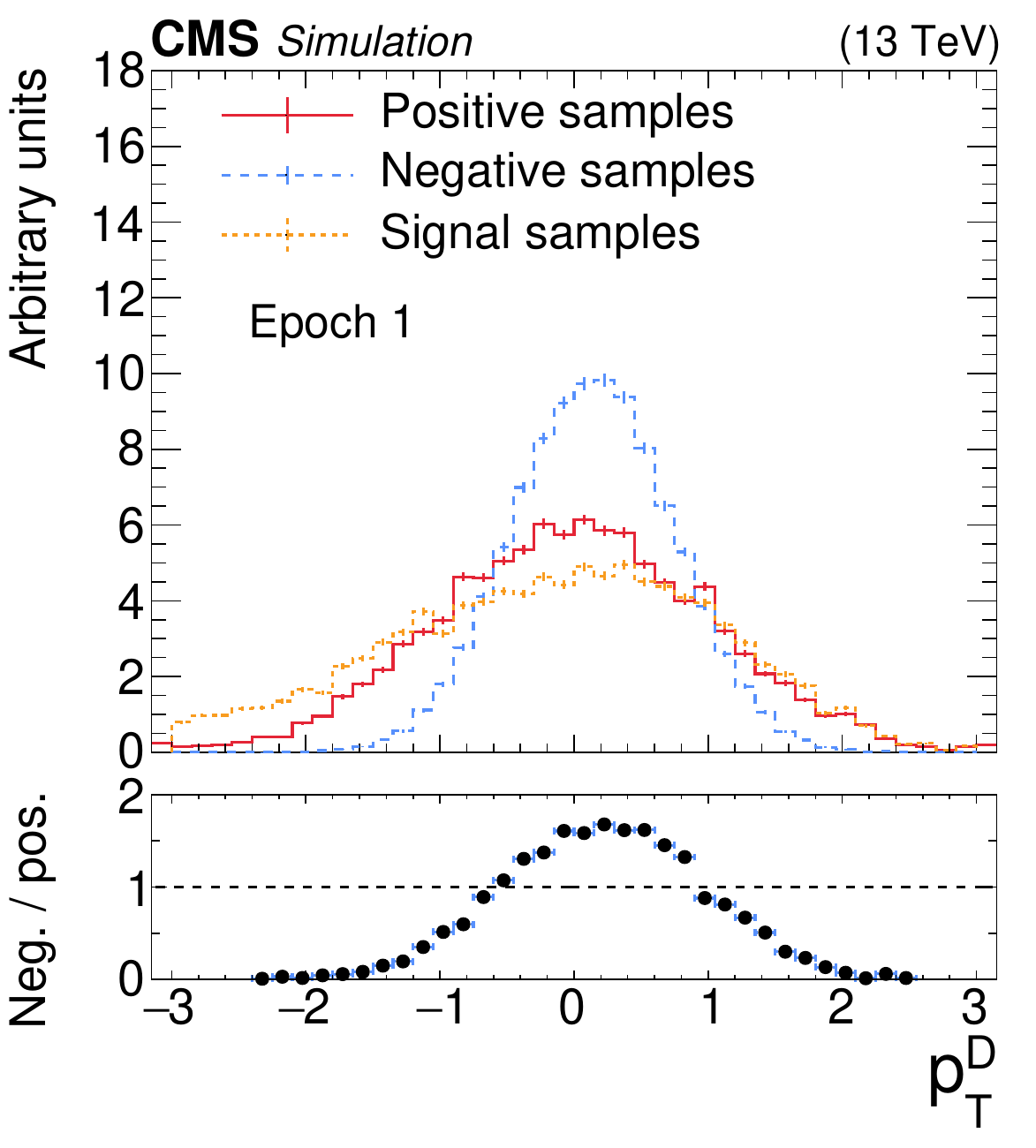}
    \includegraphics[width=0.24\textwidth]{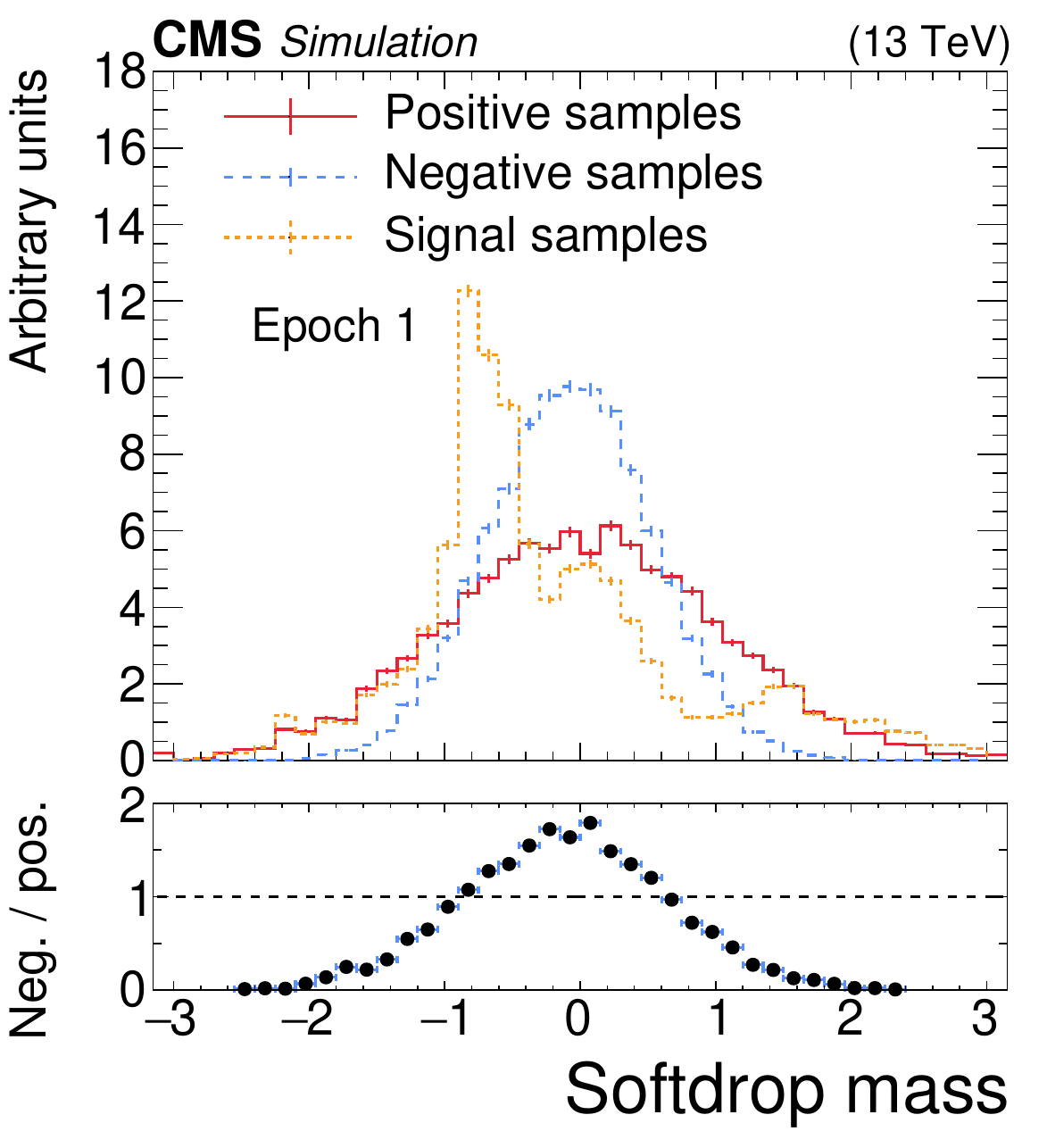}\\
    \includegraphics[width=0.24\textwidth]{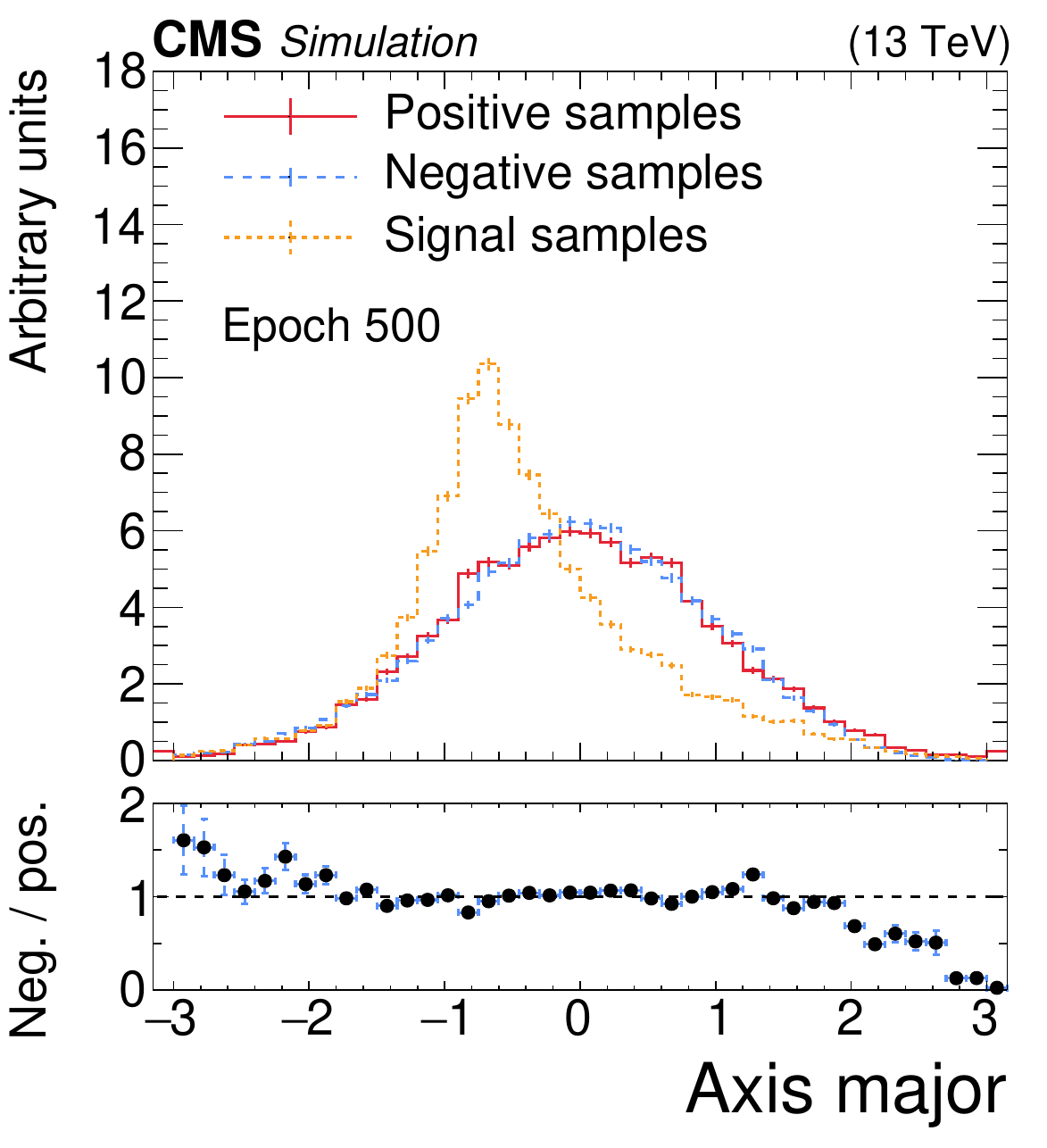}
    \includegraphics[width=0.24\textwidth]{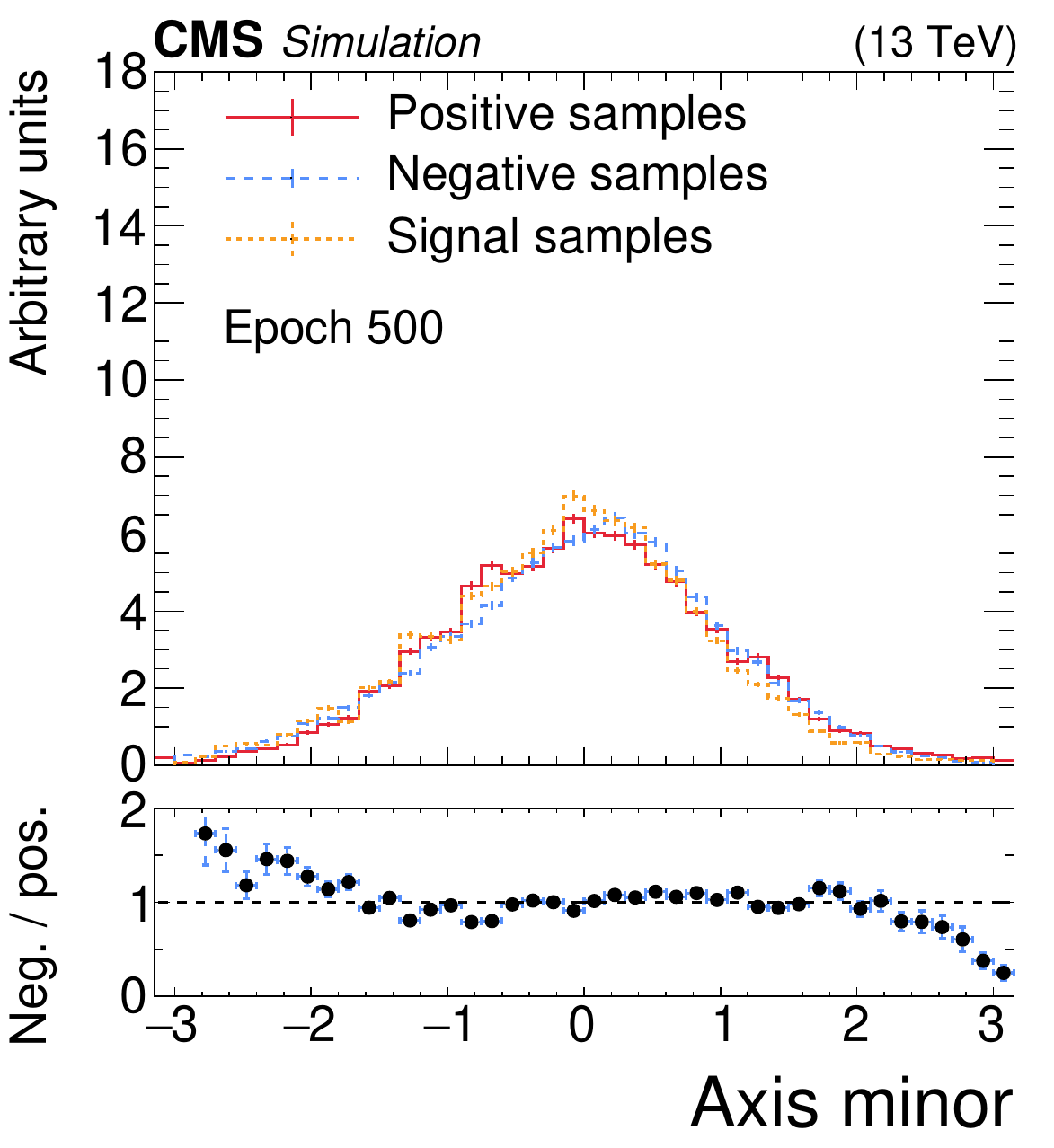}
    \includegraphics[width=0.24\textwidth]{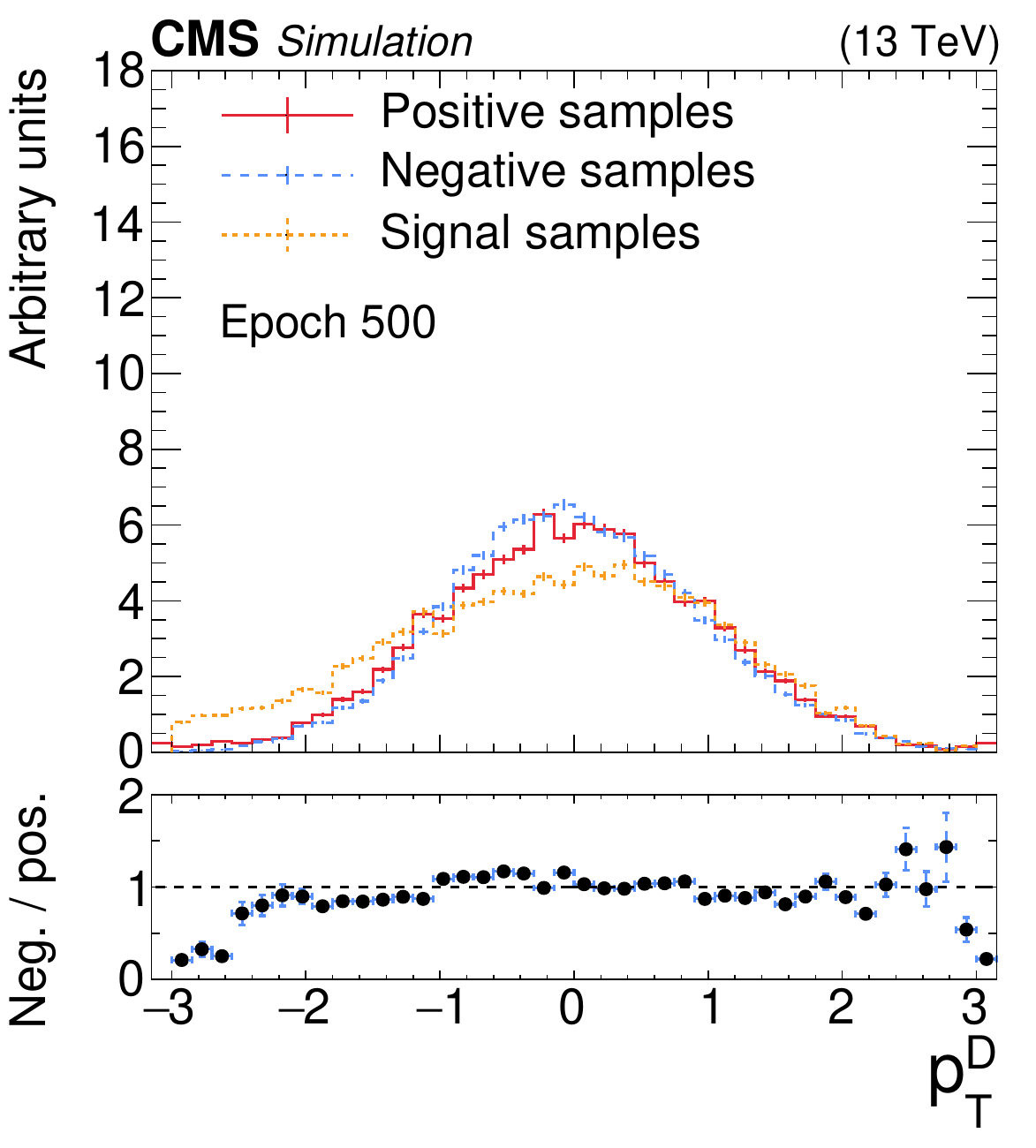}
    \includegraphics[width=0.24\textwidth]{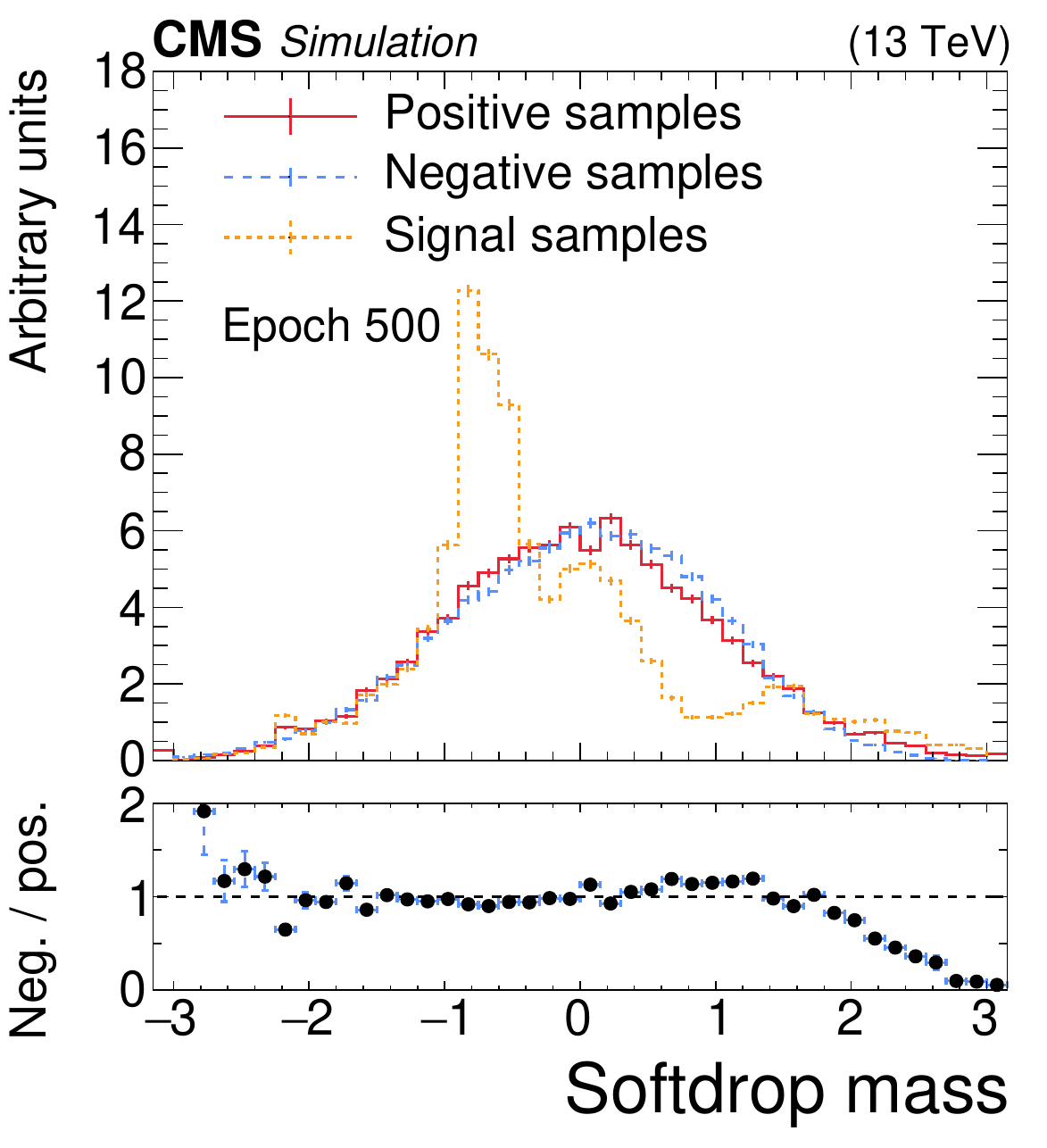}
    \caption{The distributions of the other half of the input variables, axis major, axis minor, \ptd, and softdrop mass, for the positive, negative, and signal samples, at the start (upper) and at the end (lower) of the \wnae training. The signal distributions are overlaid for illustration; signal samples are not used during the training. All distributions are normalized such that their integral is 100.}
    \label{fig:wae_input_features_epoch_500}
\end{figure}

The fact that the \wnae behaves as desired can be seen in Figs.~\ref{fig:wae_input_features_epoch_1} and~\ref{fig:wae_input_features_epoch_500}, where the distributions of the input features for the positive, negative, and signal samples are shown at the start and end of the training. Starting from an essentially multivariate distribution, induced by the random weight initialization of the network, the \wnae is able to learn the distribution of the training data, as evidenced by the distribution of negative samples eventually converging toward that of the positive samples.

\subsection{Complexity bias in the Wasserstein normalized autoencoder}

\begin{figure}[ht]
    \centering
    \includegraphics[width=0.49\textwidth]{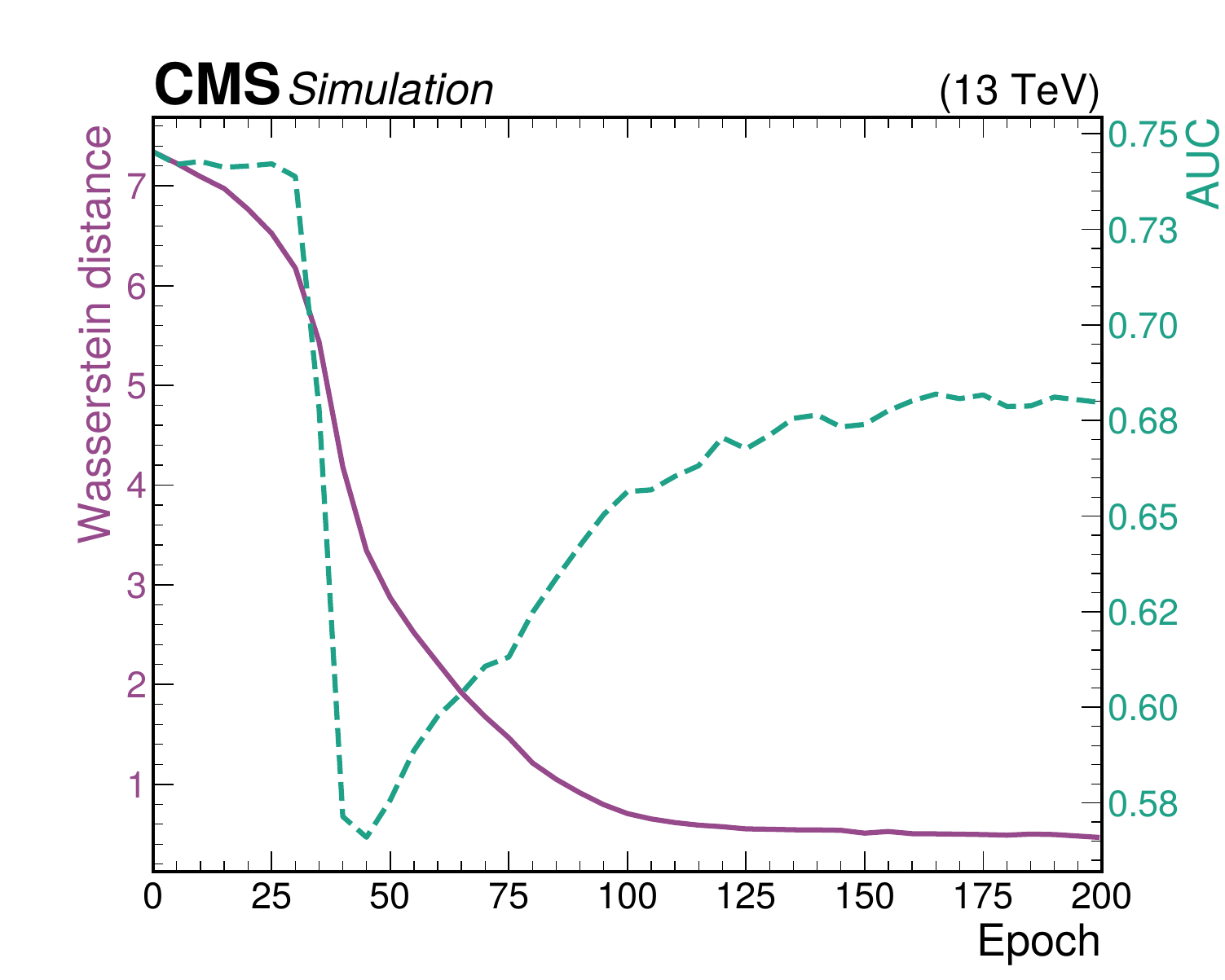}
    \caption{The Wasserstein distance between the positive and negative samples and the AUC score during the training of a \wnae on an SVJ signal ($\mMed = 2000\GeV$, $\rinv = 0.3$), with the \ttbar background used for testing.}
    \label{fig:wae_signal_training}
\end{figure}

A common weakness of plain AEs is the so-called complexity bias, in which the AE tends to be able to discriminate effectively only against anomalous examples whose distribution in input space is more complex than that of the training data. This can be demonstrated by inverting the background and signal samples and observing that, in this case, an AE trained on SM jets can tag SVJs as anomalous, but the reverse is not true. Because the \wnae learns the probability distribution of the training data in input space, this issue is mitigated, as shown in Fig.~\ref{fig:wae_signal_training}. The high AUC at the start of the training is caused by the random uniform initialization of the network weights and the different correlations between the input features of the signal and background jets. The AUC then decreases as the network learns a distribution closer to the physical input data, before increasing again as the network starts to more precisely learn the distribution of SVJ examples and thus becomes able to tag SM jets as anomalous.

\section{Conclusions}
\label{sec:summary}

Anomaly detection using autoencoders (AEs) relies on learning a reconstruction function that gives high reconstruction error to phase space regions with low probability density, such that they can be identified as anomalous. However, standard AEs are prone to learn to reconstruct outliers because they are free to minimize the reconstruction error outside the training phase space.
In addition, they may exhibit complexity bias, learning to identify examples as anomalous only if their feature distribution is more complex than the training data.
The normalized autoencoder (NAE) paradigm promotes the AE reconstruction error to an energy function in the framework of energy-based models, in order to define a normalized probabilistic model. This is achieved by minimizing the negative log-likelihood of the training data from the energy-based model probability.
In practice, this method presents a number of failure modes, such as divergence of the loss function and phase space degeneracy, leading to low reconstruction error for phase space regions distinct from the training data.

The Wasserstein normalized autoencoder (\wnae), an improvement over the NAE, is introduced to solve these failure modes. This is achieved by directly minimizing the Wasserstein distance between the probability distribution of the training data and the Boltzmann distribution of the energy function of the model.
This Wasserstein distance is found to be highly correlated with signal identification performance while still being fully signal-agnostic, preserving the unsupervised nature of the approach.
The performance is studied in the context of a search for new physics with the CMS experiment, using top-antitop quark production as the standard model background and nonresonant semivisible jet production from a strongly coupled dark sector as the proposed signal model.
The classification of the signal jets as outliers by the \wnae is shown to be on par with or better than that of the NAE.
Further, the \wnae approach is found to mitigate complexity bias, as it can effectively identify top quark jets as anomalous when trained on semivisible jets from signal events.
While the \wnae is found to be stable and effective for the anomaly detection task studied here, it may suffer from the generic failure modes inherent to anomaly detection models and energy-based models, including: overlap of the signal and background distributions in input space; training data contaminated by anomalous examples; or failure to effectively sample the model probability distribution.

Though simulated samples were used to develop the \wnae, in practice it may be preferable to use observed data directly for training, in order to limit biases arising from differences between simulation and observation.
In this case, the training data may contain anomalies, which would reduce the anomaly detection performance of the \wnae.
The \wnae can be straightforwardly trained using observed data from a control region with no anomalous examples, if such a region can be defined and follows the same probability distribution as the observed data where the \wnae will be applied.
When no assumption at all can be made about the nature of the anomalies, such as the case of triggering at a high-energy physics experiment, alternative solutions may exist.
The \wnae associates low probability density regions with high reconstruction error; because anomalies necessarily have low probability density, they still tend to have relatively high reconstruction error even when included in the training data set.
Therefore, the training data set can be iteratively refined by selecting a given fraction of examples with the lowest reconstruction error, in order to reduce the proportion of anomalous data.
This would result in a self-supervised training for the \wnae.
We leave the development of such a procedure for future work.

When applying this method to learn higher-dimensional probability distributions, the slower convergence of the Wasserstein distance with increasing dimensionality may pose practical challenges. For an arbitrary $d$-dimensional distribution $\mu$ and an independent and identically distributed sample from $\mu$ with $n$ data points $\mu_n$, the convergence rate of the 1-Wasserstein distance between $\mu$ and $\mu_n$ is $n^{-1/d}$: $W(\mu, \mu_n) \gtrsim n^{-1/d}$~\cite{Dud68, weed2019sharp}. To circumvent this behavior, approximations of the Wasserstein distance that do not suffer from this curse of dimensionality can be employed. The sliced Wasserstein distance~\cite{sw1,sw2} has been increasingly favored because of its interesting statistical properties and low computational cost. It approximates the Wasserstein distance by computing the average of one-dimensional Wasserstein distances---for which an analytical and easily computable solution is known---between projections of the two probability distributions uniformly distributed on the unit sphere. Its convergence is equivalent to the weak convergence of the exact Wasserstein distance~\cite{nadjahi2019asymptotic} and its convergence rate does not depend on the number of dimensions~\cite{nadjahi2020statistical}. The sliced Wasserstein distance therefore converges faster than the exact Wasserstein distance when the number of dimensions increases. Differentiable implementations are available, so it can straightforwardly be used in place of the exact Wasserstein distance to perform anomaly detection in very high-dimensional feature spaces.

\begin{acknowledgments}

We congratulate our colleagues in the CERN accelerator departments for the excellent performance of the LHC and thank the technical and administrative staffs at CERN and at other CMS institutes for their contributions to the success of the CMS effort. In addition, we gratefully acknowledge the computing centers and personnel of the Worldwide LHC Computing Grid and other centers for delivering so effectively the computing infrastructure essential to our analyses. Finally, we acknowledge the enduring support for the construction and operation of the LHC, the CMS detector, and the supporting computing infrastructure provided by the following funding agencies: SC (Armenia), BMBWF and FWF (Austria); FNRS and FWO (Belgium); CNPq, CAPES, FAPERJ, FAPERGS, and FAPESP (Brazil); MES and BNSF (Bulgaria); CERN; CAS, MoST, and NSFC (China); MINCIENCIAS (Colombia); MSES and CSF (Croatia); RIF (Cyprus); SENESCYT (Ecuador); ERC PRG, TARISTU24-TK10 and MoER TK202 (Estonia); Academy of Finland, MEC, and HIP (Finland); CEA and CNRS/IN2P3 (France); SRNSF (Georgia); BMFTR, DFG, and HGF (Germany); GSRI (Greece); NKFIH (Hungary); DAE and DST (India); IPM (Iran); SFI (Ireland); INFN (Italy); MSIT and NRF (Republic of Korea); MES (Latvia); LMTLT (Lithuania); MOE and UM (Malaysia); BUAP, CINVESTAV, CONACYT, LNS, SEP, and UASLP-FAI (Mexico); MOS (Montenegro); MBIE (New Zealand); PAEC (Pakistan); MES, NSC, and NAWA (Poland); FCT (Portugal);  MESTD (Serbia); MICIU/AEI and PCTI (Spain); MOSTR (Sri Lanka); Swiss Funding Agencies (Switzerland); MST (Taipei); MHESI (Thailand); TUBITAK and TENMAK (T\"{u}rkiye); NASU (Ukraine); STFC (United Kingdom); DOE and NSF (USA).

\hyphenation{Rachada-pisek} Individuals have received support from the Marie-Curie program and the European Research Council and Horizon 2020 Grant, contract Nos.\ 675440, 724704, 752730, 758316, 765710, 824093, 101115353, 101002207, 101001205, and COST Action CA16108 (European Union); the Leventis Foundation; the Alfred P.\ Sloan Foundation; the Alexander von Humboldt Foundation; the Science Committee, project no. 22rl-037 (Armenia); the Fonds pour la Formation \`a la Recherche dans l'Industrie et dans l'Agriculture (FRIA) and Fonds voor Wetenschappelijk Onderzoek contract No. 1228724N (Belgium); the Beijing Municipal Science \& Technology Commission, No. Z191100007219010, the Fundamental Research Funds for the Central Universities, the Ministry of Science and Technology of China under Grant No. 2023YFA1605804, and the Natural Science Foundation of China under Grant No. 12061141002, 12535004 (China); the Ministry of Education, Youth and Sports (MEYS) of the Czech Republic; the Shota Rustaveli National Science Foundation, grant FR-22-985 (Georgia); the Deutsche Forschungsgemeinschaft (DFG), among others, under Germany's Excellence Strategy -- EXC 2121 ``Quantum Universe" -- 390833306, and under project number 400140256 - GRK2497; the Hellenic Foundation for Research and Innovation (HFRI), Project Number 2288 (Greece); the Hungarian Academy of Sciences, the New National Excellence Program - \'UNKP, the NKFIH research grants K 131991, K 133046, K 138136, K 143460, K 143477, K 146913, K 146914, K 147048, 2020-2.2.1-ED-2021-00181, TKP2021-NKTA-64, and 2021-4.1.2-NEMZ\_KI-2024-00036 (Hungary); the Council of Science and Industrial Research, India; ICSC -- National Research Center for High Performance Computing, Big Data and Quantum Computing, FAIR -- Future Artificial Intelligence Research, and CUP I53D23001070006 (Mission 4 Component 1), funded by the NextGenerationEU program (Italy); the Latvian Council of Science; the Ministry of Education and Science, project no. 2022/WK/14, and the National Science Center, contracts Opus 2021/41/B/ST2/01369, 2021/43/B/ST2/01552, 2023/49/B/ST2/03273, and the NAWA contract BPN/PPO/2021/1/00011 (Poland); the Funda\c{c}\~ao para a Ci\^encia e a Tecnologia, grant CEECIND/01334/2018 (Portugal); the National Priorities Research Program by Qatar National Research Fund; MICIU/AEI/10.13039/501100011033, ERDF/EU, "European Union NextGenerationEU/PRTR", and Programa Severo Ochoa del Principado de Asturias (Spain); the Chulalongkorn Academic into Its 2nd Century Project Advancement Project, the National Science, Research and Innovation Fund program IND\_FF\_68\_369\_2300\_097, and the Program Management Unit for Human Resources \& Institutional Development, Research and Innovation, grant B39G680009 (Thailand); the Kavli Foundation; the Nvidia Corporation; the SuperMicro Corporation; the Welch Foundation, contract C-1845; and the Weston Havens Foundation (USA).
\end{acknowledgments}

\bibliography{auto_generated}

\appendix
\numberwithin{figure}{section}
\numberwithin{table}{section}

\section{Input features in physical space}
\label{sec:input_features}

\begin{figure}[tp!]
    \centering
    \includegraphics[width=0.24\textwidth]{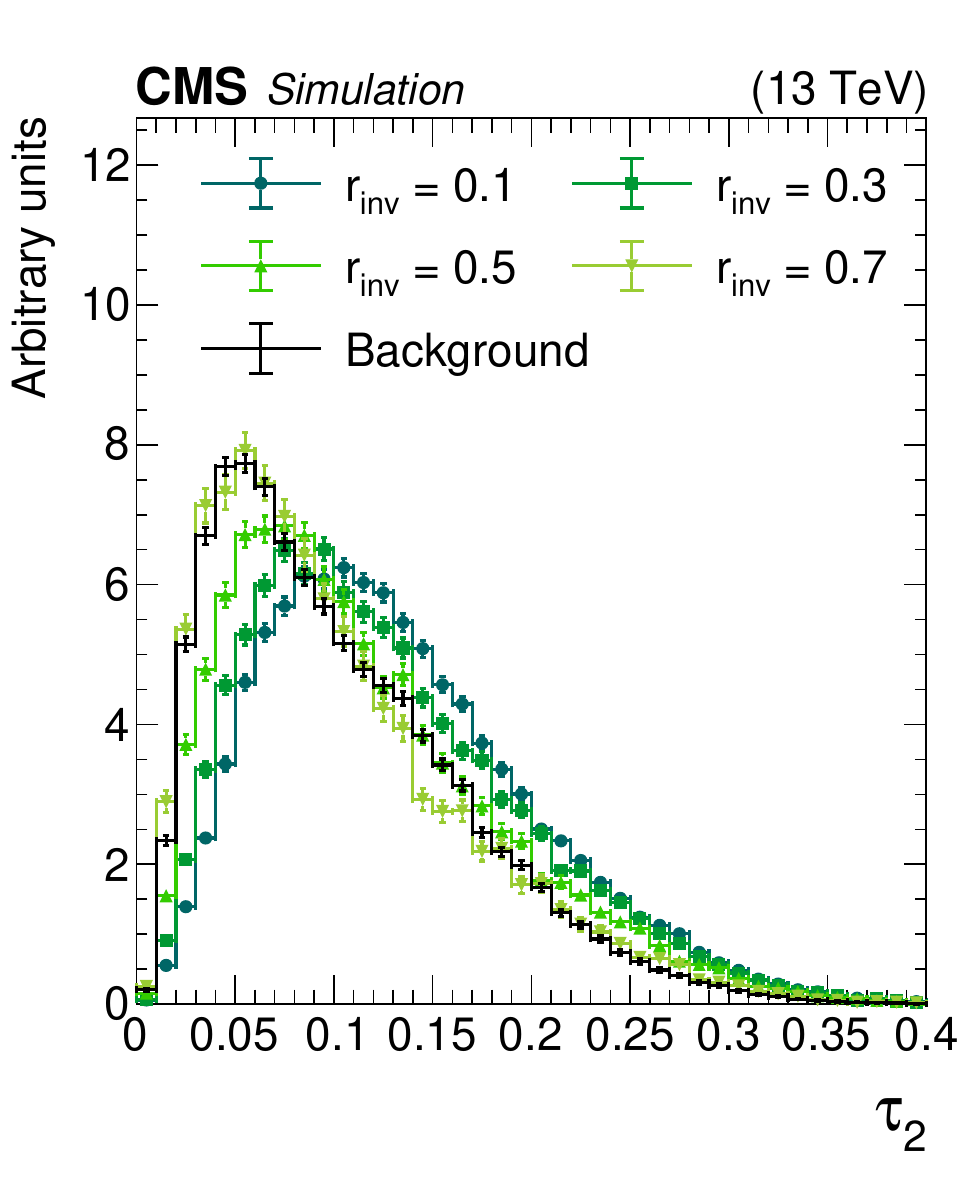}
    \includegraphics[width=0.24\textwidth]{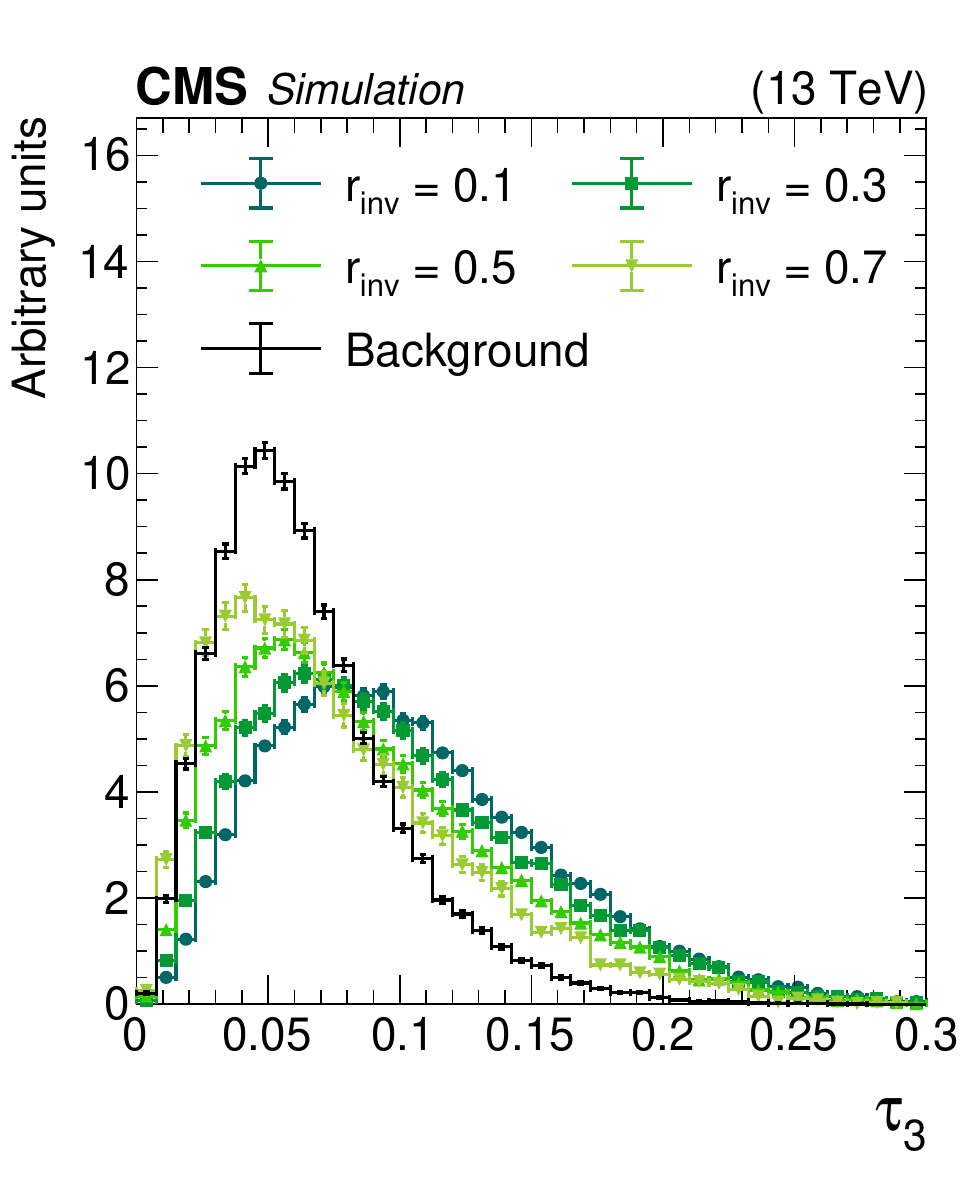}
    \includegraphics[width=0.24\textwidth]{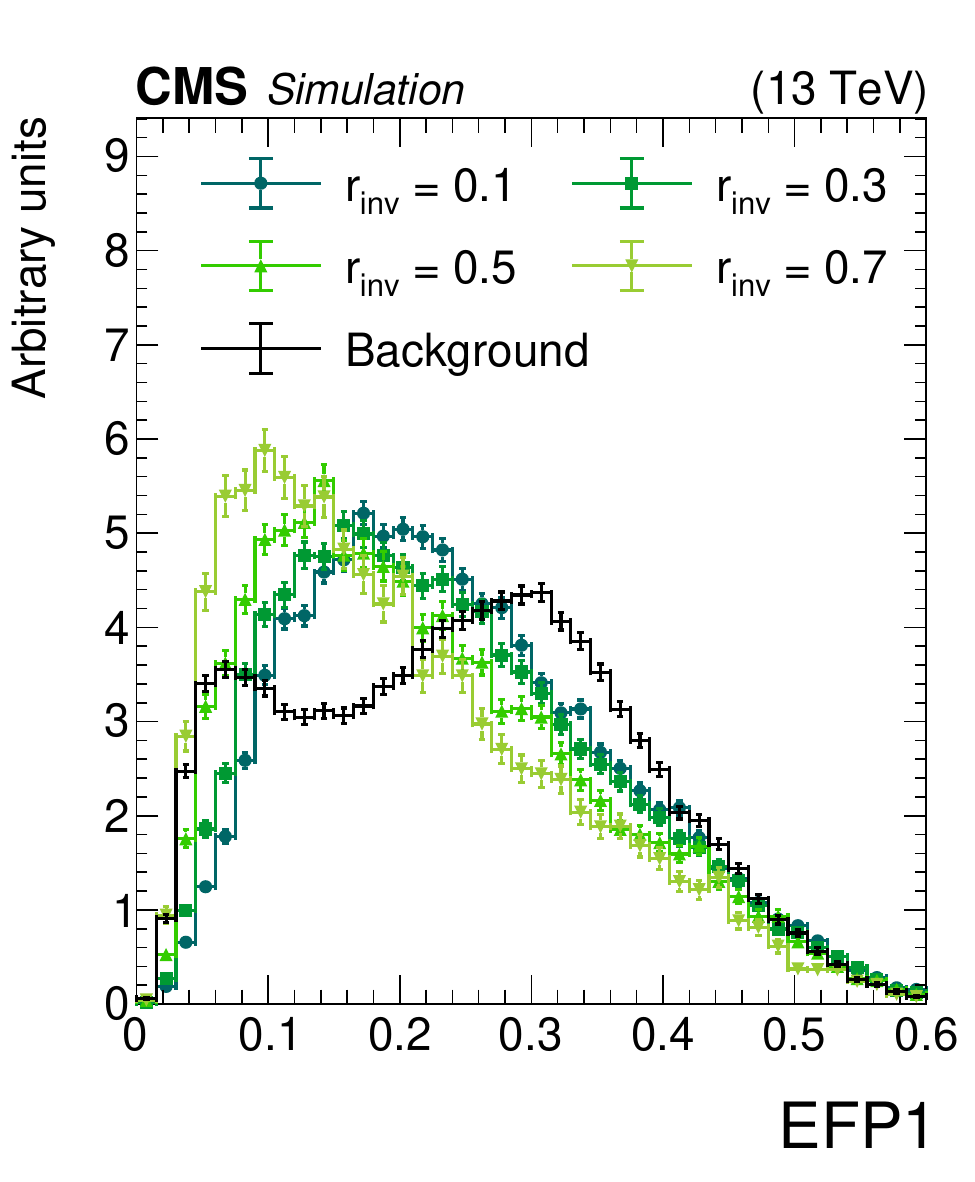}
    \includegraphics[width=0.24\textwidth]{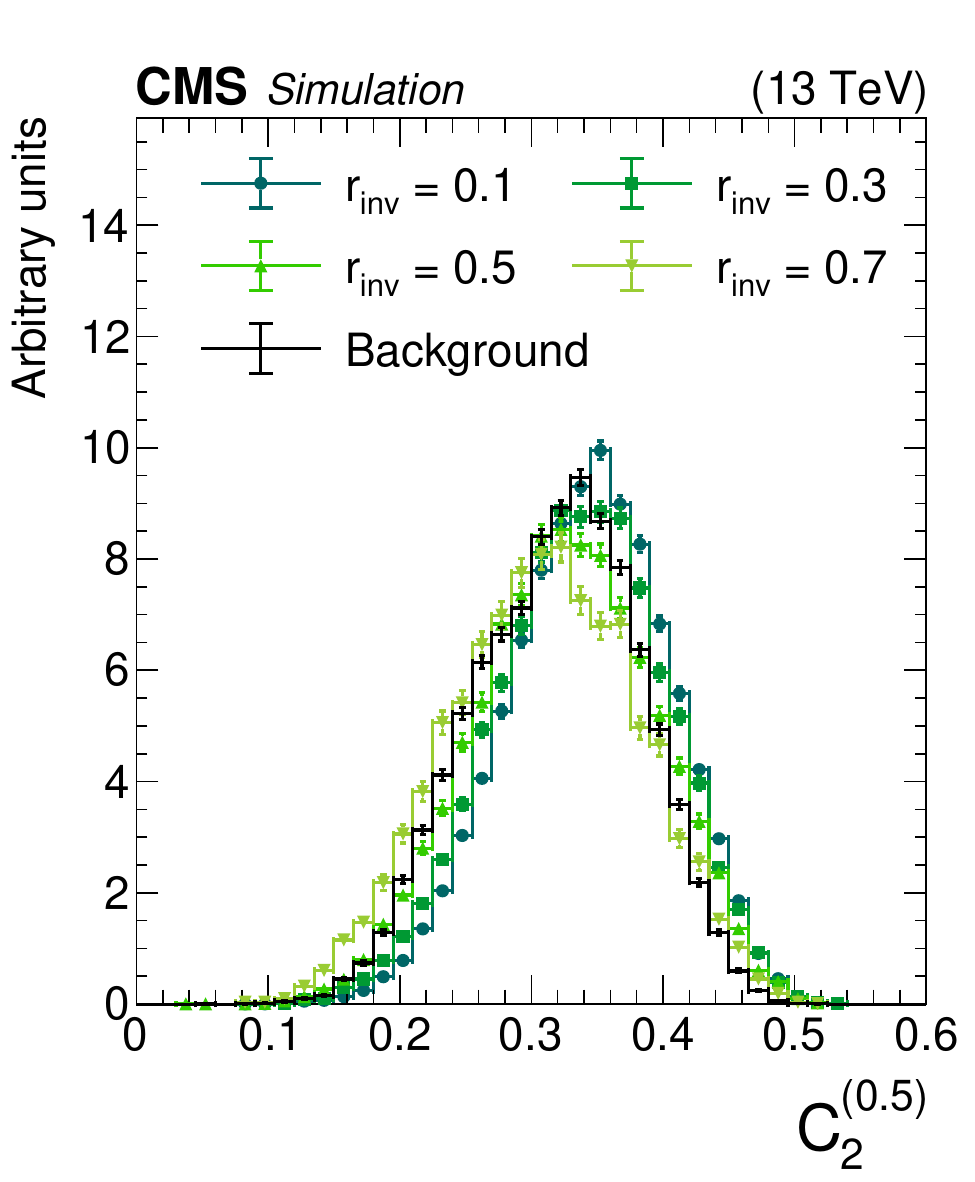}\\
    \includegraphics[width=0.24\textwidth]{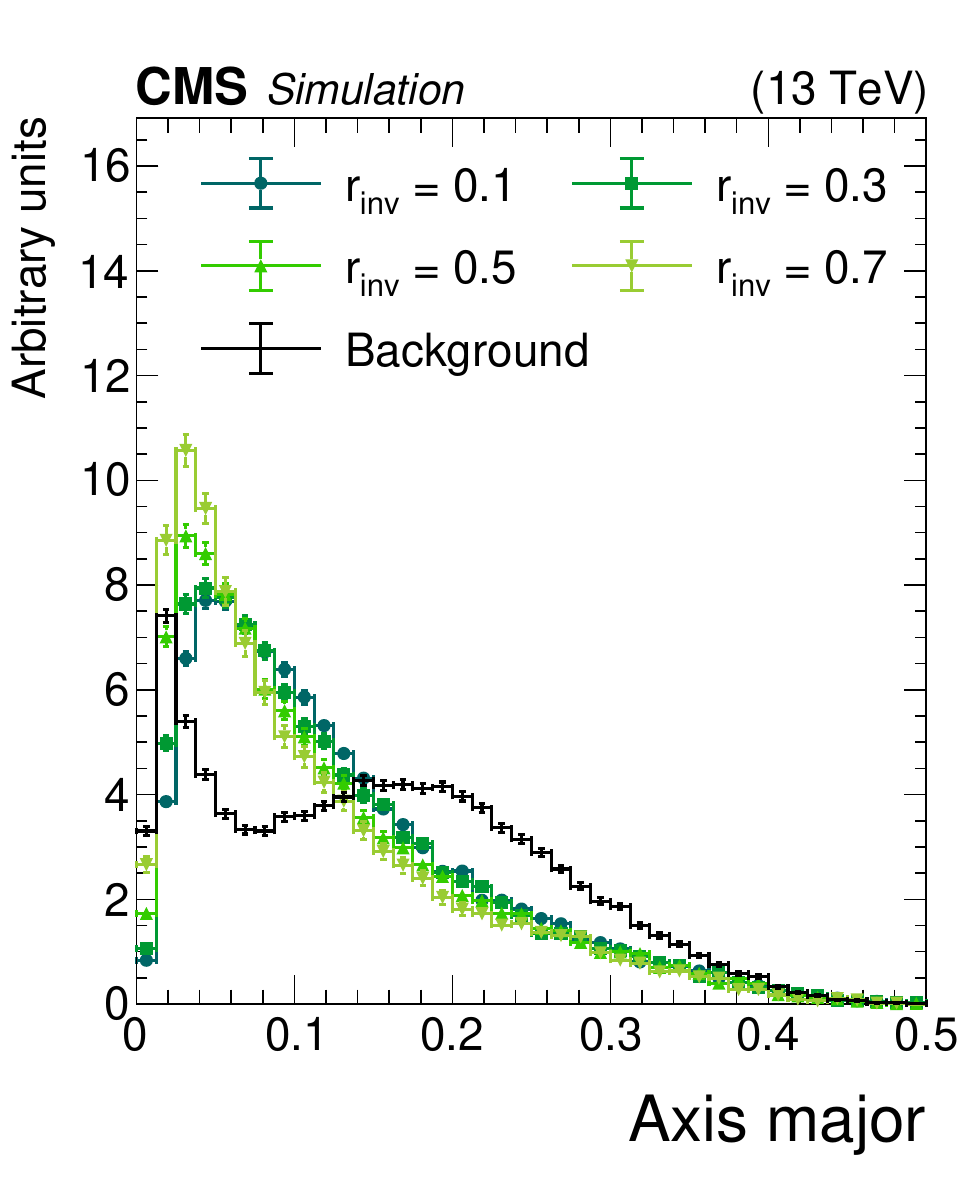}
    \includegraphics[width=0.24\textwidth]{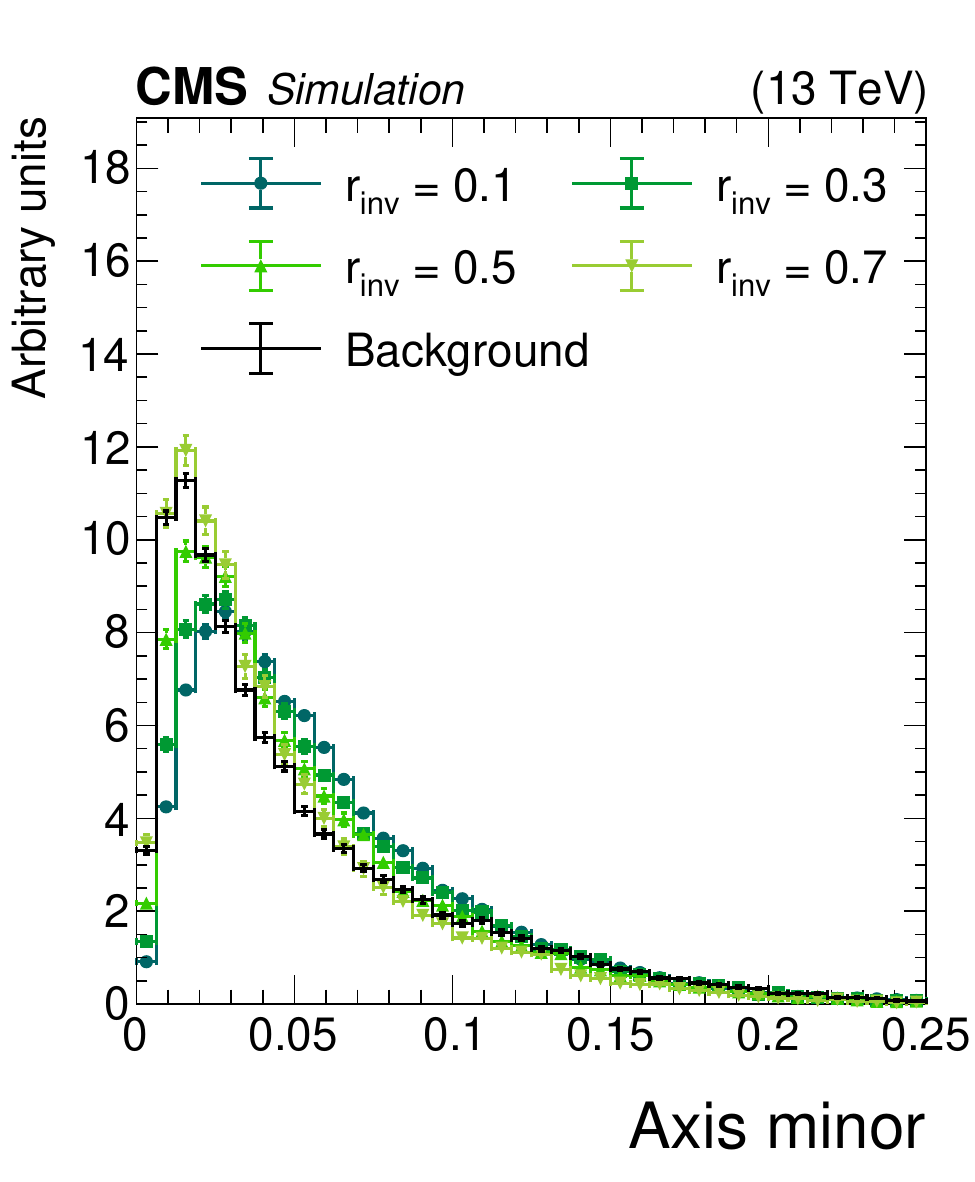}
    \includegraphics[width=0.24\textwidth]{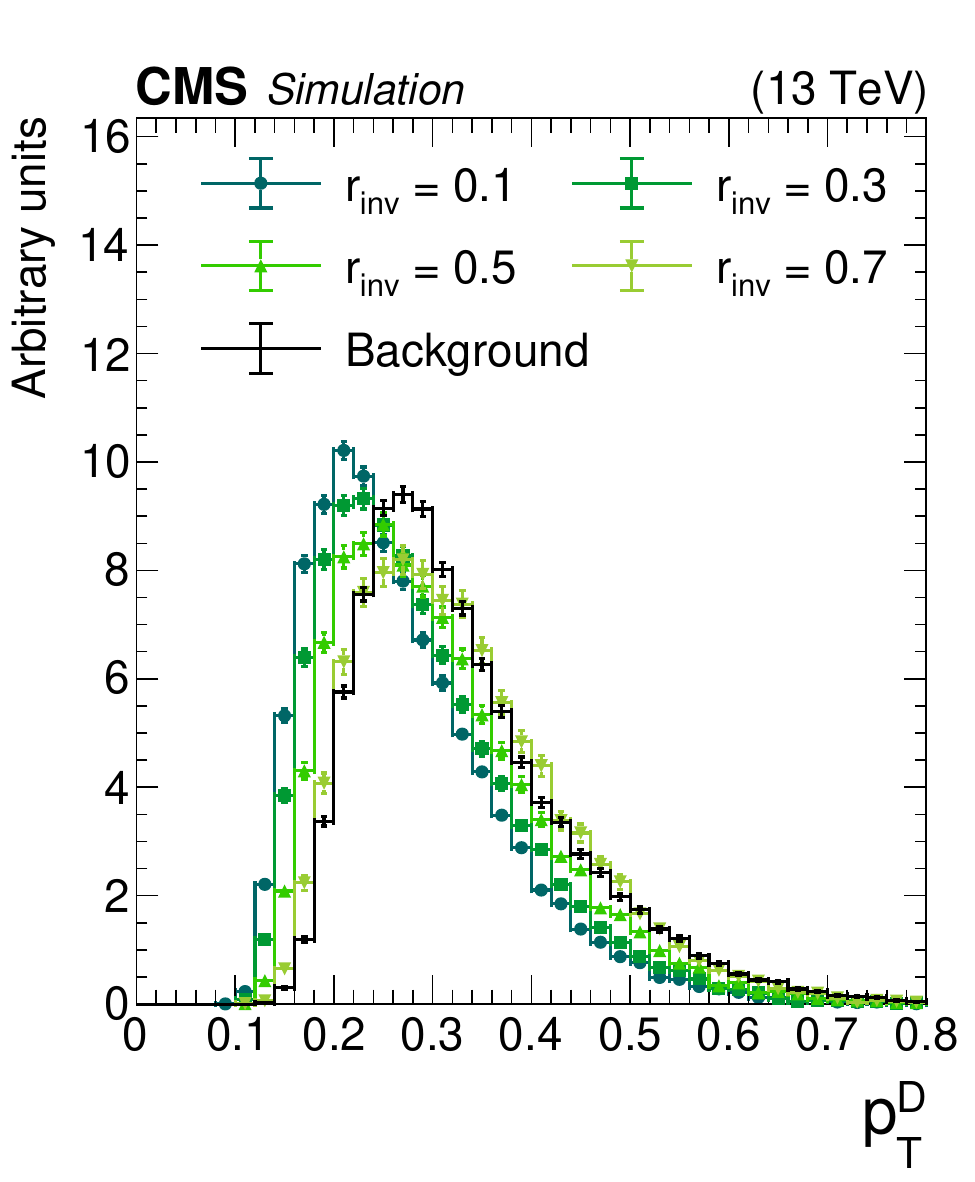}
    \includegraphics[width=0.24\textwidth]{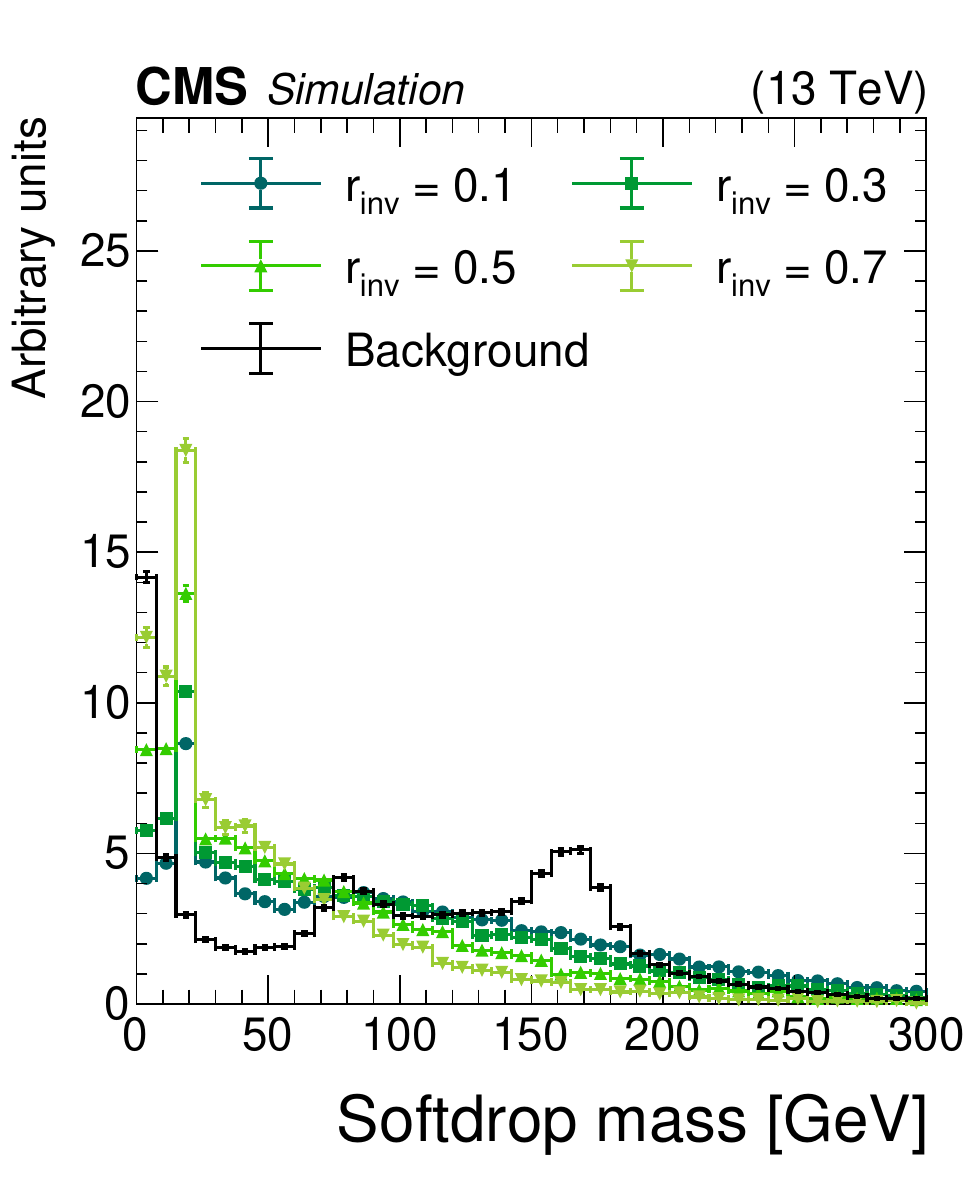}
    \caption{The distributions of the input variables in the physical space, before any preprocessing, for the \ttbar background and several signal hypotheses with fixed $\mMed = 2000\GeV$ and varying \rinv. All distributions are normalized such that their integral is 100.}
    \label{fig:input_features_physical_space}
\end{figure}

The distributions of the 8 input features used to train the models presented in this paper are presented in Fig.~\ref{fig:input_features_physical_space}, in the physical space before any preprocessing is applied. Semivisible jets tend to resemble background \ttbar jets more when \rinv increases.

\section{Gradient dependence between MCMC and AE}
\label{sec:ae_weights_dependency_in_mcmc}

During the MCMC process, the gradient with respect to the feature space needs to be calculated to compute each step.
At the same time, the dependency on the weights of the neural network must be retained in order to compute the gradient with respect to the weights for backpropagation.

For generative applications, a model $g_{\theta}$, parametrized by weights $\theta$, may be trained to minimize the Wasserstein distance $W$ between the training examples $x$ that it is trying to replicate and the samples $g_{\theta}(z)$ that it generates from random noise $z$ with prior distribution $p_z$:
\begin{equation}
    \mathcal{L}_\text{generative} = \inf_{\gamma \in \Pi(p_{\mathrm{data}}, p_z)} \mathbb{E}_{(x, z)\sim \gamma}\left[\|x - g_{\theta}(z)\|\right].
\end{equation}
In this case, the Wasserstein distance depends explicitly on the network weights $\theta$ via $g_{\theta}(z)$.

However, this is not the case for an AE with the loss function defined in Eq.~\eqref{eq:emd_loss}, since neither the positive nor the negative examples pass through the network explicitly. The dependency is in the generation of the negative examples, which is done via the MCMC described in Section~\ref{sec:nae}. Equation~\eqref{eq:langevin_dynamics_with_T} shows that the dependency on the network weights is in the gradient of the energy function with respect to the input features. As such, in order to use the Wasserstein distance as a loss function, the dependency of each step of the MCMC on the weights of the network must be retained:
\begin{equation}
    \deltheta x_{i+1,\theta} = \deltheta x_{i,\theta} - \lambda \deltheta \delx \Etheta(x_{i,\theta}).
	\label{eq:dep}
\end{equation}

The differentiable implementation of the Wasserstein distance metric is provided in the \textsc{pot} package~\cite{flamary2021pot}.
In the \PYTORCH package~\cite{Paszke:2019xhz}, the dependency retention in Eq.~\eqref{eq:dep} is accomplished by explicitly enabling the \texttt{create\_graph} flag when calling the \texttt{autograd} method on the AE energy in each MCMC step. This ensures that the computational graph, which \PYTORCH uses to perform backpropagation, includes the dependency of the negative examples on the network weights throughout the MCMC.

\section{Learning rate scheduler}
\label{sec:scheduler}

\begin{figure}[hbt]
    \centering
    \includegraphics[width=0.49\textwidth]{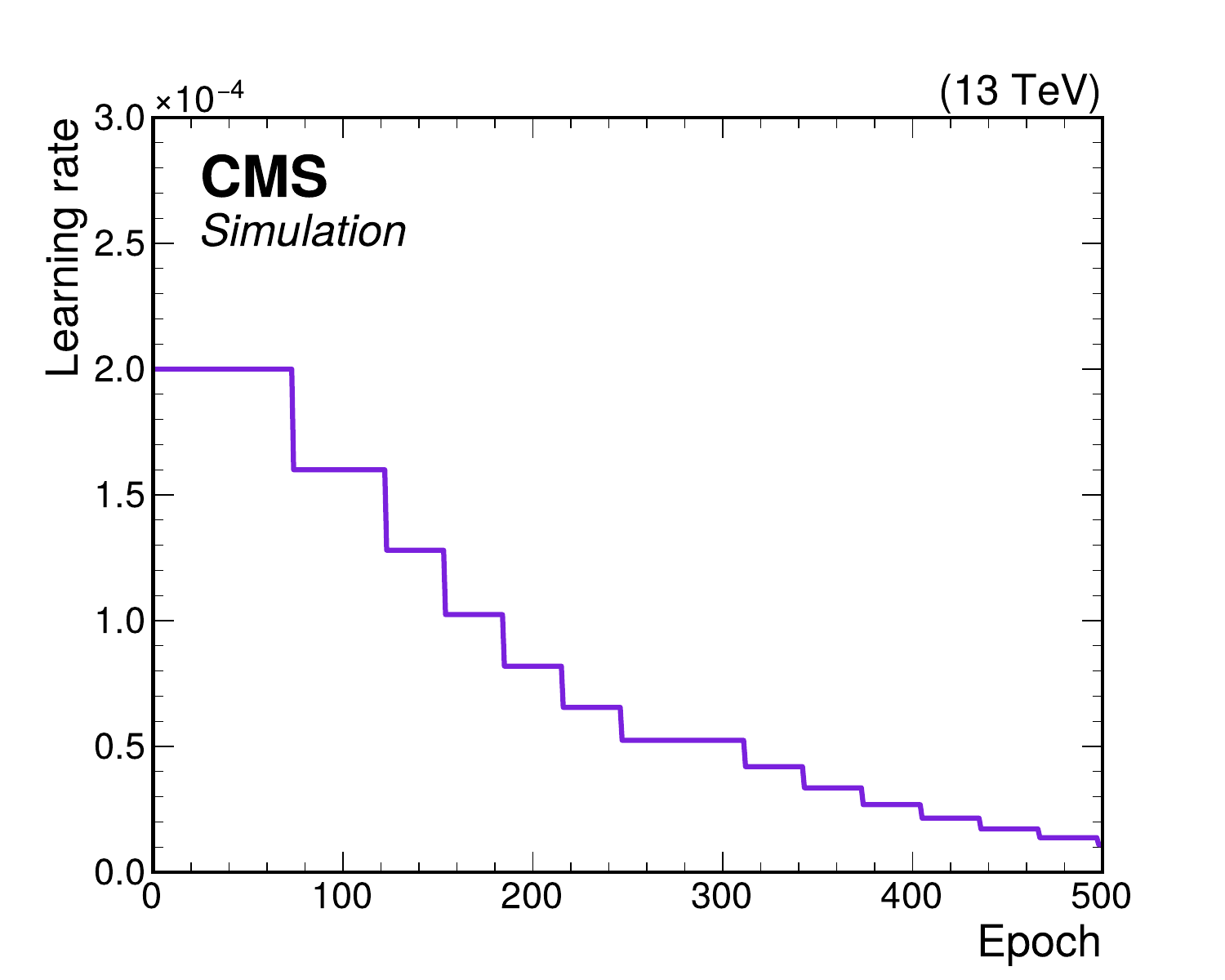}
    \caption{The learning rate during the training of the \wnae from Section~\ref{sec:training_wnae}.}
    \label{fig:lr_evolution}
\end{figure}

The \texttt{torch.optim.lr\_scheduler.ReduceLROnPlateau} method~\cite{Paszke:2019xhz} is used to reduce the learning rate by a factor of 0.8 when the Wasserstein distance does not decrease for 20 epochs. The values for all hyperparameters of the learning rate scheduler can be found in Table~\ref{table:lr_hyperparameters}. The evolution of the learning rate during the training of the \wnae is shown in Fig.~\ref{fig:lr_evolution}.

\begin{table}[htb]
    \centering
    \topcaption{The hyperparameters of the learning rate scheduler.}
    \begin{tabular}{ll}
        Hyperparameter & Value \\ 
        \hline
        Initial learning rate & $2\times 10^{-4}$     \\
        Mode                   & ``min'' \\
        Factor                 & 0.8     \\
        Patience               & 20      \\
        Cooldown               & 10      \\
        Threshold              & 0.001   \\
        Threshold mode         & ``rel'' \\
        Minimum learning rate  & 0       \\
    \end{tabular}
    \label{table:lr_hyperparameters}
\end{table}

\section{Tuning the MCMC}
\label{sec:mcmc_tuning}

Markov chains initialized with persistent contrastive divergence (PCD)~\cite{tielemanPcd2008} are found to accurately sample from the \ptheta distribution. Therefore, PCD is used to initialize the Markov chain for each batch. In PCD, the Markov chain in batch $k$ starts not from a new random set of points, but rather from the final set of points obtained from the Markov chain in batch $k-1$. This procedure can be tuned by adding a restart fraction $r$, meaning that at every batch, a fraction $r$ of the points sampled for the previous batch are discarded and replaced by new, random points, while the remaining points from the previous batch are reused~\cite{NEURIPS2019378a063b}. The restart fraction $r$ is a hyperparameter of the MCMC algorithm and is set to 5\%.

The tuning of the Markov chain hyperparameters---the coefficients of Eq.~\eqref{eq:langevin_dynamics}, the step size $\lambda$ and the noise $\sigma$---is performed as follows. The number of MCMC steps \NMCMC is fixed to 10 as a compromise between training time and sampling accuracy. The MCMC hyperparameters $\lambda$ and $\sigma$ are chosen to satisfy Eq.~\eqref{eq:mcmc_theoretical_relation}, leaving one degree of freedom, and the temperature $T$ was constrained to the interval $(0, 1]$. These parameters are optimized to maximize the AUC for discrimination between SVJs and top quark jets, using Eq.~\eqref{eq:nae_loss_logcosh} as the loss function, with two constraints to mitigate bias toward the signal hypothesis. First, at the maximum AUC, the input feature distributions of the negative and positive samples should match, as discussed in Section~\ref{sec:nae_training}. Second, the Wasserstein distance between the positive and negative samples should be minimal, as discussed in Section~\ref{sec:wassertein_distance_metric}. If the MCMC parameters are correctly set and the network is sufficiently trained, then the network learns $\ptheta = \pdata$, so the AUC is greater than 0.5, assuming the distribution of the signal is not also equal to $\pdata$, in which case no discrimination would be possible. In this case, the distribution of the negative samples is representative of the \pdata distribution. The values of the Markov chain hyperparameters obtained from this optimization are provided in Table~\ref{table:mcmc_hyperparameters}.

Trainings with a considerably larger gradient coefficient $\lambda$ compared to the optimum were found to reach larger AUCs; however, these failed to achieve a good match between the distributions of negative and positive samples. Such MCMC configurations, though appealing if only the AUC is considered, result in an inadequate NAE that has not learned ${\ptheta = \pdata}$: the AUC is large only because of the specific location of the signal hypotheses. This choice would therefore bias the performance of the NAE towards the signal hypotheses used for tuning, spoiling the unsupervised nature of the approach.

\begin{table}[hbt]
    \centering
    \topcaption{The hyperparameters of the MCMC.}
    \begin{tabular}{lll}
        Hyperparameter & Value \\ 
        \hline
        Number of steps \NMCMC & 10     \\
        Restart fraction $r$   & 0.05  \\
        Temperature $T$        & 0.063  \\
        Noise $\sigma$         & 0.22   \\
        Step size $\lambda$    & 0.0242 \\
    \end{tabular}
    \label{table:mcmc_hyperparameters}
	\end{table}
\cleardoublepage \section{The CMS Collaboration \label{app:collab}}\begin{sloppypar}\hyphenpenalty=5000\widowpenalty=500\clubpenalty=5000\input{MLG-24-002-public-authorlist.tex}\end{sloppypar}
%%% END EDITABLE REGION %%%
% skeleton_end
\end{document}

%% file: MLG-24-002-public-authorlist.tex
\cmsinstitute{Yerevan Physics Institute, Yerevan, Armenia}
{\tolerance=6000
A.~Hayrapetyan, V.~Makarenko\cmsorcid{0000-0002-8406-8605}, A.~Tumasyan\cmsAuthorMark{1}\cmsorcid{0009-0000-0684-6742}
\par}
\cmsinstitute{Institut f\"{u}r Hochenergiephysik, Vienna, Austria}
{\tolerance=6000
W.~Adam\cmsorcid{0000-0001-9099-4341}, J.W.~Andrejkovic, L.~Benato\cmsorcid{0000-0001-5135-7489}, T.~Bergauer\cmsorcid{0000-0002-5786-0293}, M.~Dragicevic\cmsorcid{0000-0003-1967-6783}, C.~Giordano, P.S.~Hussain\cmsorcid{0000-0002-4825-5278}, M.~Jeitler\cmsAuthorMark{2}\cmsorcid{0000-0002-5141-9560}, N.~Krammer\cmsorcid{0000-0002-0548-0985}, A.~Li\cmsorcid{0000-0002-4547-116X}, D.~Liko\cmsorcid{0000-0002-3380-473X}, M.~Matthewman, I.~Mikulec\cmsorcid{0000-0003-0385-2746}, J.~Schieck\cmsAuthorMark{2}\cmsorcid{0000-0002-1058-8093}, R.~Sch\"{o}fbeck\cmsAuthorMark{2}\cmsorcid{0000-0002-2332-8784}, D.~Schwarz\cmsorcid{0000-0002-3821-7331}, M.~Shooshtari\cmsorcid{0009-0004-8882-4887}, M.~Sonawane\cmsorcid{0000-0003-0510-7010}, W.~Waltenberger\cmsorcid{0000-0002-6215-7228}, C.-E.~Wulz\cmsAuthorMark{2}\cmsorcid{0000-0001-9226-5812}
% ...
\par}
\vskip\cmsinstskip
\dag:~Deceased\\
$^{1}$Also at Yerevan State University, Yerevan, Armenia\\
$^{2}$Also at TU Wien, Vienna, Austria\\
% ...
